\pdfoutput=1

\documentclass[11pt,twoside,a4paper,cmspaper,final,collab]{cms-tdr}
\def\svnVersion{494195P}\def\cmsCernNoTag{CERN-EP-2018-317}\def\cmsCernDate{\today}\def\cmsMessage{Published in the European Physical Journal C as \href{http://dx.doi.org/10.1140/epjc/s10052-019-6863-8}{\doi{10.1140/epjc/s10052-019-6863-8.}}}

\begin{document}\cmsNoteHeader{TOP-17-001}

\hyphenation{had-ron-i-za-tion}
\hyphenation{cal-or-i-me-ter}
\hyphenation{de-vices}
\RCS$HeadURL: svn+ssh://svn.cern.ch/reps/tdr2/papers/TOP-17-001/trunk/TOP-17-001.tex $
\RCS$Id: TOP-17-001.tex 494183 2019-04-30 15:19:22Z mdefranc $
\newlength\cmsFigWidth
\ifthenelse{\boolean{cms@external}}{\setlength\cmsFigWidth{0.85\columnwidth}}{\setlength\cmsFigWidth{0.4\textwidth}}
\ifthenelse{\boolean{cms@external}}{\providecommand{\cmsLeft}{top\xspace}}{\providecommand{\cmsLeft}{left\xspace}}
\ifthenelse{\boolean{cms@external}}{\providecommand{\cmsRight}{bottom\xspace}}{\providecommand{\cmsRight}{right\xspace}}
\providecommand{\CL}{CL\xspace}

\newcommand{\mumu}{\MM}
\newcommand{\ee}{\EE}
\newcommand{\pp}{\Pp\Pp\xspace}
\newcommand{\emu}{\Pe$^{\pm}$\Pgm$^{\mp}$\xspace}
\newcommand{\tW}{\ensuremath{\cPqt\PW}\xspace}
\newcommand{\Minuit} {{\textsc{Minuit}}\xspace}
\newcommand{\Minos} {{\textsc{Minos}}\xspace}
\newcommand{\xFitter} {{\textsc{xFitter}}\xspace}
\newcommand{\Hathor} {{\textsc{Hathor}}\xspace}
\newcommand{\RunDec} {{\textsc{RunDec}}\xspace}
\newcommand{\Toppp} {{\textsc{Top++}}\xspace}

\newcommand{\Wjets}{\PW+jets\xspace}
\newcommand{\Zjets}{\cPZ+jets\xspace}
\newcommand{\WW}{\PW\PW\xspace}
\newcommand{\WZ}{\PW\cPZ\xspace}
\newcommand{\ZZ}{\cPZ\cPZ\xspace}

\newcommand{\pb}{\unit{pb}}
\newcommand{\mb}{\unit{mb}}
\newcommand{\resultxsecmain}{\ensuremath{803  \pm  2 \stat \pm 25 \syst \pm 20 \lum \pb}\xspace}
\newcommand{\resultxseccheck}{\ensuremath{804 \pm  2 \stat \pm 31 \syst \pm 20 \lum \pb}\xspace}
\newcommand{\resultxsecvismain}{\ensuremath{25.61 \pm 0.05 \stat \pm 0.75 \syst \pm 0.64 \lum \pb }\xspace}
\newcommand{\xsectheo}{\ensuremath{832~^{+20}_{-29}\, (\text{scale}) \pm 35 ~(\text{PDF}+\as) \pb}\xspace}

\newcommand{\resultxsectopmass}{\ensuremath{815 \pm  2 \stat \pm 29 \syst \pm 20 \lum \pb}\xspace}
\newcommand{\resulttopmassMC}{\ensuremath{172.33 \pm  0.14 \stat \,^{+0.66}_{-0.72} \syst \GeV}\xspace}
\newcommand{\resulttopmassMCxcheck}{\ensuremath{171.92 \pm  0.13 \stat \,^{+0.76}_{-0.77} \syst \GeV}\xspace}
\newcommand{\uncorrelatedSyst}{\ensuremath{0.54 \GeV}\xspace}
\newcommand{\correlationMassXsec}{\ensuremath{12\%}\xspace}

\newcommand{\lumiv}{\ensuremath{35.9\fbinv}}

\newcommand{\mz}{\ensuremath{m_\mathrm{\cPZ}}\xspace}
\newcommand{\as}{\ensuremath{\alpS}\xspace}
\newcommand{\asq}{\ensuremath{\as(Q)}\xspace}
\newcommand{\asmz}{\ensuremath{\as(\mz)}\xspace}
\newcommand{\asmin}{\ensuremath{\as^\text{min}}\xspace}
\newcommand{\mur}{\ensuremath{\mu_\mathrm{r}}\xspace}
\newcommand{\muf}{\ensuremath{\mu_\mathrm{f}}\xspace}
\newcommand{\stt}{\ensuremath{\sigma_\ttbar}\xspace}
\newcommand{\sttvis}{\ensuremath{\sigma_{\ttbar}^{\text{vis}}}\xspace}
\newcommand{\stttheo}{\ensuremath{\sigma_{\ttbar}^{\text{theo}}}\xspace}
\newcommand{\msbar}{\ensuremath{\mathrm{\overline{MS}}}\xspace}
\newcommand{\mt}{\ensuremath{m_\mathrm{\cPqt}}\xspace}
\newcommand{\mtmt}{\ensuremath{\mt(\mt)}\xspace}
\newcommand{\mtp}{\ensuremath{\mt^{\text{pole}}}\xspace}
\newcommand{\mtmc}{\ensuremath{\mt^{\mathrm{MC}}}\xspace}
\newcommand{\mlb}{\ensuremath{m_{\ell\cPqb}^{\text{min}}}\xspace}
\newcommand{\chisq}{\ensuremath{\chi^2}\xspace}
\newcommand{\sqrts}{\ensuremath{\sqrt{s}}\xspace}

\cmsNoteHeader{TOP-17-001}

\title{Measurement of the \ttbar production cross section, the top quark mass, and the strong coupling constant using dilepton events in \pp collisions at $\sqrt{s}=13\TeV$}
\titlerunning{Measurement of the \ttbar production cross section, the top quark mass, and the strong coupling constant\ldots}

\date{\today}

\abstract{
A measurement of the top quark-antiquark pair production cross section \stt in proton-proton collisions at a centre-of-mass energy of 13\TeV is presented. The data correspond to an integrated luminosity of \lumiv, recorded by the CMS experiment at the CERN LHC in 2016. Dilepton events (\emu, \mumu, \ee) are selected and the cross section is measured from a likelihood fit. For a top quark mass parameter in the simulation of $ \mtmc = 172.5 \GeV$ the fit yields a measured cross section $\stt = \resultxsecmain$, in agreement with the expectation from the standard model calculation at next-to-next-to-leading order. A simultaneous fit of the cross section and the top quark mass parameter in the \POWHEG simulation is performed. The measured value of $\mtmc = \resulttopmassMC$ is in good agreement with previous measurements. The resulting cross section is used, together with the theoretical prediction, to determine the top quark mass and to extract a value of the strong coupling constant with different sets of parton distribution functions.}

\hypersetup{
pdfauthor={CMS Collaboration},
pdftitle={Measurement of the ttbar production cross section, the top quark mass, and the strong coupling constant using dilepton events in pp collisions at sqrt(s) = 13 TeV},
pdfsubject={CMS},
pdfkeywords={CMS, physics, ttbar, top quark}}

\maketitle

\hyphenation{MINUIT}

\section{Introduction}
\label{sec:introduction}

Measurements of the top quark-antiquark pair cross section \stt in proton-proton (\pp) collisions provide important tests of the standard model (SM). At the CERN LHC, measurements with increasing precision have been performed by the ATLAS and CMS Collaborations in several different decay
channels and at four \pp collision energies~\cite{Aad:2014kva,Aaboud:2016pbd,Sirunyan:2017ule,Khachatryan:2016mqs,Sirunyan:2017uhy}.
Precise theoretical predictions of \stt have been performed in perturbative quantum chromodynamics (QCD) at next-to-next-to-leading
order (NNLO)~\cite{PhysRevLett.109.132001,Czakon:2012zr,Czakon:2012pz,Czakon:2013goa}. The calculations depend on several fundamental parameters: the
top quark mass \mt, the strong coupling constant \as, and the parton distribution functions (PDFs) of the proton.
The measurements of \stt have been used to determine the top
quark pole mass~\cite{Abazov:2011pta,Chatrchyan:2013haa,Aad:2014kva,Khachatryan:2016mqs,Abazov:2016ekt}, \as~\cite{Khachatryan:2016mqs,Klijnsma:2017eqp}, and the PDFs~\cite{Czakon:2013tha,Guzzi:2014wia,Sirunyan:2017azo,Alekhin:2017kpj}.

The value of \mt significantly affects the prediction for many observables, either directly or via radiative corrections.
It is a key input to electroweak precision fits~\cite{Baak:2014ora} and, together with the value of the Higgs boson mass and \as, it has direct implications on the SM predictions for the stability of the electroweak vacuum~\cite{Degrassi:2012ry}. In QCD calculations beyond leading order, \mt depends on the renormalization scheme.
In the context of the \stt predictions, the pole (on-shell) definition for the top quark mass \mtp has wide applications; however, it suffers from the renormalon problem that introduces a theoretical ambiguity in its definition.
The minimal subtraction (\msbar) renormalization scheme has been shown to have a faster convergence than other schemes~\cite{Dowling:2013baa}.
The relation between the
pole and \msbar masses is known to the four-loop level in QCD~\cite{Marquard:2015qpa}. Experimentally, the most precise measurements of the top quark
mass are obtained in so-called direct measurements performed at the Tevatron and LHC~\cite{TevatronElectroweakWorkingGroup:2016lid,Aaboud:2016igd,ATLAS:2014wva,Khachatryan:2015hba}. Except for a few cases such as Ref.~\cite{Chatrchyan:2013boa}, the measurements rely on Monte Carlo (MC) generators to provide
the relation between the top quark mass and an experimental observable.
Current MC generators implement matrix elements at leading or next-to-leading order (NLO), while
higher orders are simulated through parton showering. Studies suggest that the top quark mass parameter \mtmc, as implemented in current MC generators, corresponds to \mtp to an uncertainty on the order of 1\GeV~\cite{Buckley:2011ms,Hoang:2018zrp}. A theoretically well-defined mass can be determined by comparing the measured \ttbar cross section to the fixed-order
theoretical predictions~\cite{Abazov:2011pta,Chatrchyan:2013haa,Aad:2014kva,Khachatryan:2016mqs,Abazov:2016ekt}.

With the exception of the quark masses, \as is the only free parameter in the QCD Lagrangian. While the renormalization
group equation predicts the energy dependence of \as, \ie it gives a functional form for \asq, where $Q$ is the energy scale of the
process, actual values of \as can only be obtained from experimental data. By convention and to facilitate comparisons, \as values measured at
different energy scales are typically evolved to $Q = \mz$, the mass of the \PZ boson. The current world-average value for \asmz is $0.1181 \pm 0.0011$~\cite{PDG2018}. In spite of this relatively precise result, the uncertainty in \as still contributes significantly to many QCD predictions, including cross sections for top quark or Higgs boson production. Very few measurements allow \as to be tested at
high $Q$, and the precision on the world-average value for \asq is driven by low-$Q$ measurements. A determination of \stt was used by the CMS Collaboration to extract the value of \asmz at NNLO for the
first time~\cite{Chatrchyan:2013haa}.
In the prediction for \stt, \as appears not only
in the expression for the parton-parton interaction but also in the QCD evolution of the PDFs. Varying the value of \asmz in the \stt calculation therefore requires a consistent modification of the PDFs.
The full correlation between the gluon PDF, \as, and \mt in the prediction for \stt has to be accounted for.

The analysis uses events in the dileptonic decay channels in which the two \PW\ bosons from the electroweak decays of the two top quarks each produce an electron or a muon, leading to three event categories: \emu, \mumu, and \ee. The data set was recorded by CMS in 2016 at a centre-of-mass energy of 13\TeV, corresponding to an integrated luminosity of \lumiv. The measurement is performed using a maximum-likelihood fit in which the sources of systematic uncertainty are treated as nuisance parameters.
Distributions of observables are chosen as input to the fit so as to further constrain the uncertainties. The fitting procedure largely follows the approach of Ref.~\cite{Khachatryan:2016mqs}.
In this analysis, the number of events is significantly larger than in previous data sets, thus providing tighter constraints. The dominant uncertainties come from the integrated luminosity and the efficiency to identify the two leptons. The correlation between the three decay channels is used to constrain the overall lepton identification uncertainty to that of the better-constrained lepton, which is the muon.

Experimentally, the measured value of \stt has a residual dependence on the value of \mtmc used in the simulation
to estimate the detector efficiency and acceptance. In contrast, the experimental dependence of \stt on the value of \asmz used in the simulation is
negligible~\cite{Chatrchyan:2013haa}. For the extraction of a theoretically well-defined \mt, the dependence of the cross section
on the assumption of a \mtmc value can be reduced by including \mtmc as an additional free parameter in the fit~\cite{Kieseler:2015jzh}.
In this paper, the cross section \stt is first measured for a fixed value of $\mtmc = 172.5 \GeV$, and then determined simultaneously with \mtmc. In the simultaneous fit, input distributions sensitive to the top quark mass are introduced in order to constrain \mtmc. For the measured parameter \mtmc, the same systematic uncertainties are taken into account as in Ref.~\cite{Sirunyan:2018gqx}. Finally, the measured
value of \stt at the experimentally constrained value of \mtmc is used to extract \asmz and \mt in the \msbar scheme, using different PDF sets. For \mt, the pole mass scheme is also considered.

The paper is structured as follows. After a brief description of the CMS experiment and the MC event generators in Section~\ref{sec:detector},
the event selection is presented in Section~\ref{sec:eventselection}. The event categories and the maximum-likelihood fit are explained in Section~\ref{sec:fit}.
The systematic uncertainties in the measurement are discussed in Section~\ref{sec:systematics}. The result of the cross section measurement at a fixed value of $\mtmc = 172.5 \GeV$ is presented in Section~\ref{sec:crosssection}, and the simultaneous measurement of \stt and \mtmc is presented in Section~\ref{sec:simultaneous_fit}.
The extraction of \mt and \as in the \msbar scheme and the top quark pole mass are described in Sections~\ref{sec:alphas} and~\ref{sec:mtp}, respectively, and a summary is given in Section~\ref{sec:conclusions}.

\section{The CMS detector and Monte Carlo simulation}
\label{sec:detector}

The central feature of the CMS apparatus~\cite{Chatrchyan:2008zzk} is a superconducting solenoid of 6\unit{m} internal diameter, providing a magnetic field of 3.8\unit{T}. Within the solenoid volume are a silicon pixel and strip tracker, a lead tungstate crystal electromagnetic calorimeter, and a brass and scintillator hadron calorimeter, each composed of a barrel and two endcap sections. These are used to identify electrons, photons, and jets. Forward calorimeters extend the pseudorapidity coverage provided by the barrel and endcap detectors. Muons are detected in gas-ionization chambers embedded in the steel flux-return yoke outside the solenoid. The detector is nearly hermetic, providing reliable measurement of the momentum imbalance in the plane transverse to the beams. A two-level trigger system selects interesting events for offline analysis~\cite{Khachatryan:2016bia}. A more detailed description of the CMS detector, together with a definition of the coordinate system used and the relevant kinematic variables, can be found in Ref.~\cite{Chatrchyan:2008zzk}.

The \POWHEG~v2~\cite{bib:powheg2,Frixione:2007vw,Nason:2004rx} NLO MC generator is used to simulate \ttbar events~\cite{Frixione:2007nw} and its model dependencies on \mtmc, the PDFs~\cite{Frixione:2007nw}, and the renormalization and factorization scales, $\mur=\muf=\mT=\sqrt{\smash[b]{\mt^2+\pt^2}}$, where \mt is the pole mass and \pt is the transverse momentum of the top quark.
The PDF set NNPDF3.0~\cite{Ball:2012cx} is used to describe the proton structure.
The parton showers are modelled using \PYTHIA~8.2~\cite{Sjostrand:2014zea}
with the CUETP8M2T4 underlying event (UE) tune~\cite{CMS-PAS-TOP-16-021,Skands:2014pea}.
In this analysis, \ttbar events are split into a signal and a background component. The signal consists of dilepton events and includes contributions from leptonically decaying \Pgt\xspace leptons. All other \ttbar events are considered as background.

{\tolerance=1700
Contributions to the background include single top quark processes (\tW), Drell--Yan (DY) events ($\cPZ/\cPgg^*$+jets), and \Wjets production, as well as diboson (VV) events (including \WW, \WZ, and \ZZ) with multiple jets, while the contribution from QCD multijet production is found to be negligible.
The DY and \tW processes are simulated in \POWHEG~v2~\cite{Kardos:2014dua,Re:2010bp,Alioli:2010qp} with the NNPDF3.0 PDF and interfaced to \PYTHIA~8.202 with the UE tune CUETP8M2T4~\cite{Khachatryan:2015pea} for hadronization and fragmentation.
The \Wjets events are generated at NLO using \MGvATNLO~2.2.2~\cite{Alwall:2014hca,Frixione:2002ik} with the NNPDF3.0 PDF and \PYTHIA~8.2
with the UE tune CUETP8M1. Events with \WW, \WZ, and \ZZ diboson processes are generated at leading order using \PYTHIA~8.2 with the NNPDF2.3 PDF and the CUETP8M1 tune. \par
}

To model the effect of additional \pp interactions within the same or nearby bunch crossing (pileup), simulated minimum bias interactions are added to the simulated data.
Events in the simulation are then weighted to reproduce the pileup distribution in the data, which is estimated from the measured bunch-to-bunch instantaneous
luminosity, assuming a total inelastic \pp cross section of 69.2\mb~\cite{Aaboud:2016mmw}.

{\tolerance=900
For comparison with the measured distributions, the event yields in the simulated samples are normalized to their cross section predictions. These are obtained from calculations at NNLO (for \Wjets and $\cPZ/\cPgg^*$+jets~\cite{Gavin:2010az}), NLO plus next-to-next-to-leading logarithms (NNLL) (for \tW production~\cite{bib:twchan}), and NLO (for diboson processes~\cite{bib:mcfm:diboson}).
For the simulated \ttbar sample, the full NNLO+NNLL calculation, performed with the \Toppp~2.0 program, is used~\cite{Czakon:2011xx}. The proton structure is described by the CT14nnlo~\cite{Dulat:2015mca} PDF set, where the PDF and \as uncertainties are estimated using the prescription by the authors. These are added in quadrature to the uncertainties originating from the scale variation $\mt/2<\mur, \muf<2\mt$. The cross section prediction is $\stttheo = \xsectheo$, assuming a top quark pole mass of 172.5\GeV. \par
}

\section{Event selection}
\label{sec:eventselection}

Events with at least two leptons (electron or muon) of opposite charge are selected.
In events with more than two leptons, the two leptons of opposite charge with the highest \pt are used. An event sample of three mutually exclusive event categories \emu, \mumu, and \ee is
obtained.

A combination of single and dilepton triggers is used to collect the events. Each event is required to pass
at least one of the triggers described below. Events in the \emu channel are required to contain either one electron with $\pt > 12\GeV$ and one muon with $\pt > 23\GeV$, or one electron with $\pt> 23\GeV$ and one muon with $\pt > 8\GeV$. Events in the same-flavour channels are required to have $\pt > 23\,(17)\GeV$ for the electron (muon) with the higher \pt, referred to in the following as the leading lepton, and $\pt > 12\,(8) \GeV$ for the other electron (muon), referred to as the subleading lepton. For all channels, single-lepton triggers with one electron (muon) with $\pt > 27\,(24)\GeV$\, are also used.

The particle-flow (PF) algorithm aims to reconstruct and identify each individual particle in an event, and to form PF candidates by combining information from the various components of the CMS detector~\cite{Sirunyan:2017ulk}. The reconstructed vertex with the largest value of summed physics-object $\pt^2$ is taken to be the primary \pp interaction vertex.

Electron and muon candidates are identified through their specific signatures in the detector~\cite{Khachatryan:2015hwa,Sirunyan:2018fpa}. Lepton candidates are required to have $\pt > 25\,(20)\GeV$ for the leading (subleading) lepton, in the range $\abs{\eta} < 2.4$. Electron candidates in the transition region between the barrel and endcap
calorimeters, corresponding to $1.4442 < \abs{\eta}< 1.5660$, are rejected because the reconstruction of electrons in this region is not optimal.

Lepton isolation requirements are based on the ratio of the scalar sum of the \pt of neighbouring PF candidates to the \pt of the lepton candidate, which is referred to as the lepton isolation variable. These PF candidates are the ones falling within a cone of size $\DR=0.3 \,(0.4)$ for electrons (muons),
centred on the lepton direction, excluding the contribution from the lepton candidate itself. The cone size \DR\xspace is defined as the square root of the quadrature sum of the differences in the azimuthal angle and pseudorapidity. The value of the isolation variable is required to be smaller than 6\% for electrons and 15\% for muons. Events with dilepton invariant mass $m_{\ell\ell} < 20\GeV$ ($\ell=\Pe,\Pgm$) are rejected to suppress backgrounds due to QCD multijet production and decays of low mass resonances. Additionally, leptons are required to be consistent with originating from the primary interaction vertex.

Jets are reconstructed from the PF candidates using the anti-\kt clustering algorithm with a distance parameter of 0.4~\cite{Cacciari:2008gp, Cacciari:2011ma}. The jet momentum is determined from the vectorial sum of all particle momenta in the jet, and is found from simulation to be within 5 to 10\% of the true momentum over the relevant phase space of this analysis~\cite{Khachatryan:2016kdb}.
Pileup interactions can contribute additional tracks and calorimetric energy depositions to the jet momentum. To mitigate this effect, charged particles identified as originating from pileup vertices are discarded and an offset correction is applied to correct for remaining contributions~\cite{Khachatryan:2016kdb}. The jet energy corrections are determined from measurements of the energy balance in dijet, multijet, photon+jet, and leptonically decaying \Zjets events, and are applied as a function of the jet \pt and $\eta$ to both data and simulated events~\cite{Khachatryan:2016kdb}. For this measurement, jets are selected if they fulfill the criteria $\pt> 30\GeV$ and $\abs{\eta}< 2.4$.

Jets originating from the hadronization of \cPqb~quarks (\cPqb~jets) are identified (\cPqb~tagged) using the combined secondary vertex~\cite{Sirunyan:2017ezt} algorithm, which combines lifetime information from tracks and secondary vertices. To achieve high purity, a working point is chosen such that the fraction of light-flavour jets with $\pt > 30 \GeV$ that are falsely identified as \cPqb~jets is $0.1\%$, resulting in an average efficiency of about 41\% for genuine \cPqb~jets and 2.2\% for \cPqc\ jets~\cite{Sirunyan:2017ezt}.

In the same-flavour channels, \mumu and \ee, DY events are suppressed by excluding the region of the \PZ boson mass through the requirement $76 < m_{\ell\ell} < 106\GeV$. In these channels, events are also required to contain at least one \cPqb-tagged jet.

\begin{figure*}[htb!]
  \centering
    \includegraphics[width=0.49\textwidth]{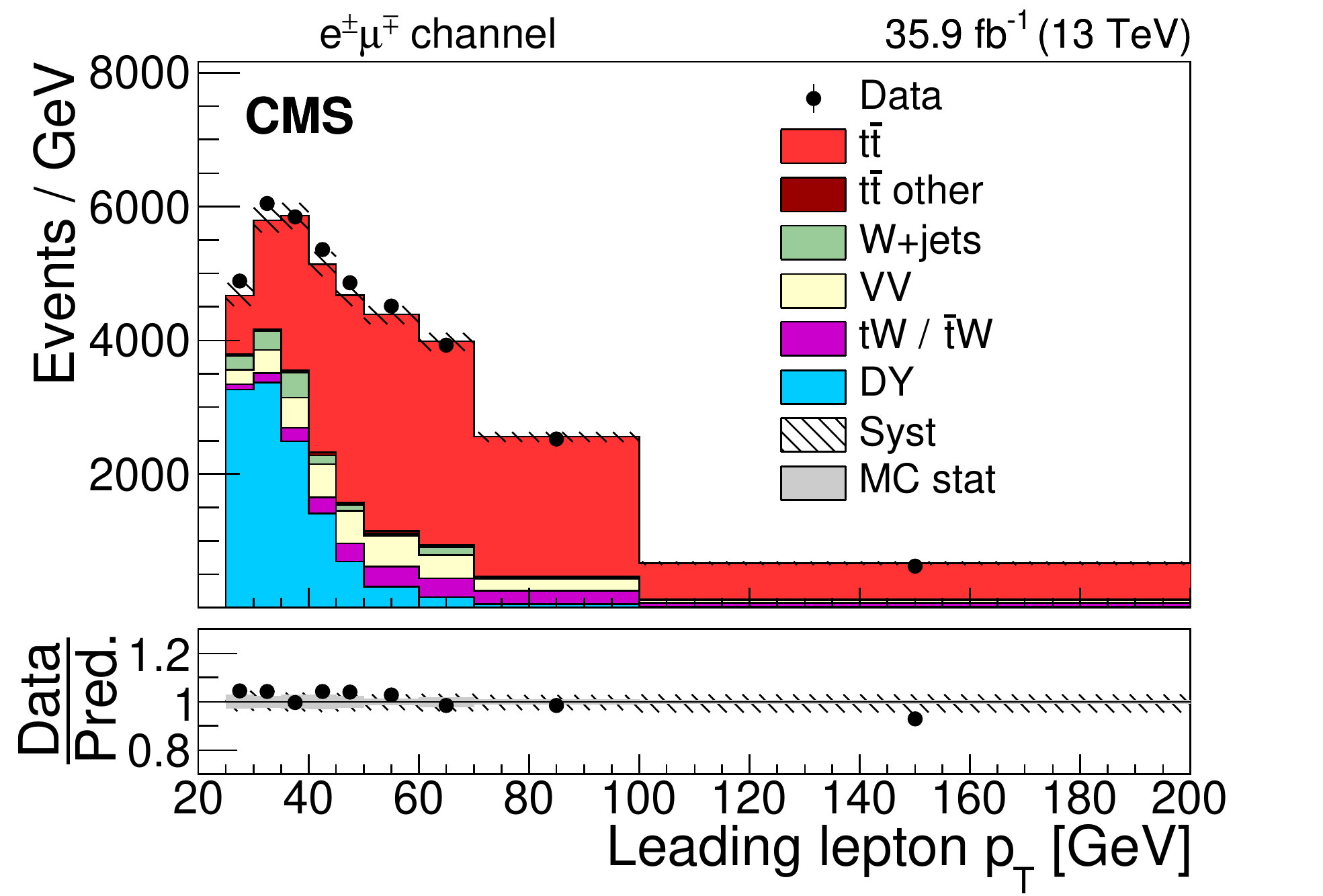}
    \includegraphics[width=0.49\textwidth]{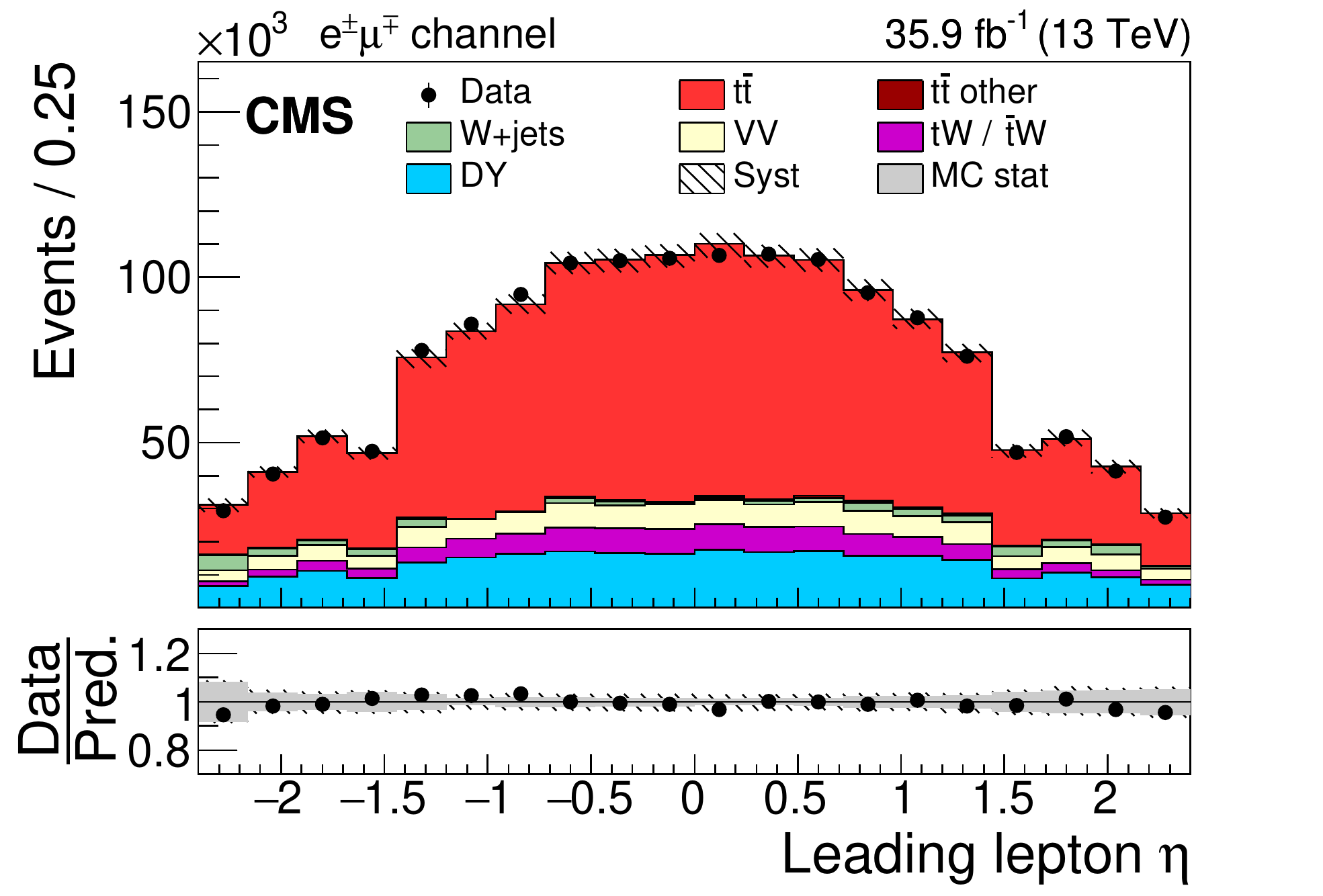}
    \includegraphics[width=0.49\textwidth]{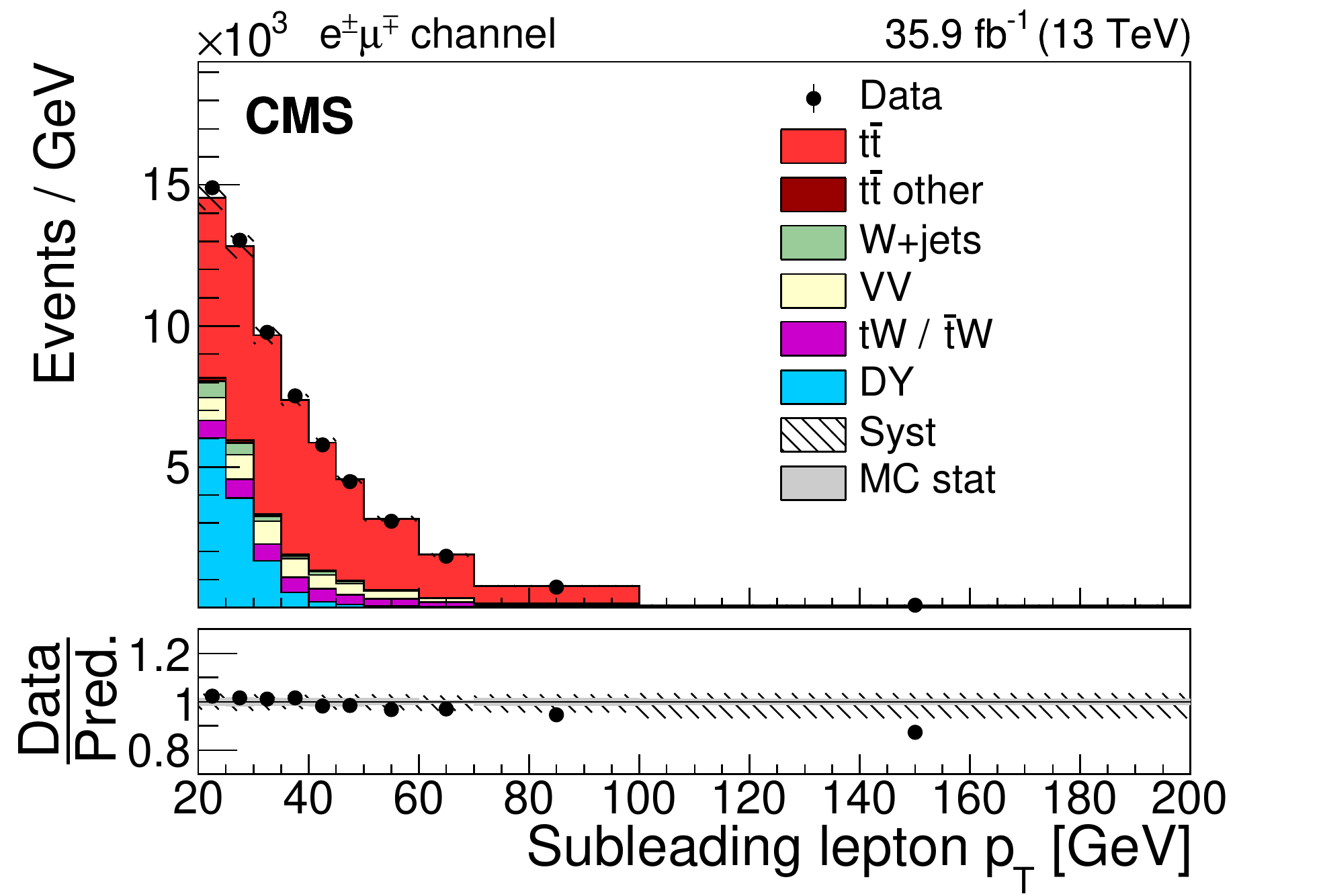}
    \includegraphics[width=0.49\textwidth]{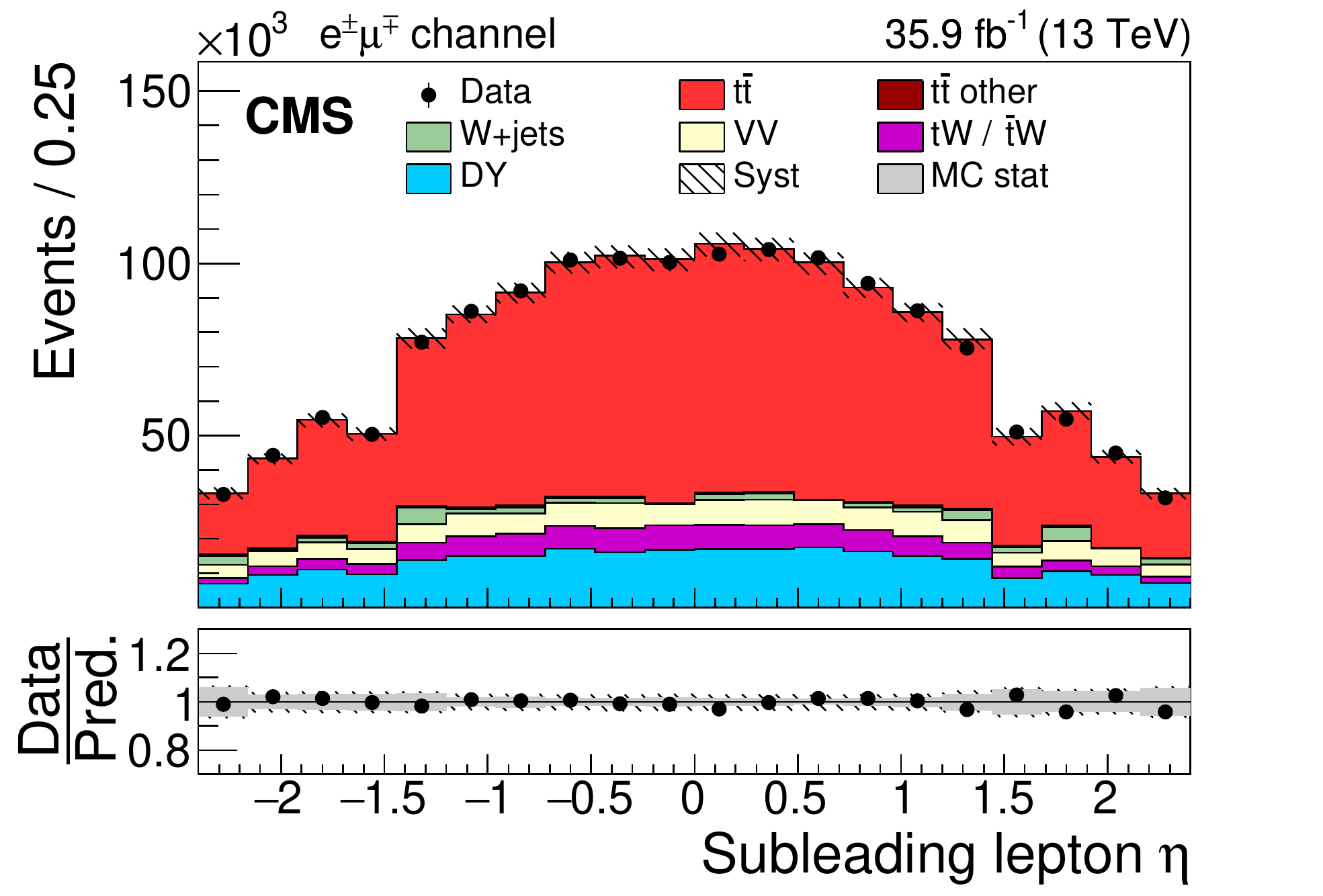}
    \includegraphics[width=0.49\textwidth]{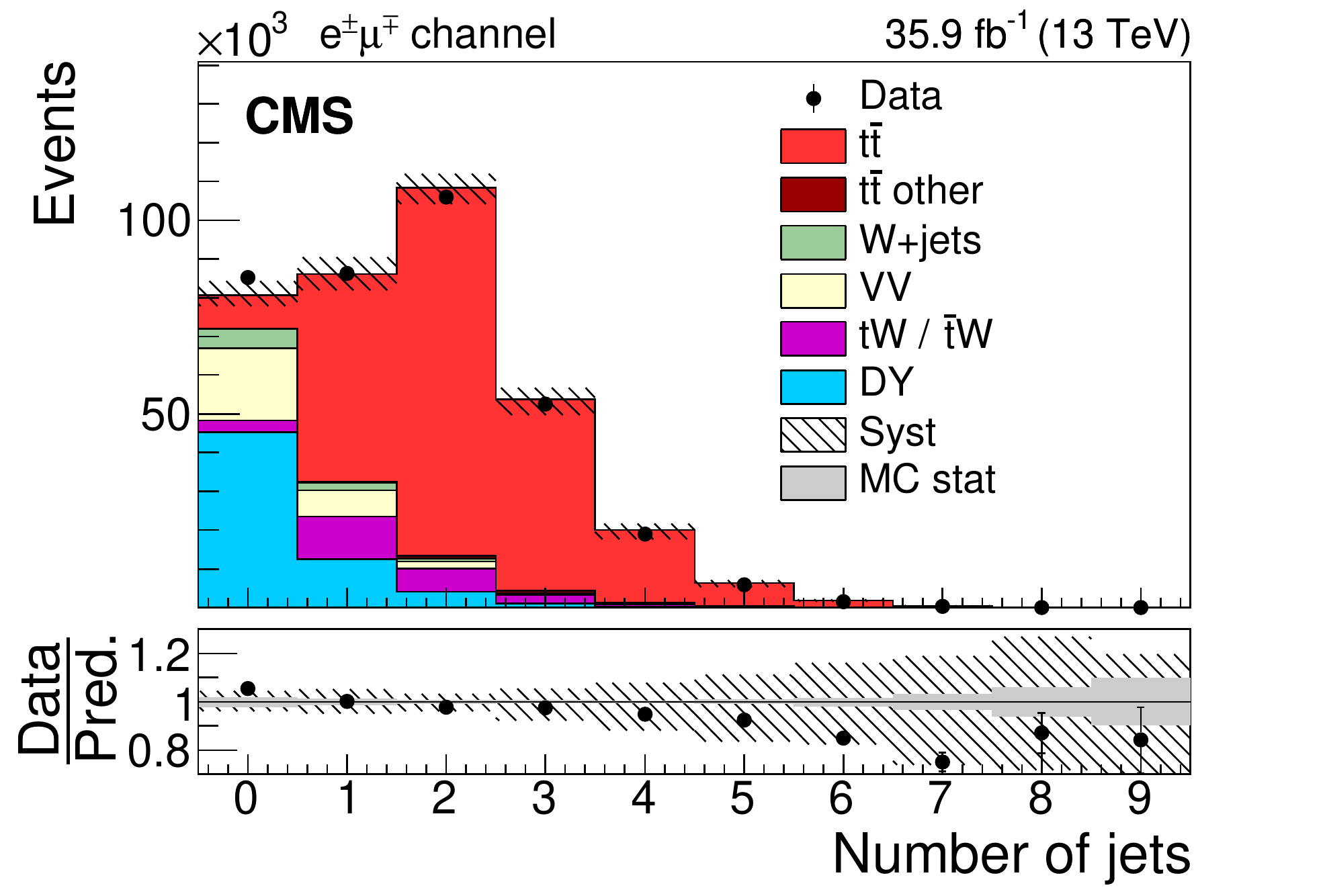}
    \includegraphics[width=0.49\textwidth]{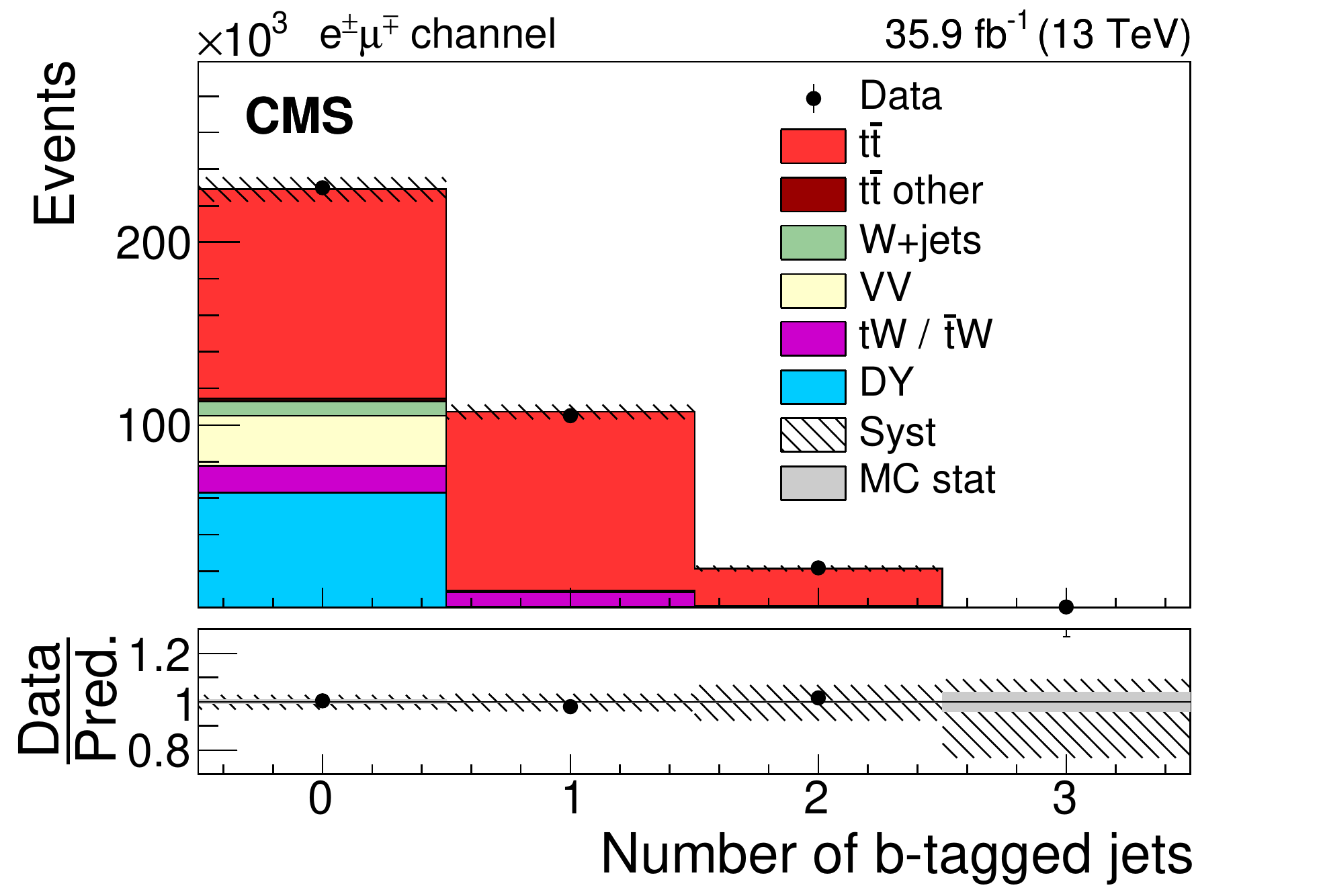}
      \caption{Distributions of the transverse momentum (left) and pseudorapidity (right)
        of the leading (upper) and subleading (middle) leptons in the \emu channel after the
        event selection for the data (points) and the predictions for the signal and various backgrounds from the simulation (shaded histograms).
        The lower row shows the jet (left) and \cPqb-tagged jet (right) multiplicity distributions. The vertical bars on the points represent the statistical uncertainties in the data.
        The hatched bands correspond to the systematic uncertainty in the \ttbar signal MC simulation.
        The uncertainties in the integrated luminosity and background contributions are not included.
        The ratios of the data to the sum of the predicted yields are
        shown in the lower panel of each figure. Here, the solid gray band
        represents the contribution of the statistical uncertainty in the MC simulation.}
       \label{fig:lh_emu_ctrplots}
\end{figure*}

\begin{figure*}[htbp!]
  \centering
    \includegraphics[width=0.49\textwidth]{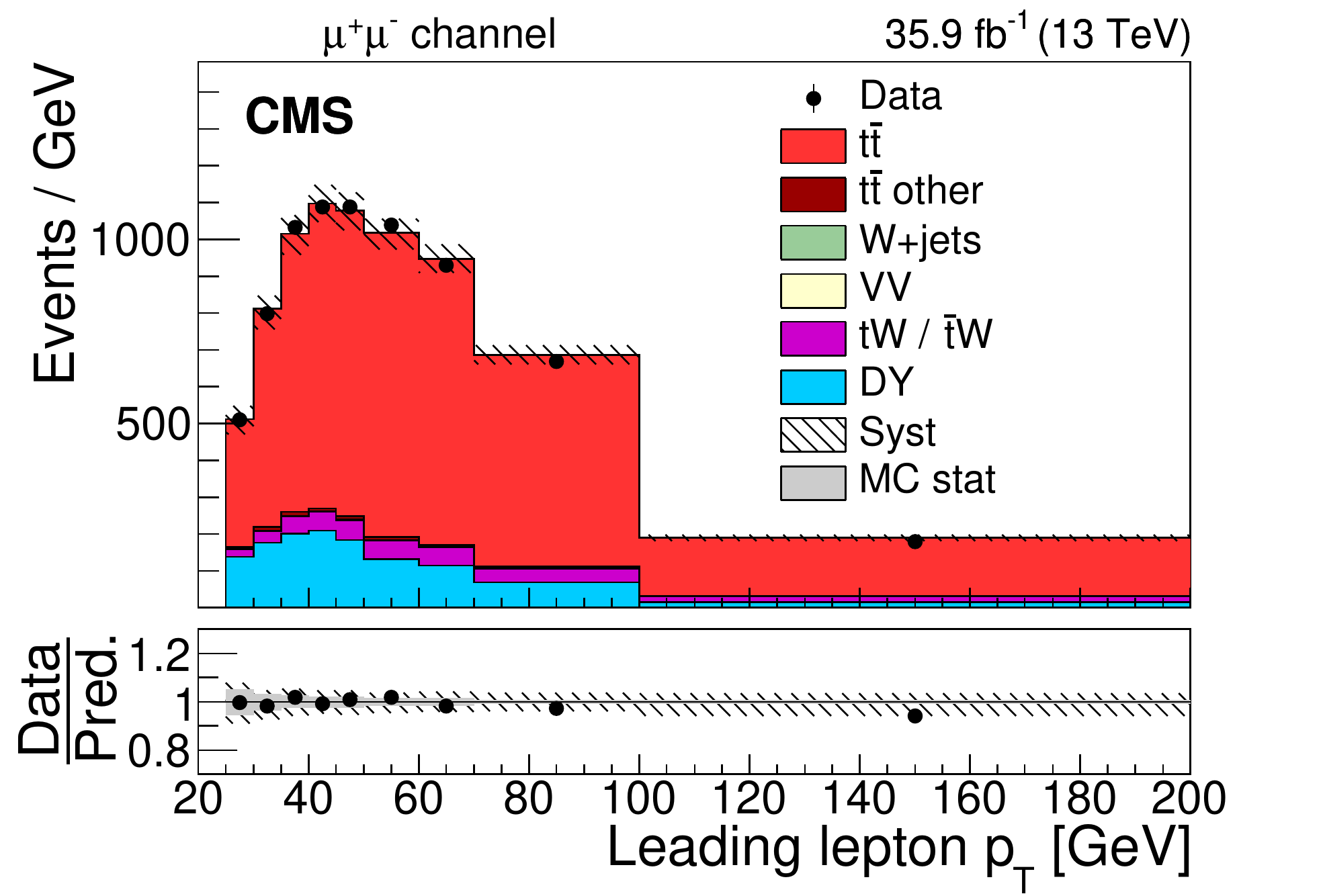}
    \includegraphics[width=0.49\textwidth]{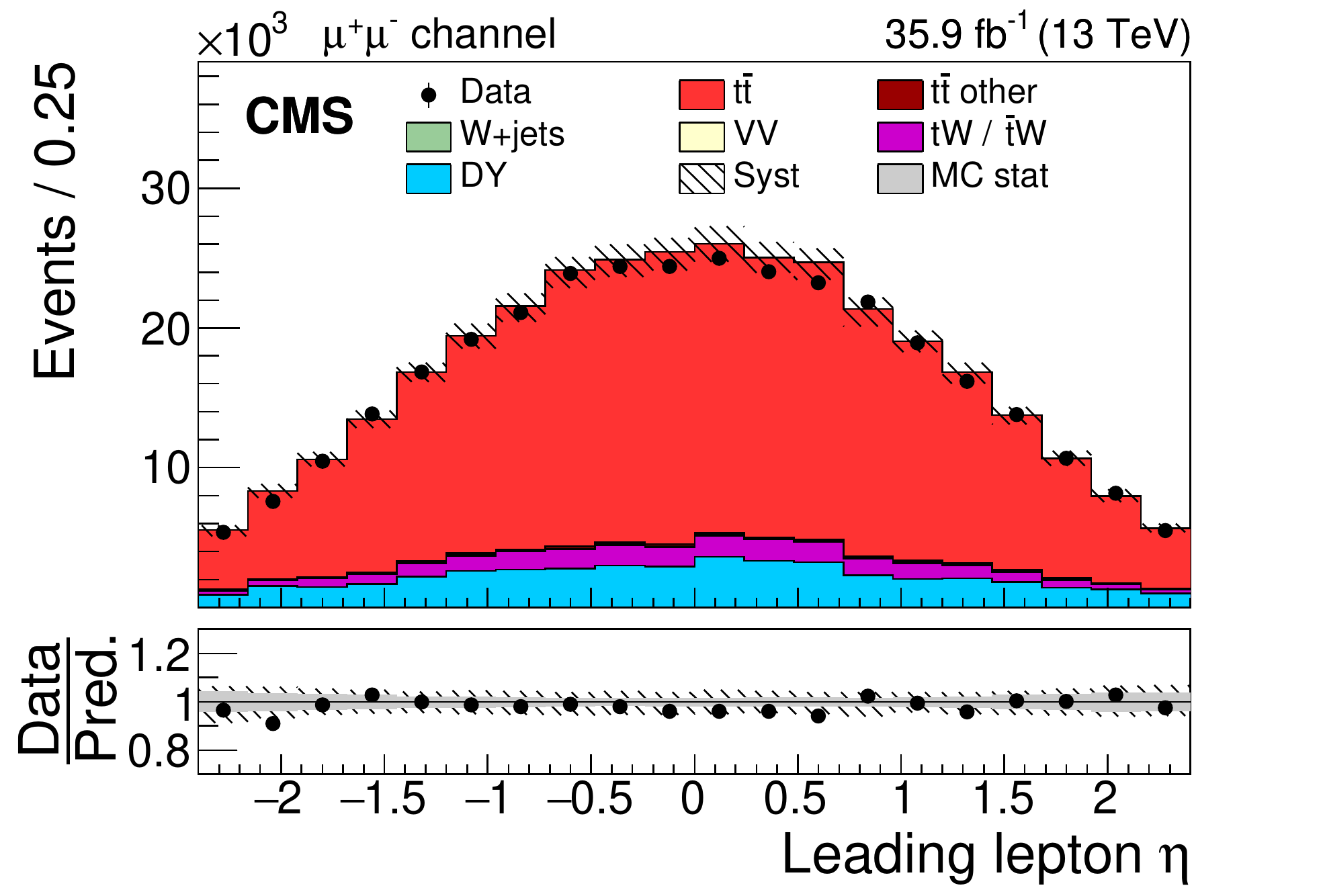}
    \includegraphics[width=0.49\textwidth]{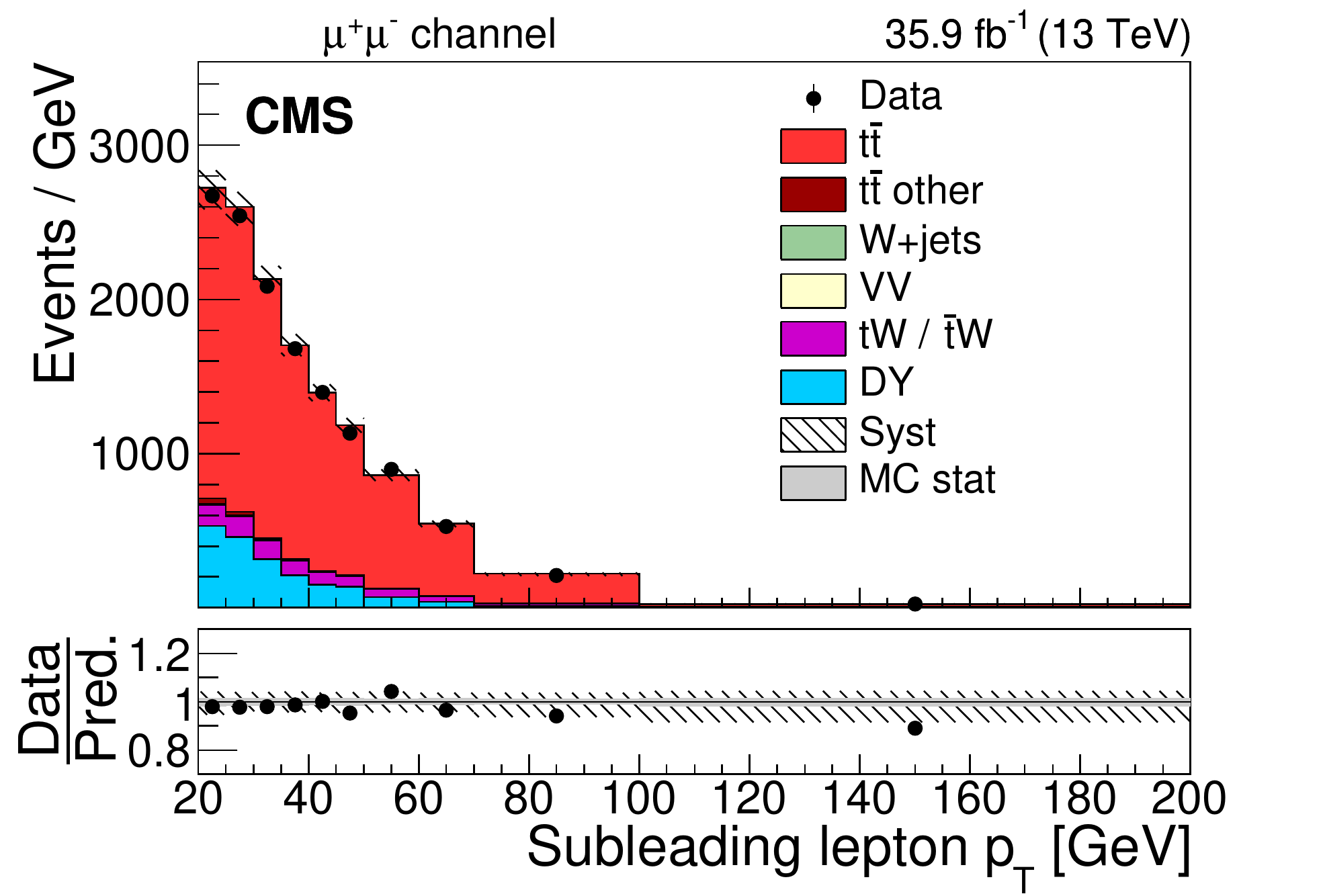}
    \includegraphics[width=0.49\textwidth]{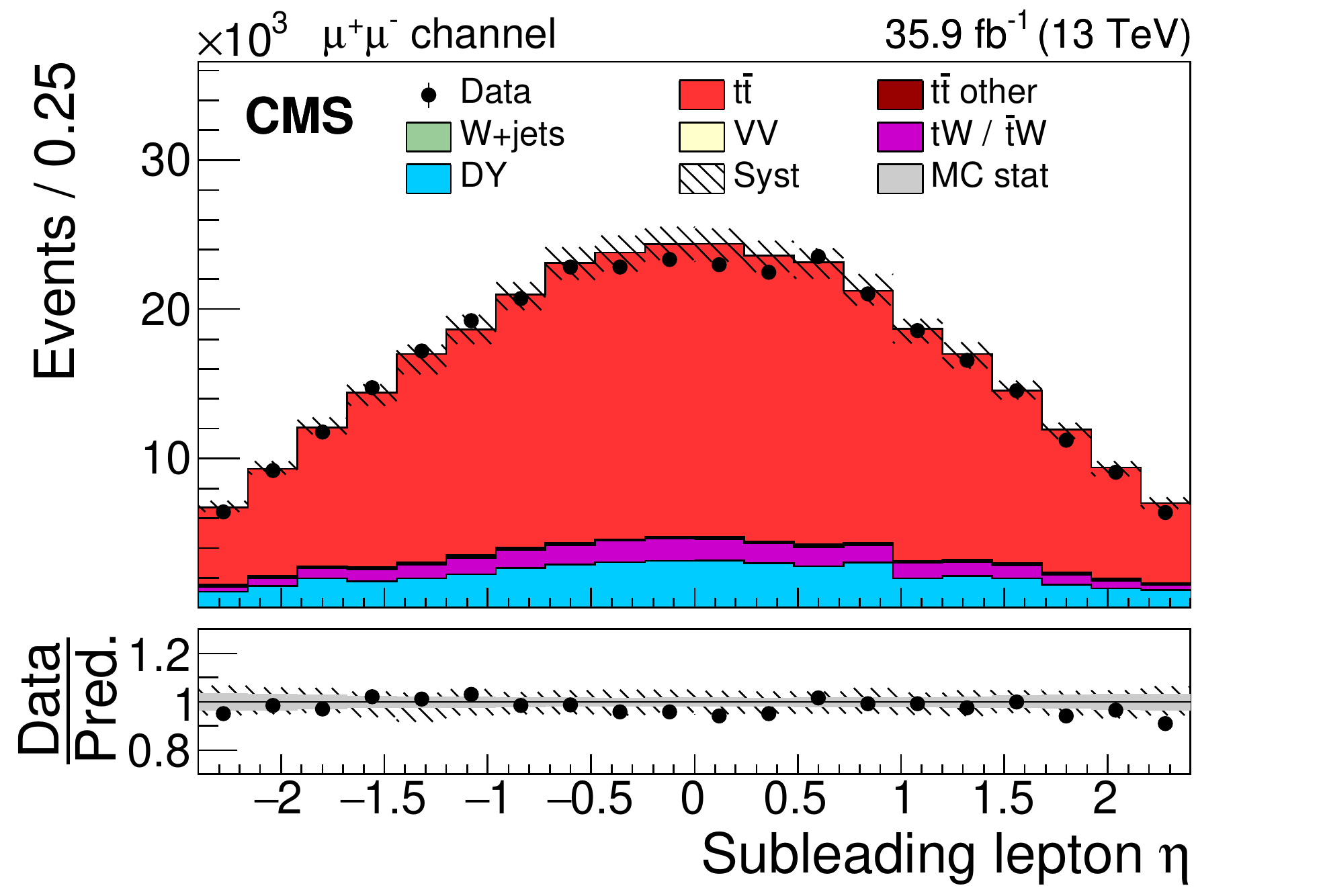}
    \includegraphics[width=0.49\textwidth]{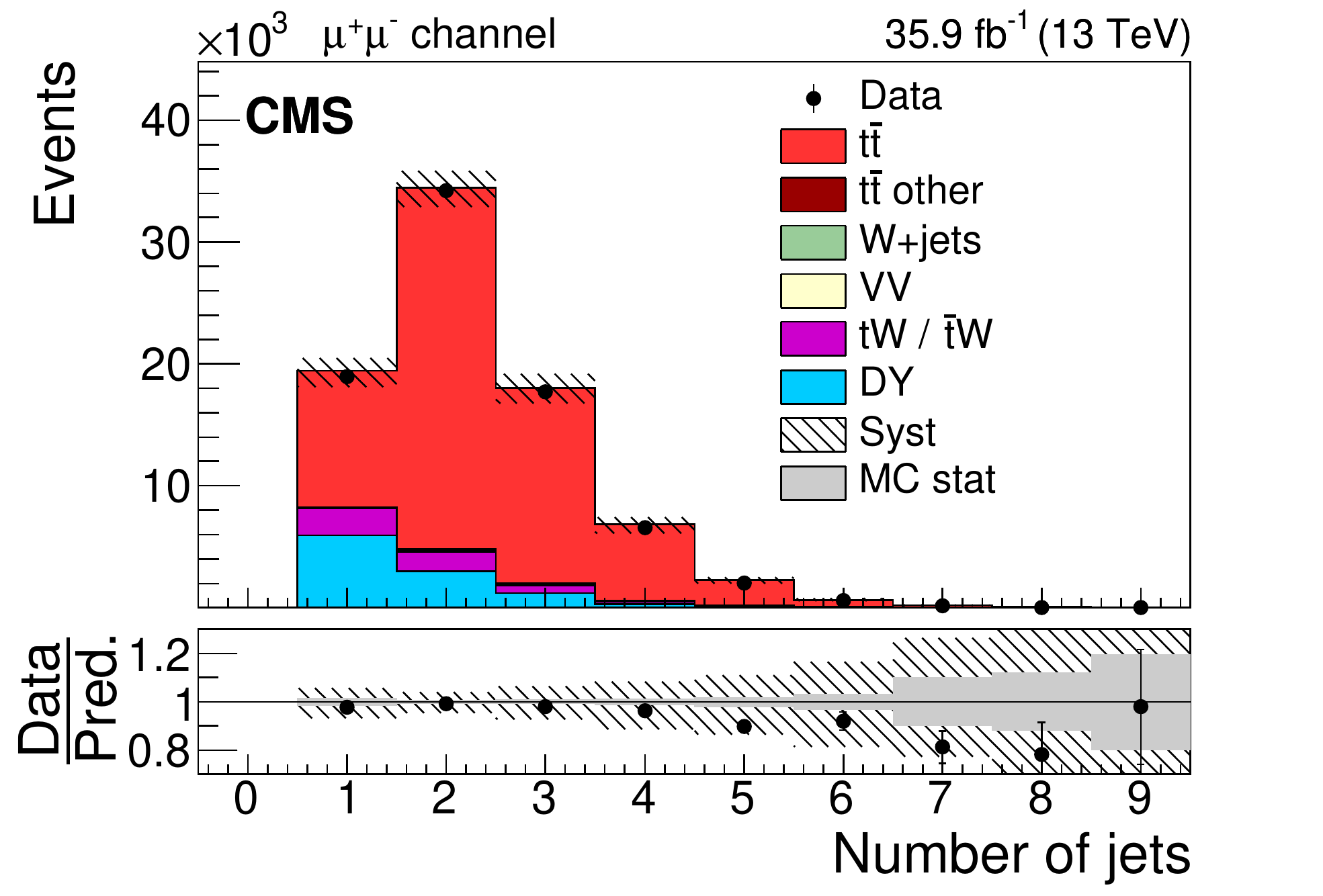}
    \includegraphics[width=0.49\textwidth]{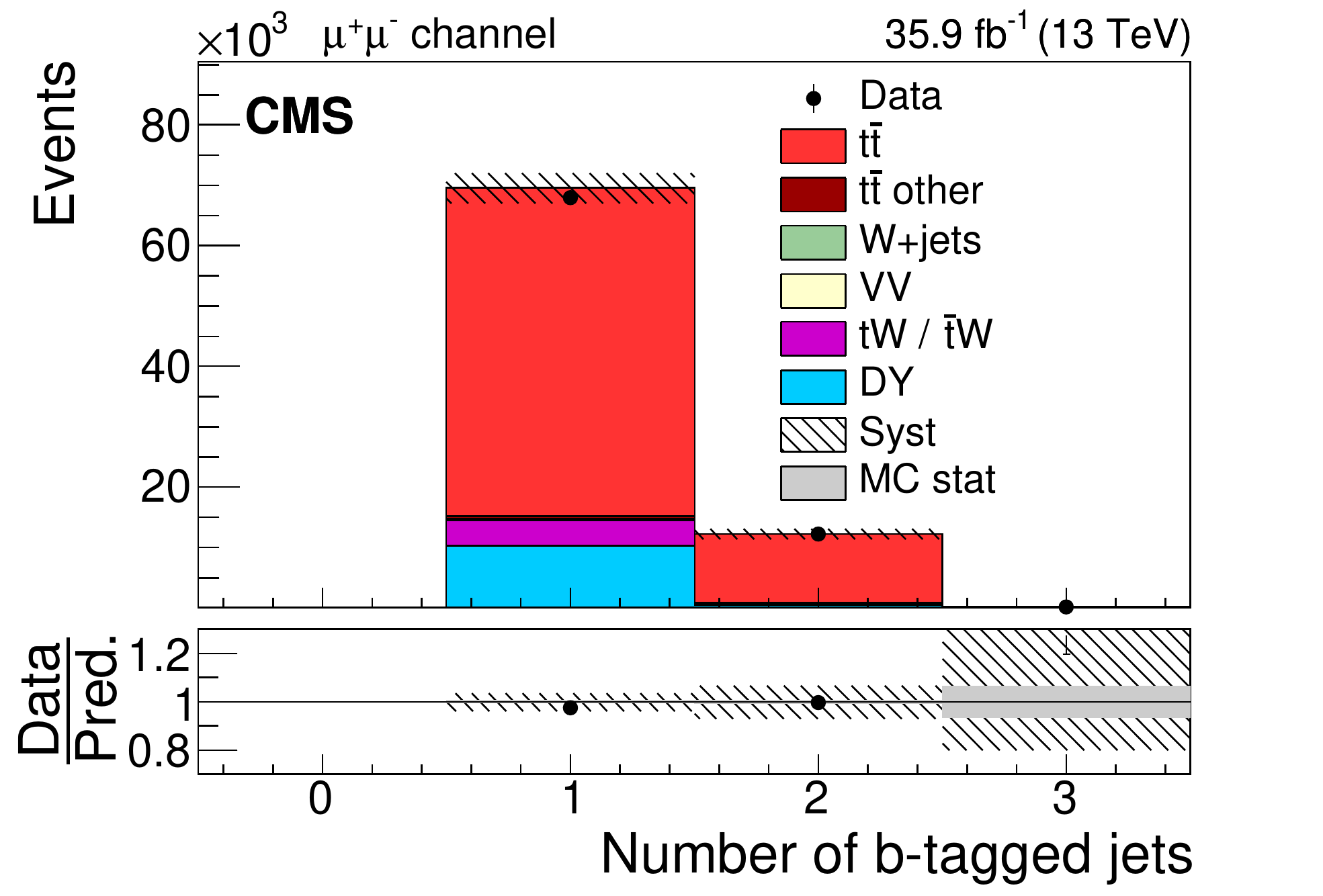}
      \caption{The same distributions as in Fig.~\ref{fig:lh_emu_ctrplots}, but for the \mumu channel.}
       \label{fig:lh_mumu_ctrplots}
\end{figure*}

\begin{figure*}[htbp!]
  \centering
    \includegraphics[width=0.49\textwidth]{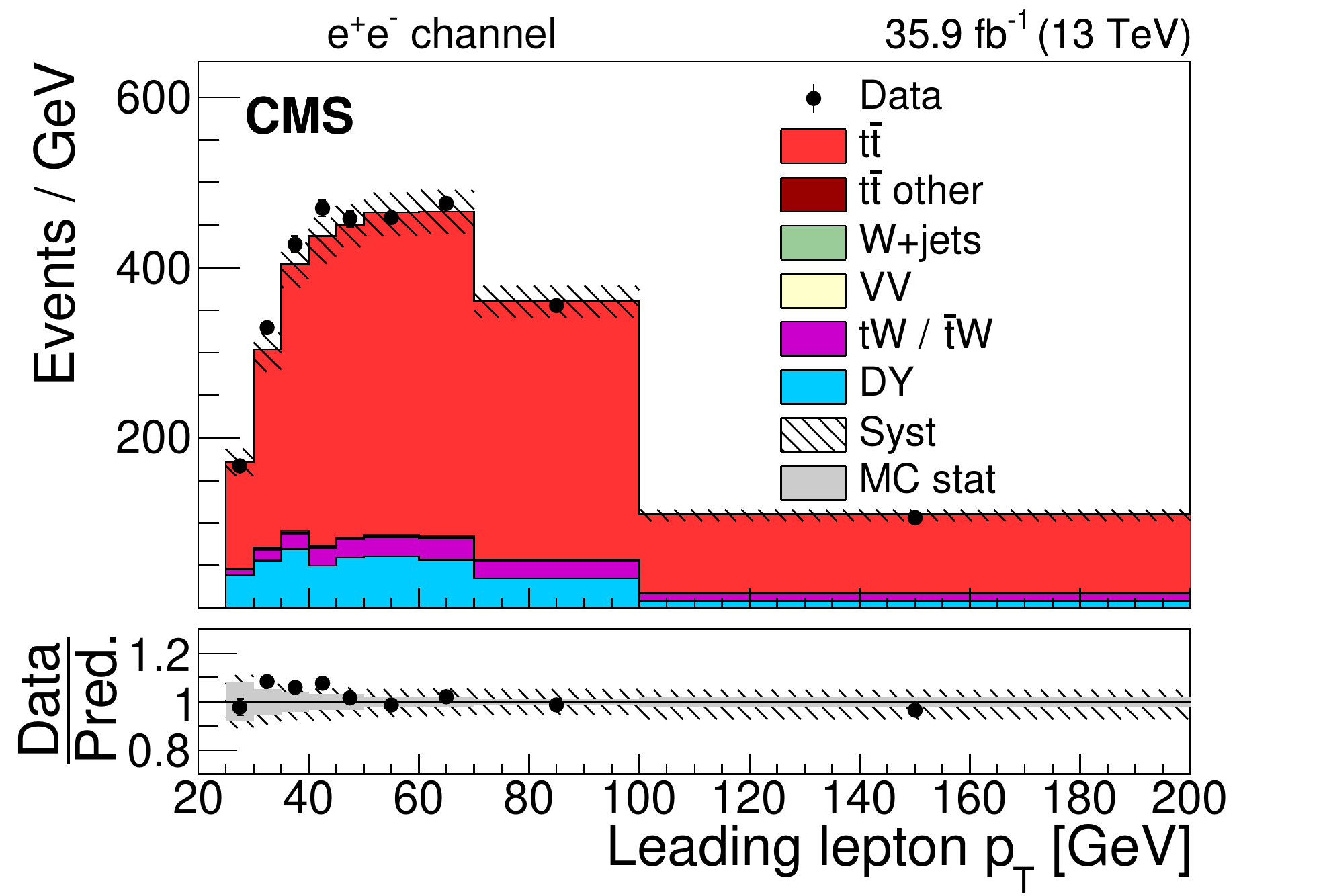}
    \includegraphics[width=0.49\textwidth]{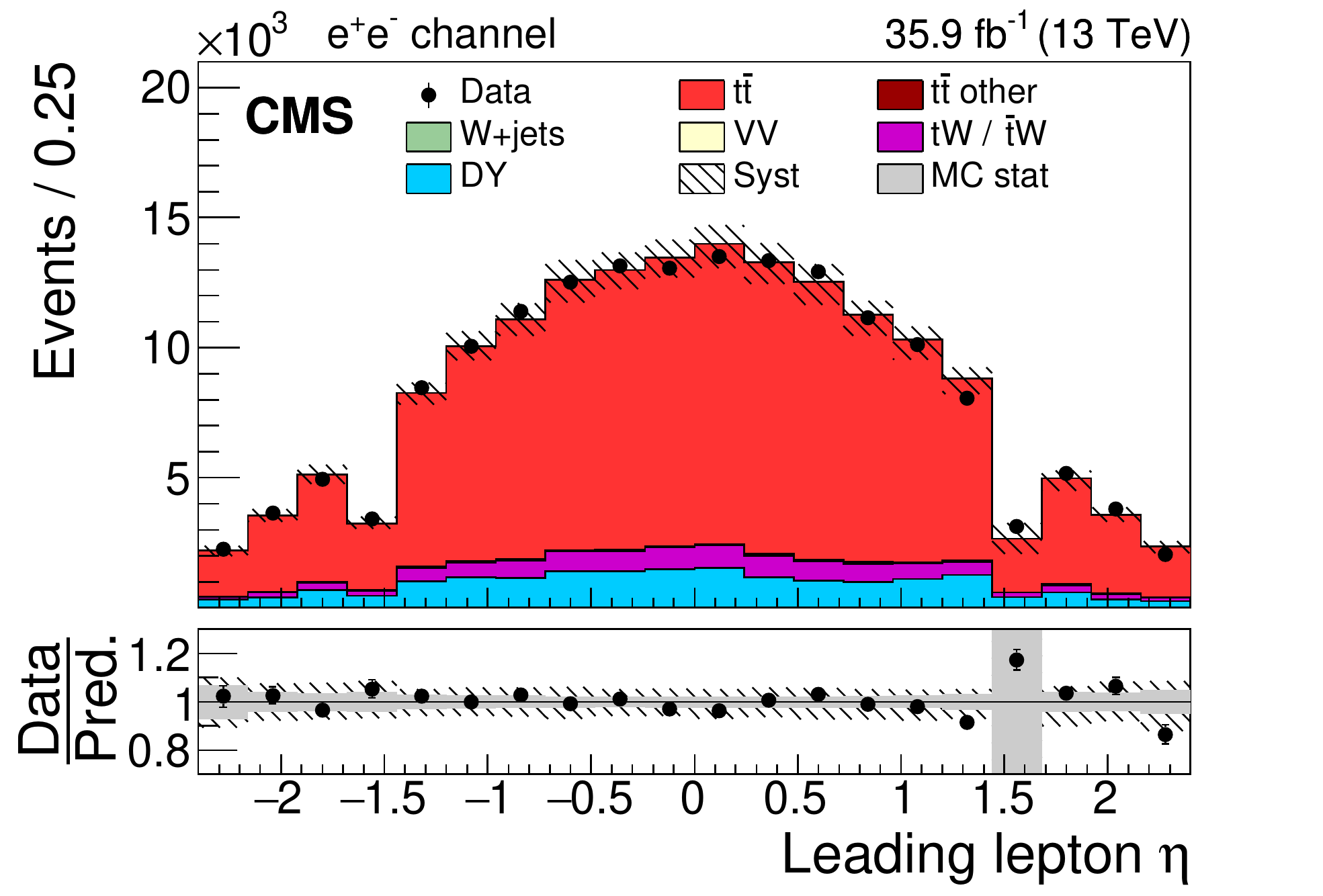}
    \includegraphics[width=0.49\textwidth]{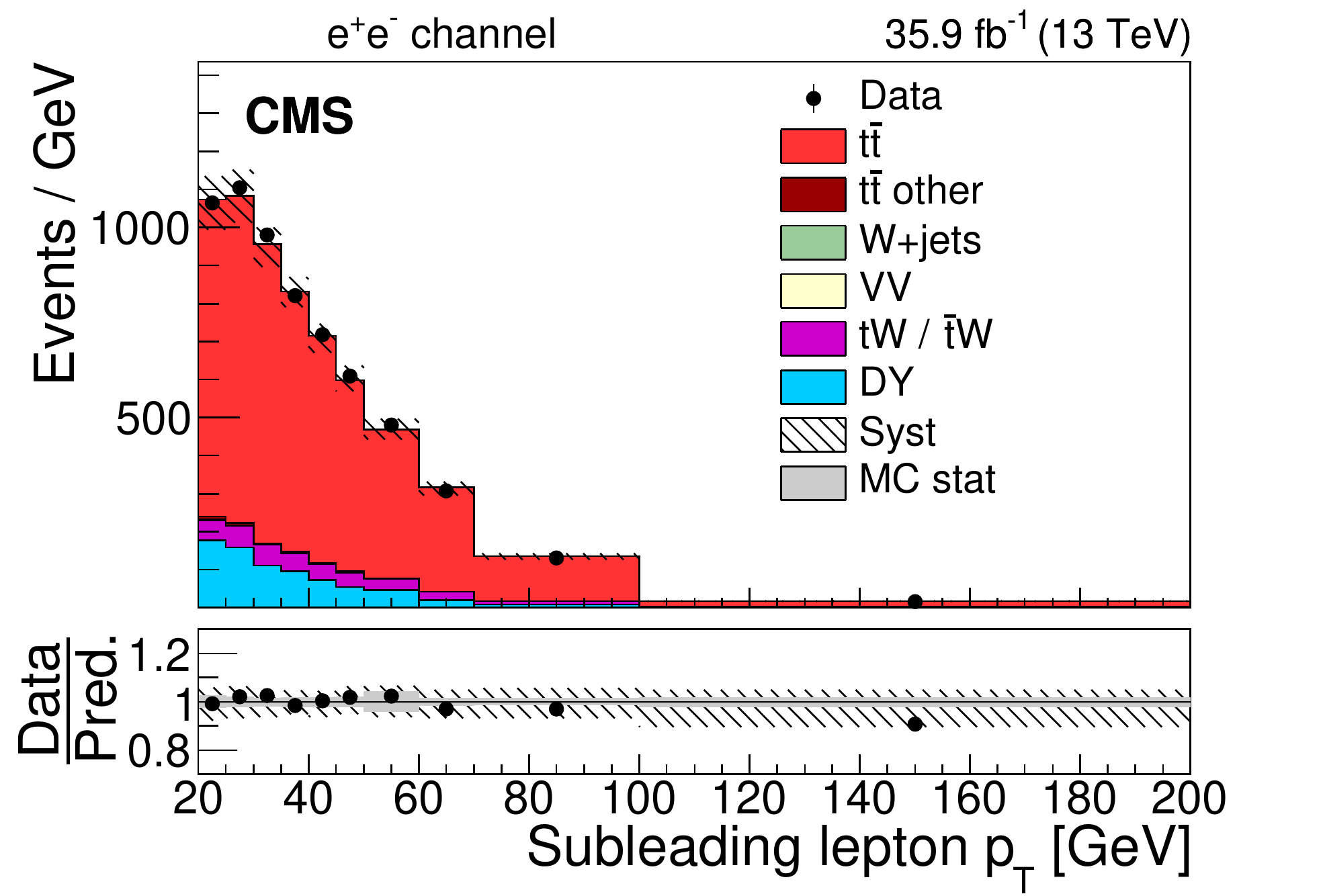}
    \includegraphics[width=0.49\textwidth]{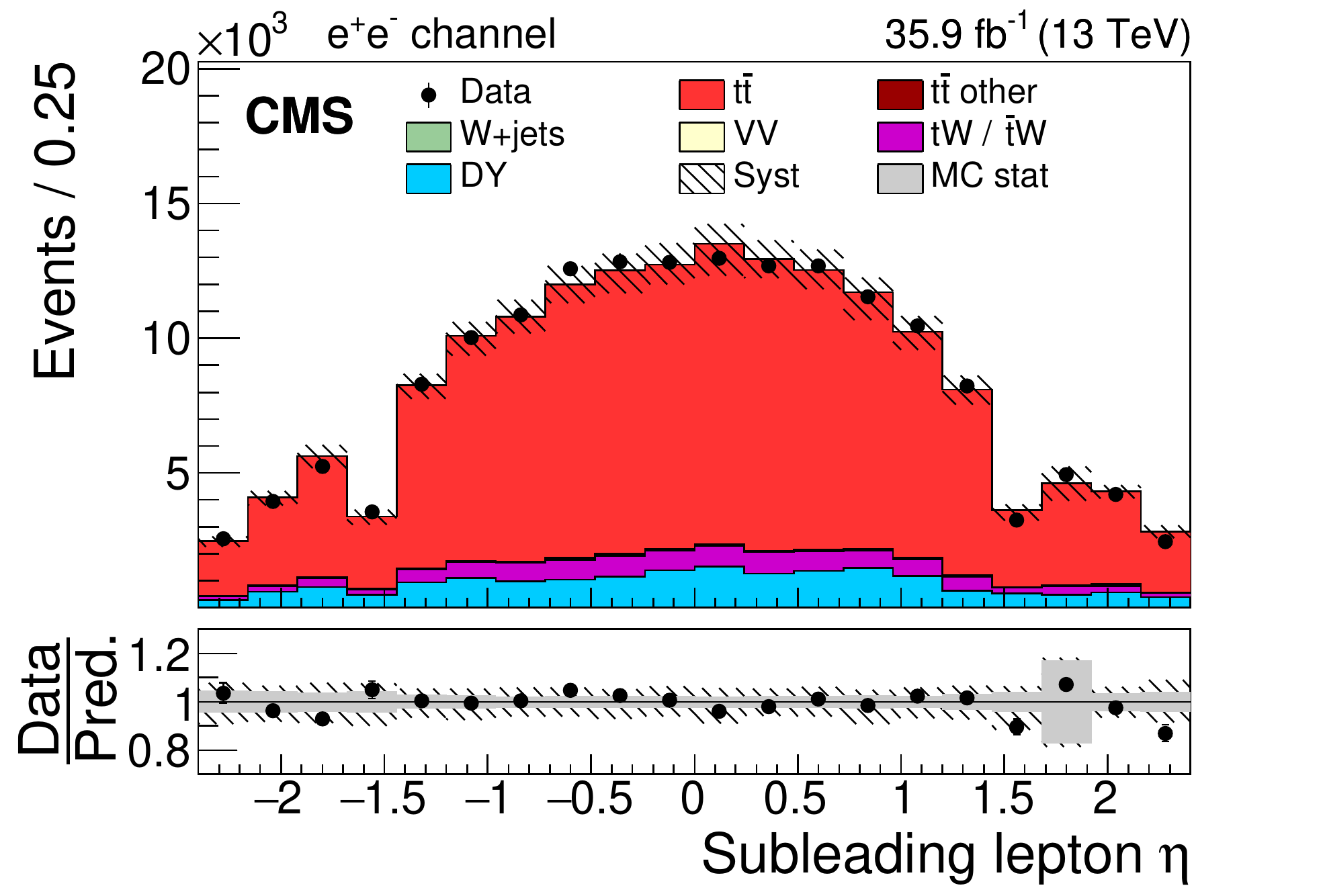}
    \includegraphics[width=0.49\textwidth]{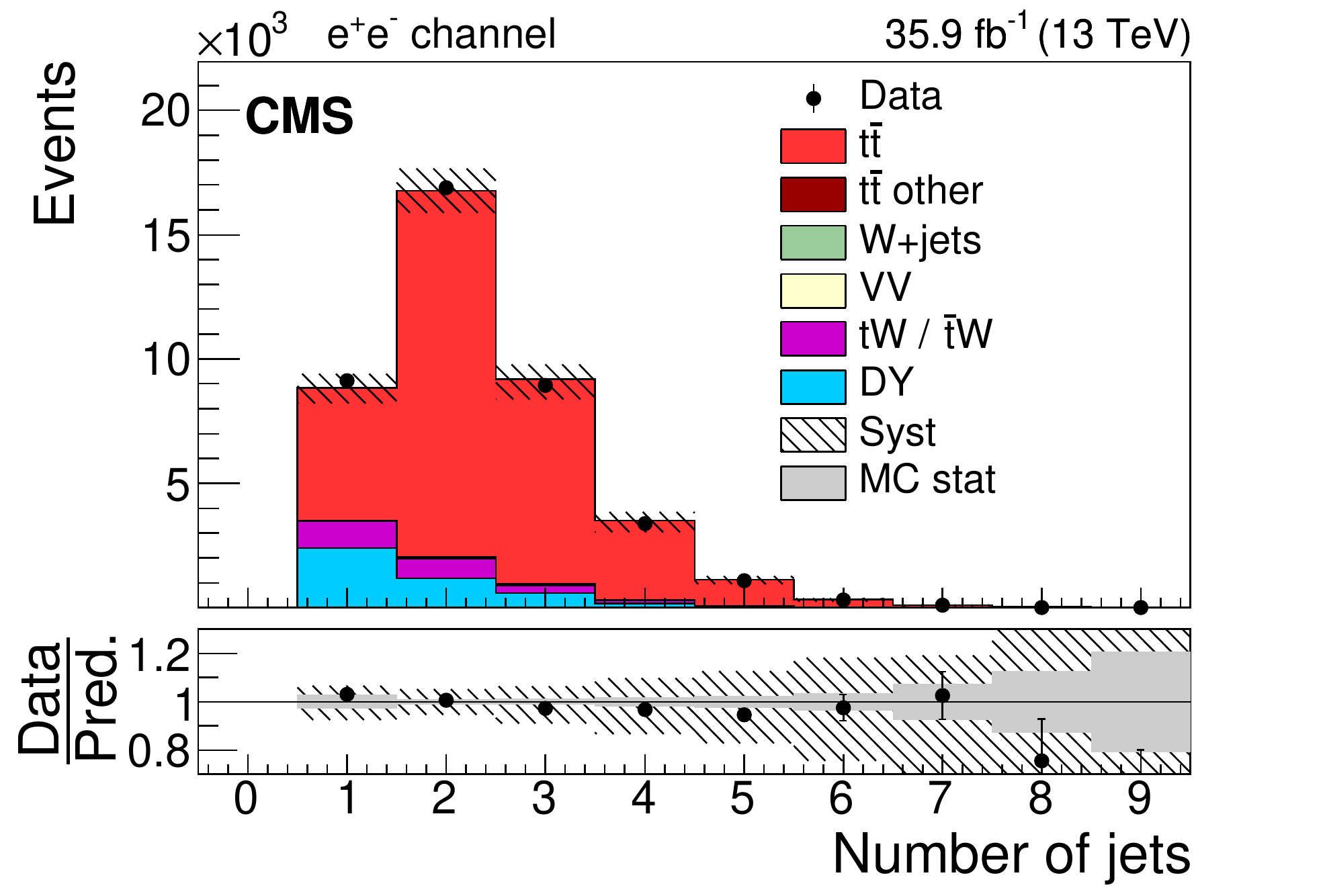}
    \includegraphics[width=0.49\textwidth]{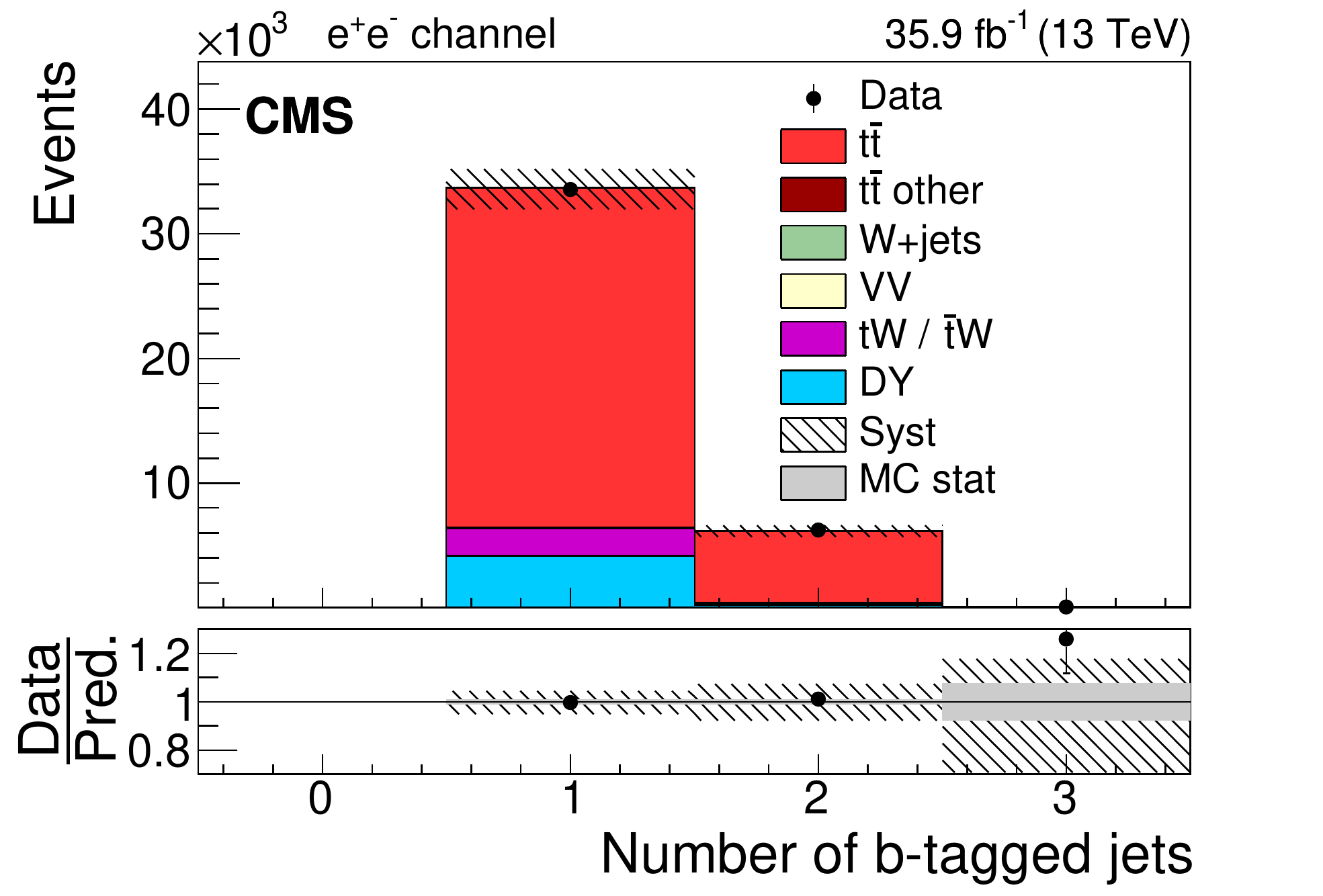}
      \caption{The same distributions as in Fig.~\ref{fig:lh_emu_ctrplots}, but for the \ee channel.}
       \label{fig:lh_ee_ctrplots}
\end{figure*}

Distributions of the leading and subleading lepton \pt and $\eta$, and the jet and \cPqb-tagged jet multiplicities in events fulfilling the above selection criteria are shown
in Figs.~\ref{fig:lh_emu_ctrplots}-\ref{fig:lh_ee_ctrplots}
for the \emu, \mumu, and \ee channels, respectively. The event yields in the simulations are normalized to the corresponding cross section predictions, as explained in Section~\ref{sec:detector}.
Selected events include a very small contribution from \ttbar processes in the lepton+jets decay channel (referred to as ``\ttbar other'' in the figures) in
which one of the charged leptons originates from
heavy-flavour hadron decay, misidentified hadrons, muons from light-meson
decays, or electrons from unidentified photon conversions. Such leptons
also lead to dilepton background in this analysis via \Wjets processes.

In all categories, the simulation is found to describe the data well within the systematic uncertainties, indicated by the bands in the
figures.

\section{Event categories and fit procedure}
\label{sec:fit}

The measurement is performed using a template fit to multidifferential distributions, divided into distinct event categories using the \cPqb-tagged jet multiplicity, similar to the method utilized in a previous measurement~\cite{Khachatryan:2016mqs}.
In each of the same-flavour channels, two categories are defined, corresponding to events
having 1 or 2 \cPqb-tagged jets.  Events with zero \cPqb-tagged jets are not included since they are dominated by the DY background
process.  In the \emu channel, three categories are defined,  corresponding to events having 1, 2, or 0 or $\geq$3 \cPqb-tagged jets. The templates describing the distributions for the signal and background events are taken from simulation.
Categorizing the events by their \cPqb-tagged jet multiplicity allows the efficiency $\epsilon_{\cPqb}$ for selecting and identifying a \cPqb~jet to be constrained.
Previous measurements that used a template fit with dilepton events were restricted to the \emu channel~\cite{Khachatryan:2016mqs,Aad:2014kva}. In this analysis, the decay channels with two electrons and two muons are also included in the fit. In this way, additional constraints on the lepton identification efficiencies are obtained.

First, a visible \ttbar cross section \sttvis, defined for a phase space corresponding to the experimentally accessible fiducial volume, as described in Section~\ref{sec:crosssection}, is determined. For the visible cross section, the fit is used to constrain the systematic uncertainties from the data. Using the relation
\begin{equation}
\linenomath
\stt =  \frac{\sttvis}{A_{\ell\ell}},
\label{eq:extrapol}
\end{equation}
the measured visible cross section is then extrapolated to the full phase space to obtain $\stt$. Here, $A_{\ell\ell}$ denotes the acceptance, which is defined as the fraction of \ttbar events that fulfill the selection criteria for the visible cross section. The acceptance incorporates the combined branching fraction for the \cPqt\xspace and \cPaqt\xspace quarks to decay to two charged leptons~\cite{PDG2018}. Apart from the free parameter of interest \sttvis, the parameters of the fit are the $J$ nuisance parameters $\vec{\lambda} = (\lambda_1, \lambda_2, ..., \lambda_J)$ corresponding to the various sources of systematic uncertainty, discussed in detail in Section~\ref{sec:systematics}.

The likelihood function $L$ is based on Poisson statistics:
\begin{equation}
\linenomath
 L  =  \prod_{i} \frac{\re^{ -\nu_i } \nu_i^{n_i}}{n_i!}\,  \prod_{j} \pi(\lambda_j),
 \label{eq:lhfunct}
 \end{equation}
where $i$ denotes the bin of the respective final-state distribution, and $\nu_i$ and $n_i$ are the expected and observed number of events in bin $i$, respectively.
The symbol $\pi(\lambda_j)$ denotes
a penalty term for the deviation of the nuisance parameter $\lambda_j$  from its nominal value according to its
prior density distribution. A Gaussian prior density distribution is assumed for all nuisance parameters.
The expectation values $\nu_i$ can be written as
\begin{equation}
\linenomath
\nu_i = s_i(\sttvis,\vec{\lambda})
+ \sum_{k} b_{k,i}^{\mathrm{MC}}(\vec{\lambda}).
\label{eq:expectev}
\end{equation}
Here, $s_i$ denotes the expected number of \ttbar signal events in bin $i$ and the quantity $b_{k,i}^{\mathrm{MC}}$ represents the prediction of the number of background events in bin $i$ from source $k$. The \Minuit
program~\cite{James:1975dr} is used to minimize $-2 \ln{(L)}$ with $L$ given in Eq.~(\ref{eq:lhfunct}), and the \Minos~\cite{James:1975dr} algorithm is used to estimate the uncertainties.

For the determination of the \cPqb~tagging efficiencies, multinomial probabilities are used to describe the expected number of signal events with one \cPqb-tagged jet, $s_{1\cPqb}$,
two \cPqb-tagged jets, $s_{2\cPqb}$, and zero or more than two \cPqb-tagged jets, $s_{\text{other}}$:
\begin{eqnarray}
\label{eq:nb1}
s_{1\cPqb}  & = & \lumi \sttvis\epsilon_{\ell\ell}  2 \epsilon_{\cPqb}(1-C_{\cPqb}\epsilon_{\cPqb}),   \\
s_{2\cPqb}  & = &{\lumi \sttvis\epsilon_{\ell\ell}    \epsilon_{\cPqb}^2 C_{\cPqb} \label{eq:nb2}},\\
s_{\text{other}} & = &\lumi\sttvis \epsilon_{\ell\ell}  (1-2\epsilon_{\cPqb}(1-C_{\cPqb} \epsilon_{\cPqb})-\epsilon_{\cPqb}^2C_{\cPqb}),
\end{eqnarray}
where \lumi denotes the integrated luminosity and $\epsilon_{\ell\ell}$ is the efficiency for events in the visible phase space to pass the full selection described in Section~\ref{sec:eventselection}. The quantity $C_{\cPqb}$ corrects for any small correlations between the tagging of two \cPqb~jets in an event, expressed as
$ C_{\cPqb} = 4 s_{\text{all}} s_{2\cPqb}/ (s_{1\cPqb}+2 s_{2\cPqb})^2$, where $s_{\text{all}}$ denotes the total number of signal events.
The values for $\epsilon_{\ell\ell}$, $\epsilon_{\cPqb}$, and $C_{\cPqb}$ are directly determined
from the \ttbar signal simulation, expressing $\epsilon_{\cPqb}$\/ as $(s_{1\cPqb} + 2 s_{2\cPqb})/2 s_{\text{all}}$.
The values of these parameters for the nominal signal simulation in the \emu channel are $\epsilon_{\Pe\Pgm} = 0.49$, $\epsilon_{\cPqb} = 0.30$, and $C_{\cPqb} = 1.00$.

The overall selection efficiency $\epsilon_{\ell\ell}$ is a linear combination of the efficiencies
$\epsilon_{\Pe\Pgm}$, $\epsilon_{\Pe\Pe}$, and $\epsilon_{\Pgm\Pgm}$, in the three different dilepton channels, each given by the product of the two efficiencies for identifying a single lepton of the respective flavour.
Prior to the fit, the muon identification uncertainty is smaller than that for electrons. By fitting the three dilepton decay channels simultaneously, the ratio of single-lepton efficiencies $\epsilon_\Pe$ and $\epsilon_\Pgm$ is constrained. In the fit, the electron identification uncertainty is constrained to that for muons.

The values for $\epsilon_{\ell\ell}$, $\epsilon_{\cPqb}$, $C_{\cPqb}$, the number of signal events in each category,
and the background rates depend on the nuisance parameters $\vec{\lambda}$.
The dependence on the parameter $\lambda_j$ is modelled by a second-order polynomial that describes the quantity at the three values $\lambda_j=0,1,-1$, corresponding to the nominal value of the
parameter and to a variation by +1 and -1 standard deviation, respectively.
If a variation is only possible in one direction, a linear function is used to model the dependence on $\lambda_j$.

The events are further categorized by the number of additional non-\cPqb-tagged jets in the event.  Each of
the seven previously described event categories is further
divided by grouping together events with 0, 1, 2, or $\geq$3
additional non-\cPqb-tagged jets, thus producing 28 disjoint event
categories. For those categories that have events with at least
one additional non-\cPqb-tagged jet, the smallest \pt among those
jets is used as the observable in the fit.  For those categories
containing events with zero additional non-\cPqb-tagged jets, the
total number of events in the category is used as the observable
in the fit.  The further division of events into these categories
and the observable distributions from each category provide the
sensitivity to constrain the modelling systematic uncertainties,
such as those coming from variations in the scales for the matrix
element (ME) and parton shower (PS) matching.
For events with no additional jets, the total event yield is used.

The statistical uncertainty in the templates from simulation is taken into account by using pseudo-experiments. At each iteration, templates are varied within their statistical uncertainty.
Templates created from different simulations are treated as statistically uncorrelated, while templates derived by varying weights in the simulation are treated as correlated. The template dependencies are rederived and the fit to data is repeated. Repeating this 30~000 times yields an approximately Gaussian distribution of the fitted value of the \ttbar cross section (and of \mtmc in the combined fit) and of the vast majority of the nuisance parameters. The root-mean-square of each distribution is considered as an additional uncertainty from the event counts in the simulated samples for the corresponding nuisance parameter.

\begin{figure*}[htbp!]
  \centering
    \includegraphics[width=0.325\textwidth]{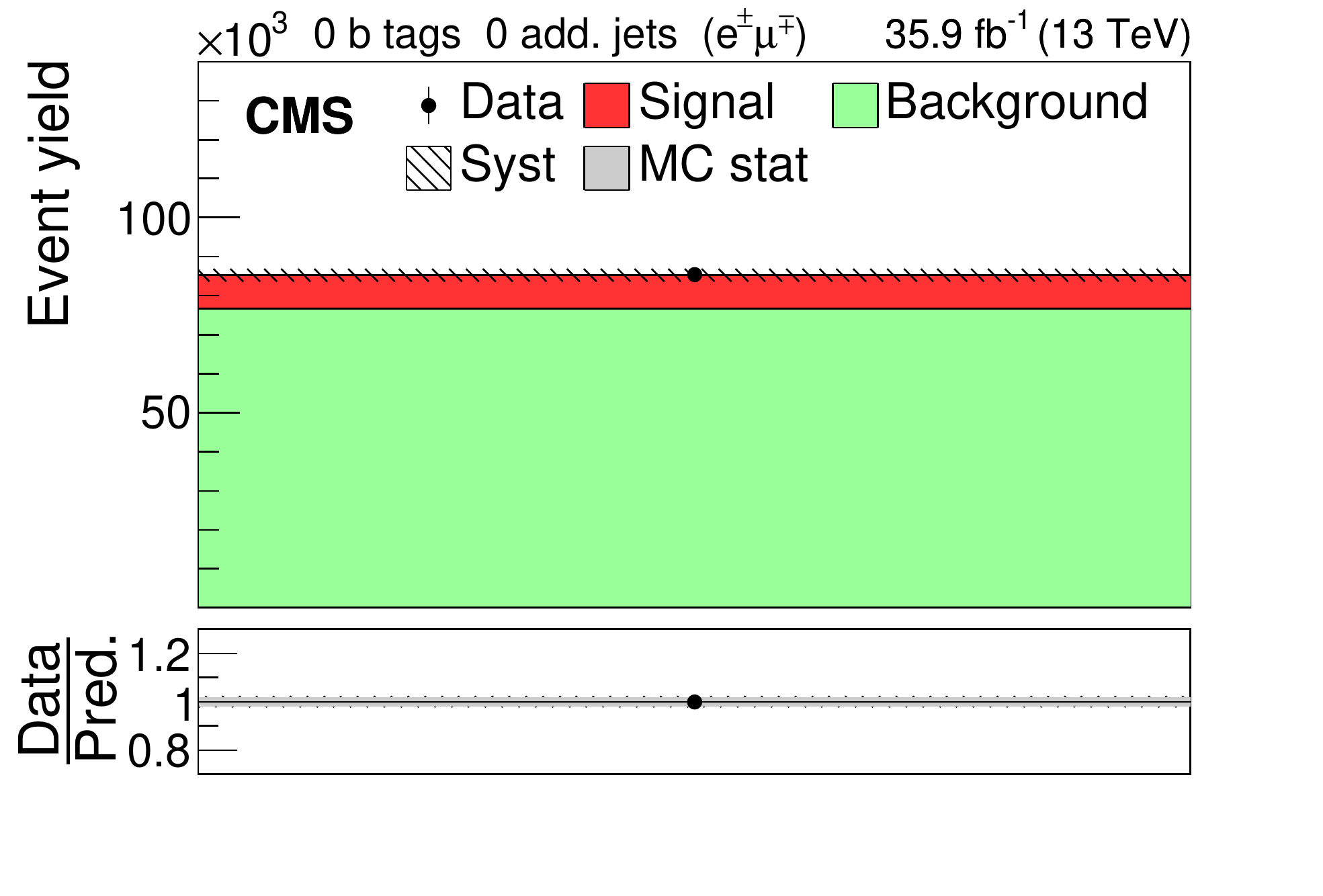}
    \includegraphics[width=0.325\textwidth]{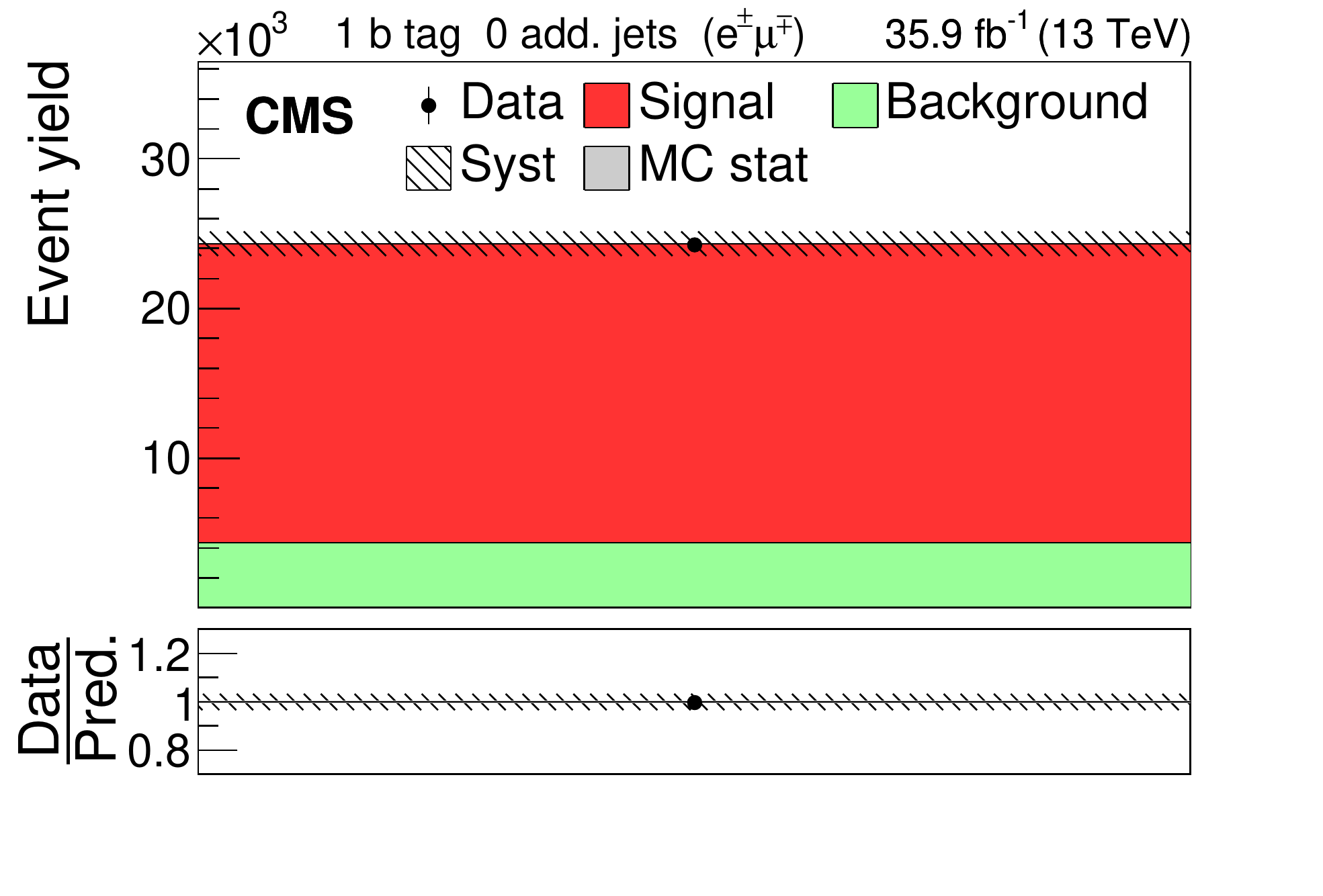}
    \includegraphics[width=0.325\textwidth]{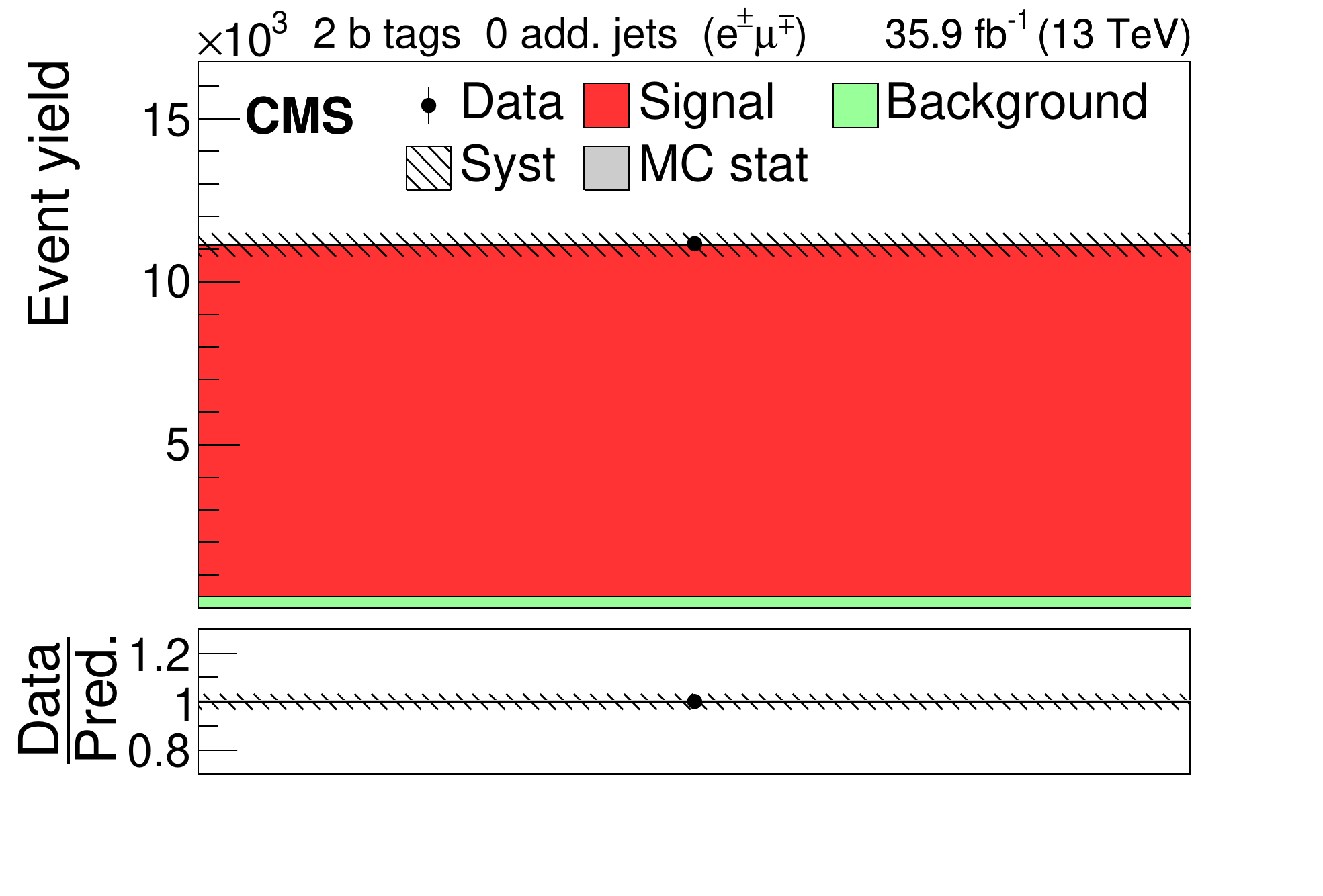}

    \includegraphics[width=0.325\textwidth]{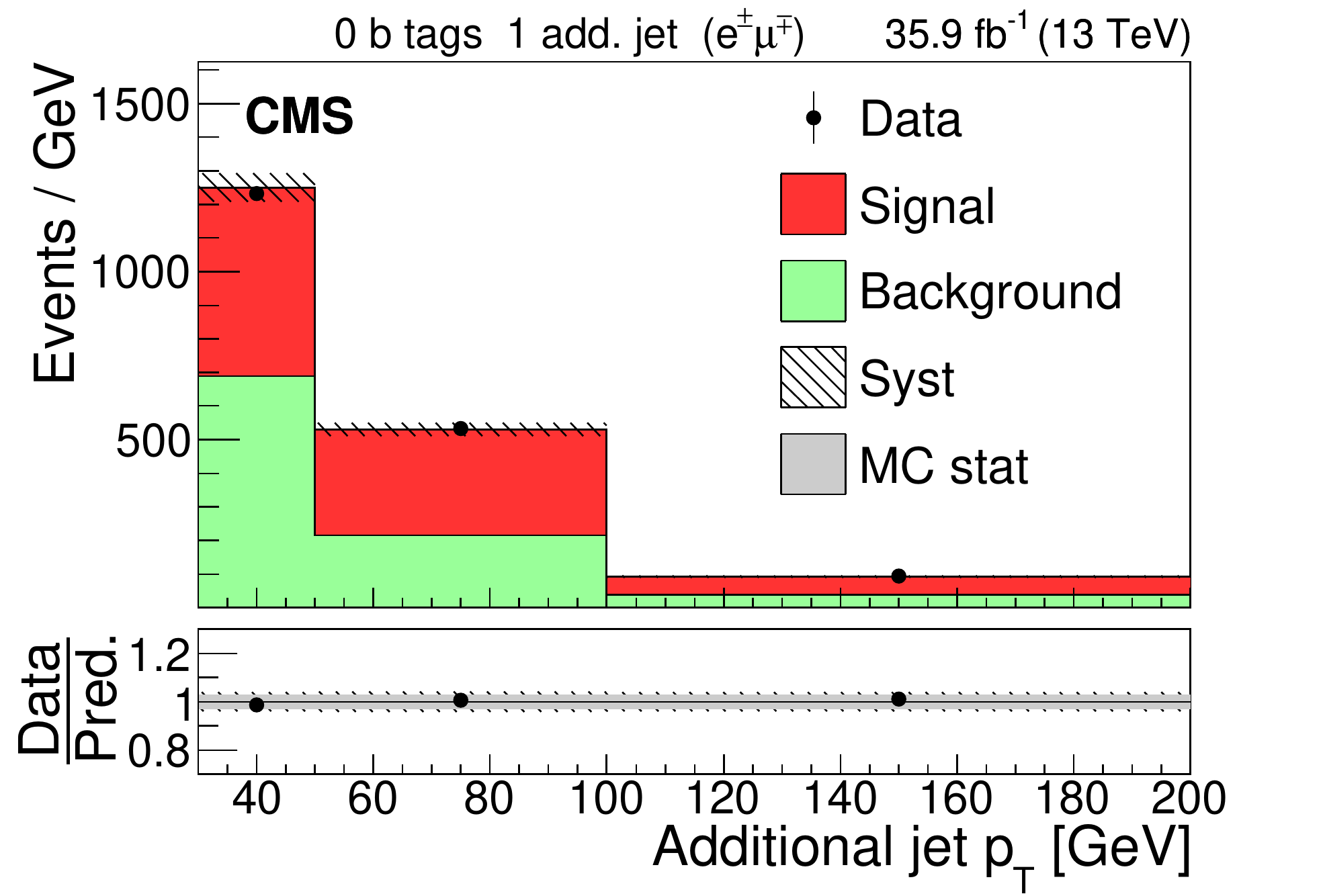}
    \includegraphics[width=0.325\textwidth]{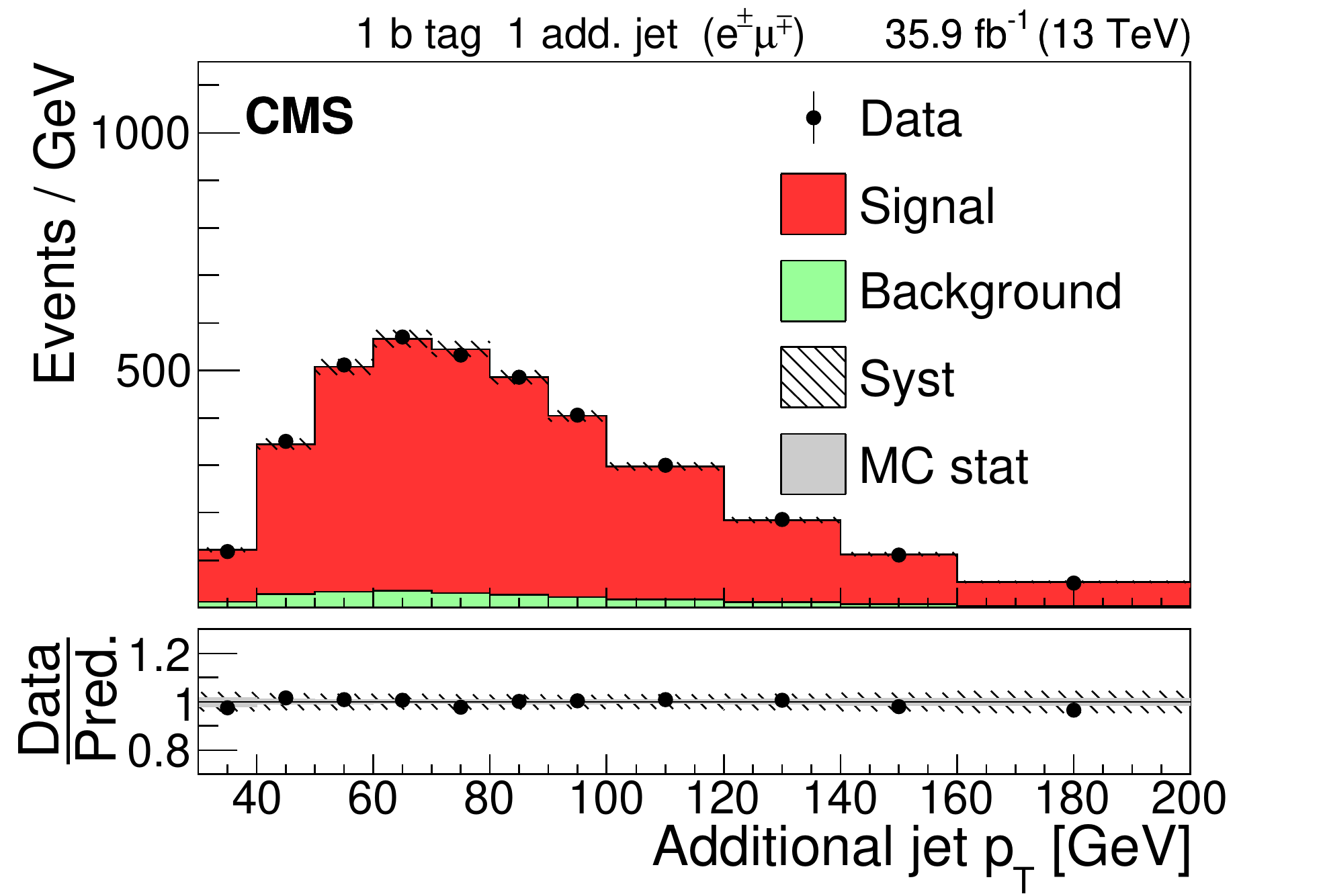}
    \includegraphics[width=0.325\textwidth]{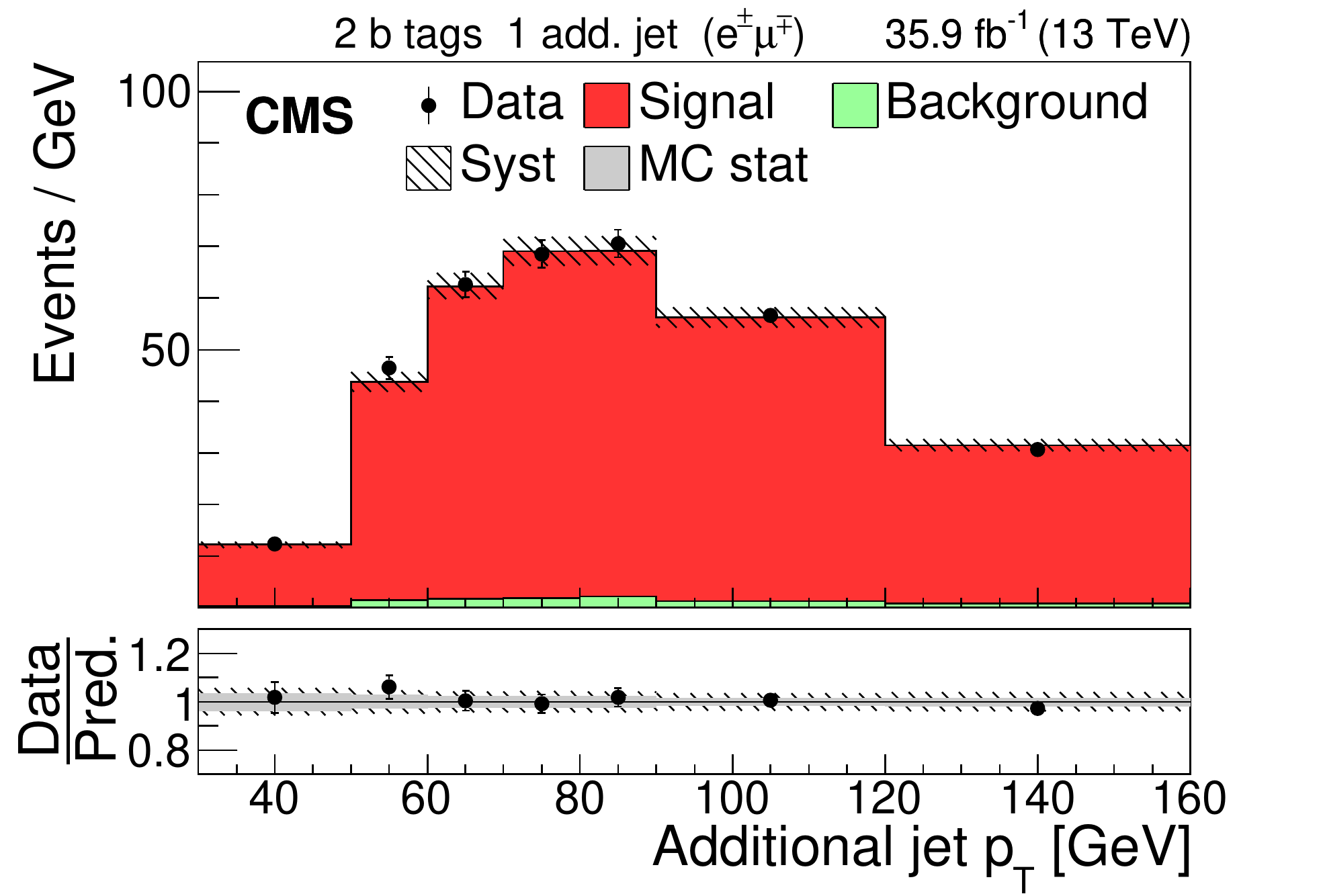}

    \includegraphics[width=0.325\textwidth]{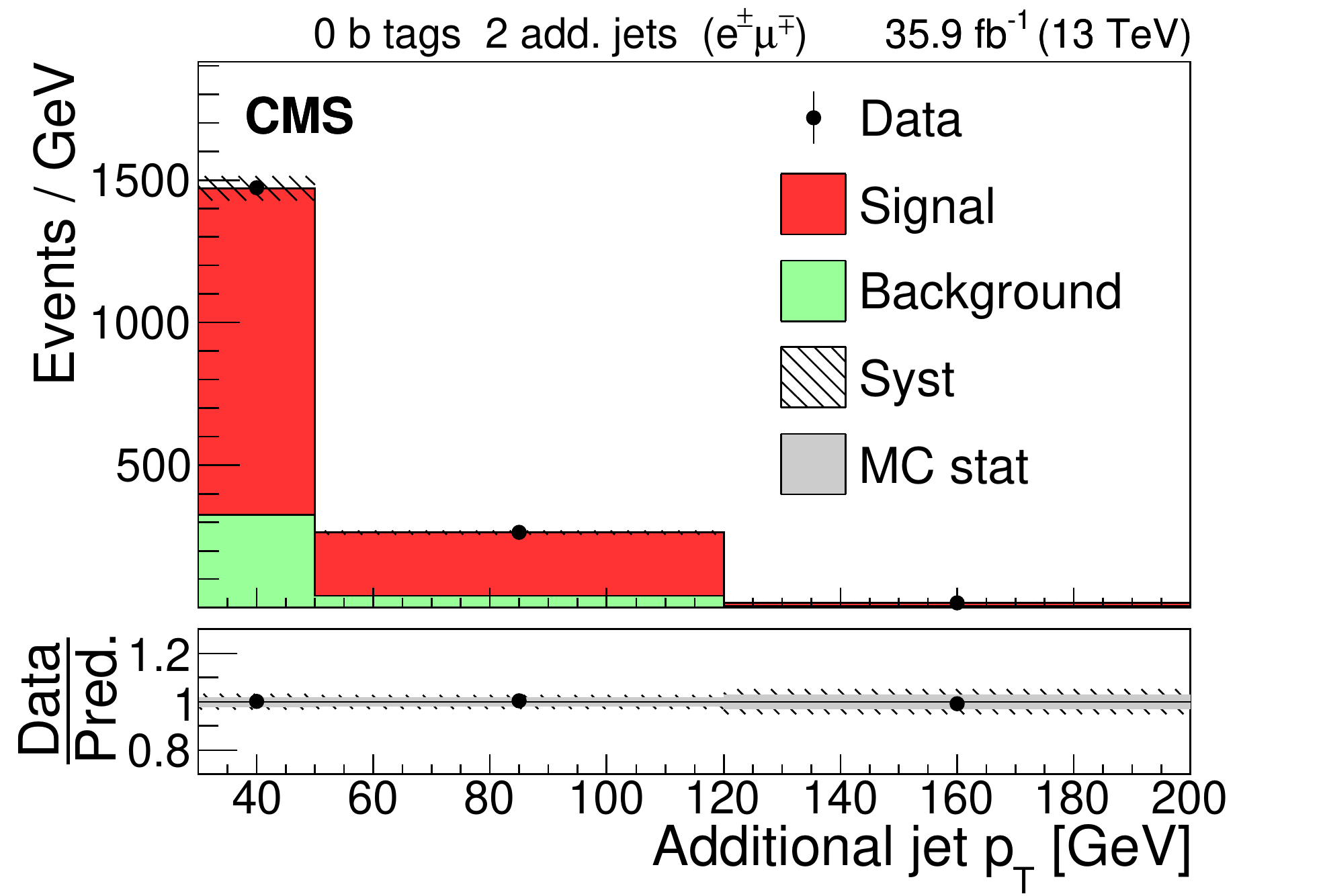}
    \includegraphics[width=0.325\textwidth]{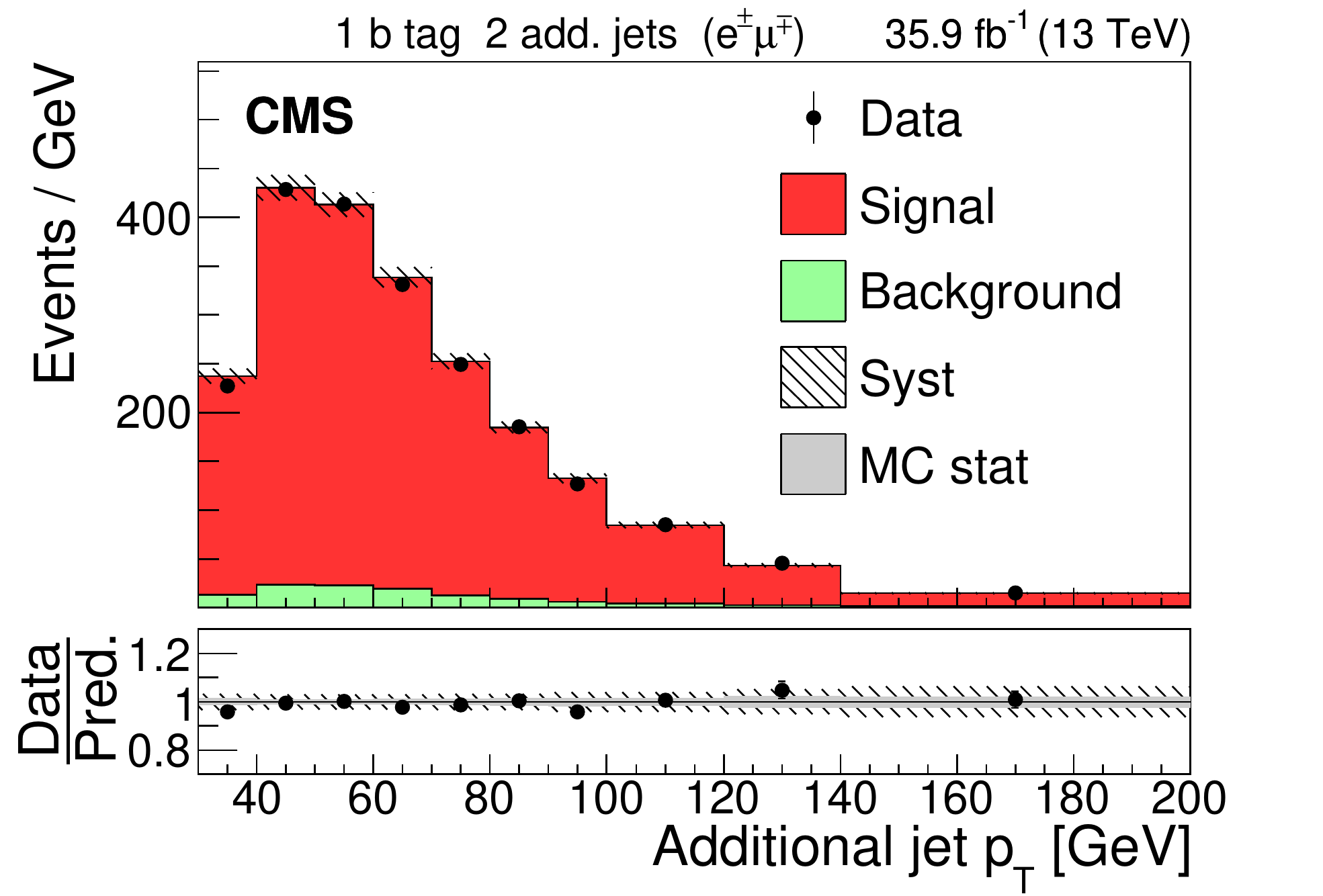}
    \includegraphics[width=0.325\textwidth]{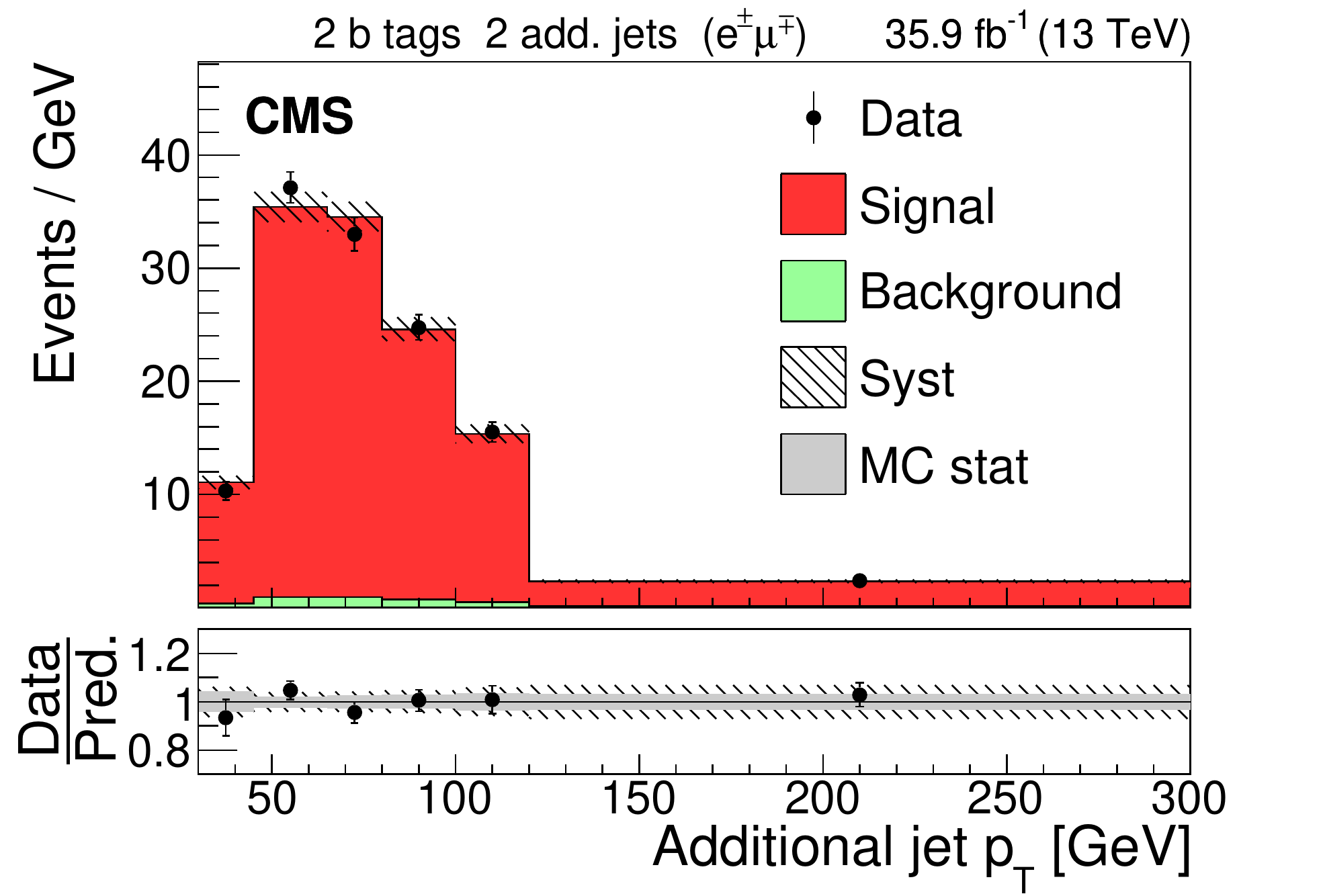}

    \includegraphics[width=0.325\textwidth]{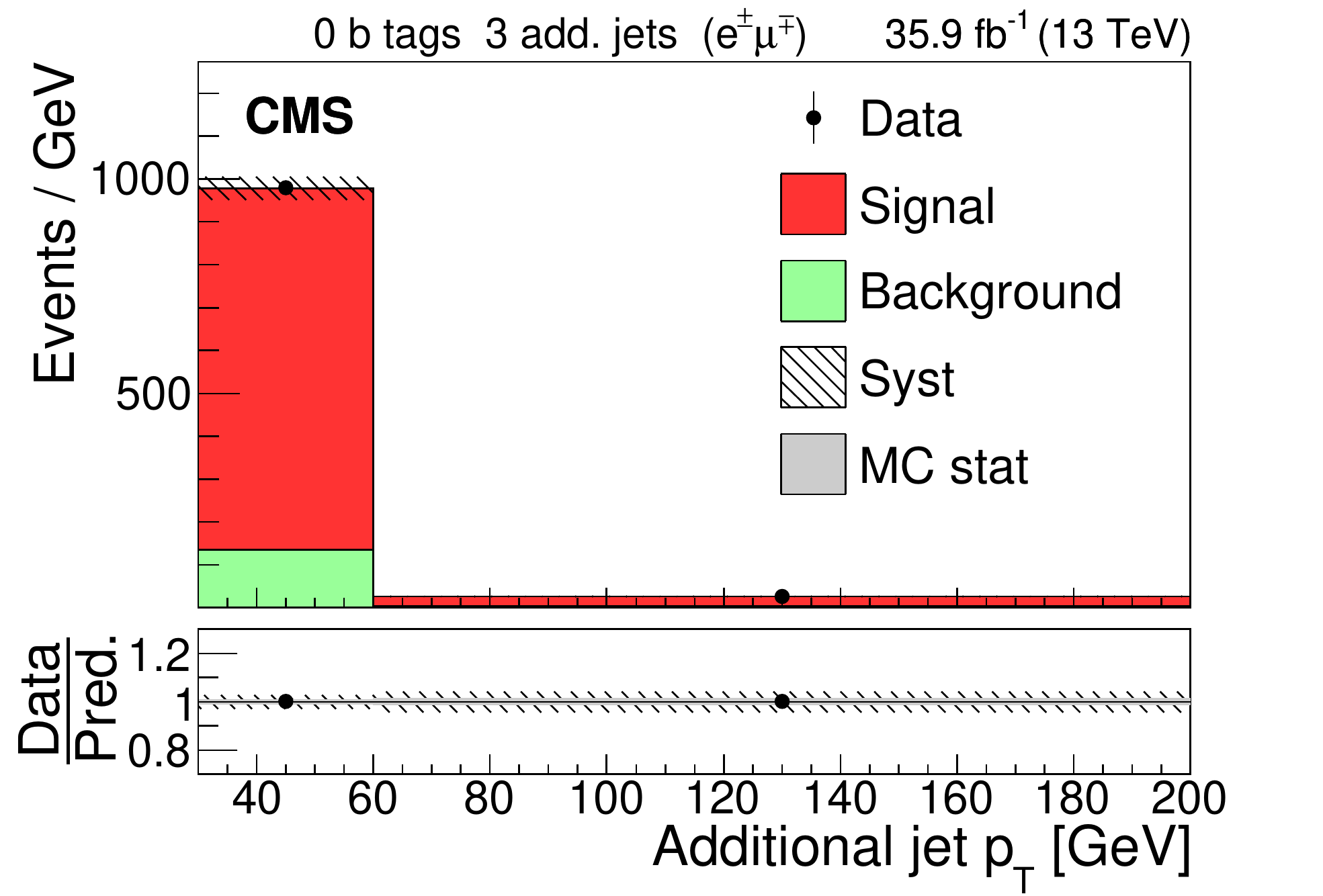}
    \includegraphics[width=0.325\textwidth]{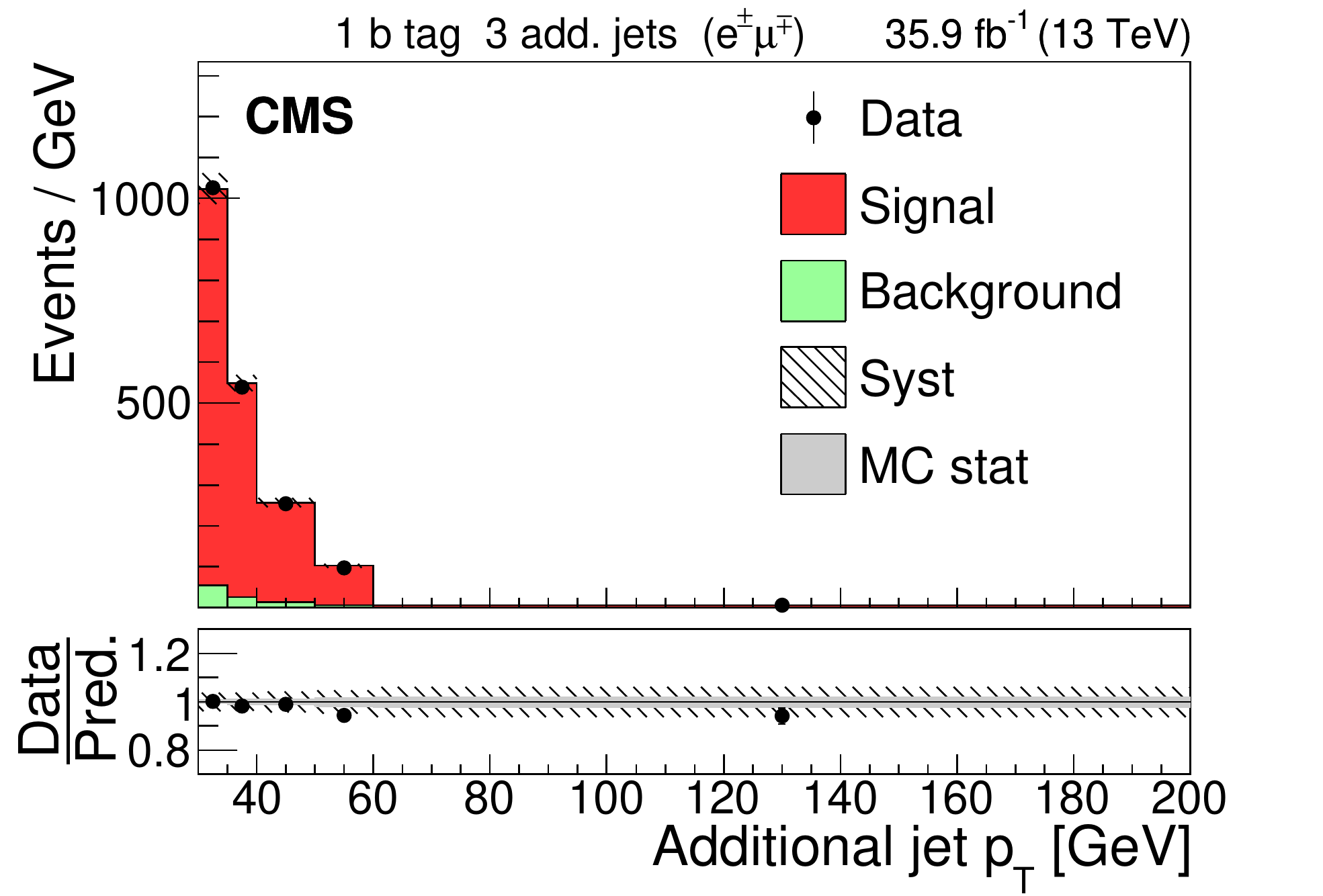}
    \includegraphics[width=0.325\textwidth]{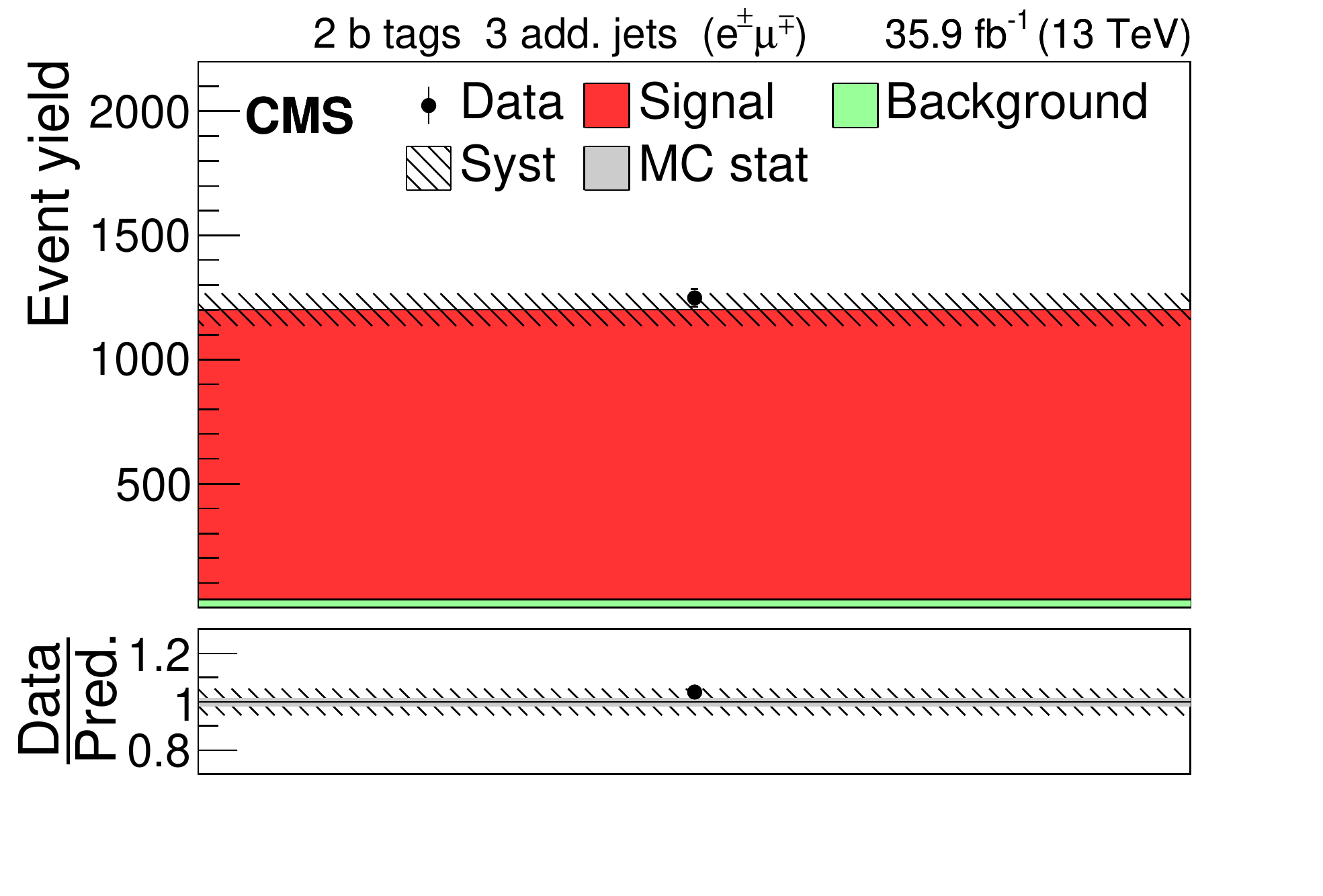}

\caption{Distributions in the \emu channel after the fit to the data.
  In the left column events with zero or three or more \cPqb-tagged jets are shown. The middle (right)
  column shows events with exactly one (two) \cPqb-tagged jets.
  Events with zero, one, two, or three or more additional non-\cPqb-tagged jets are shown in the first, second, third,
  and fourth row, respectively.
  The hatched bands correspond to the total uncertainty in the sum of the predicted yields
  including all correlations. The ratios of the data to the sum of the simulated
  yields after the fit are shown in the lower panel of each figure. Here, the solid gray
  band
  represents the contribution of the statistical uncertainty in the MC simulation.
  \label{fig:lh_emu_postfitdistr8}}
\end{figure*}

\begin{figure*}[htbp!]
  \centering

    \includegraphics[width=0.325\textwidth]{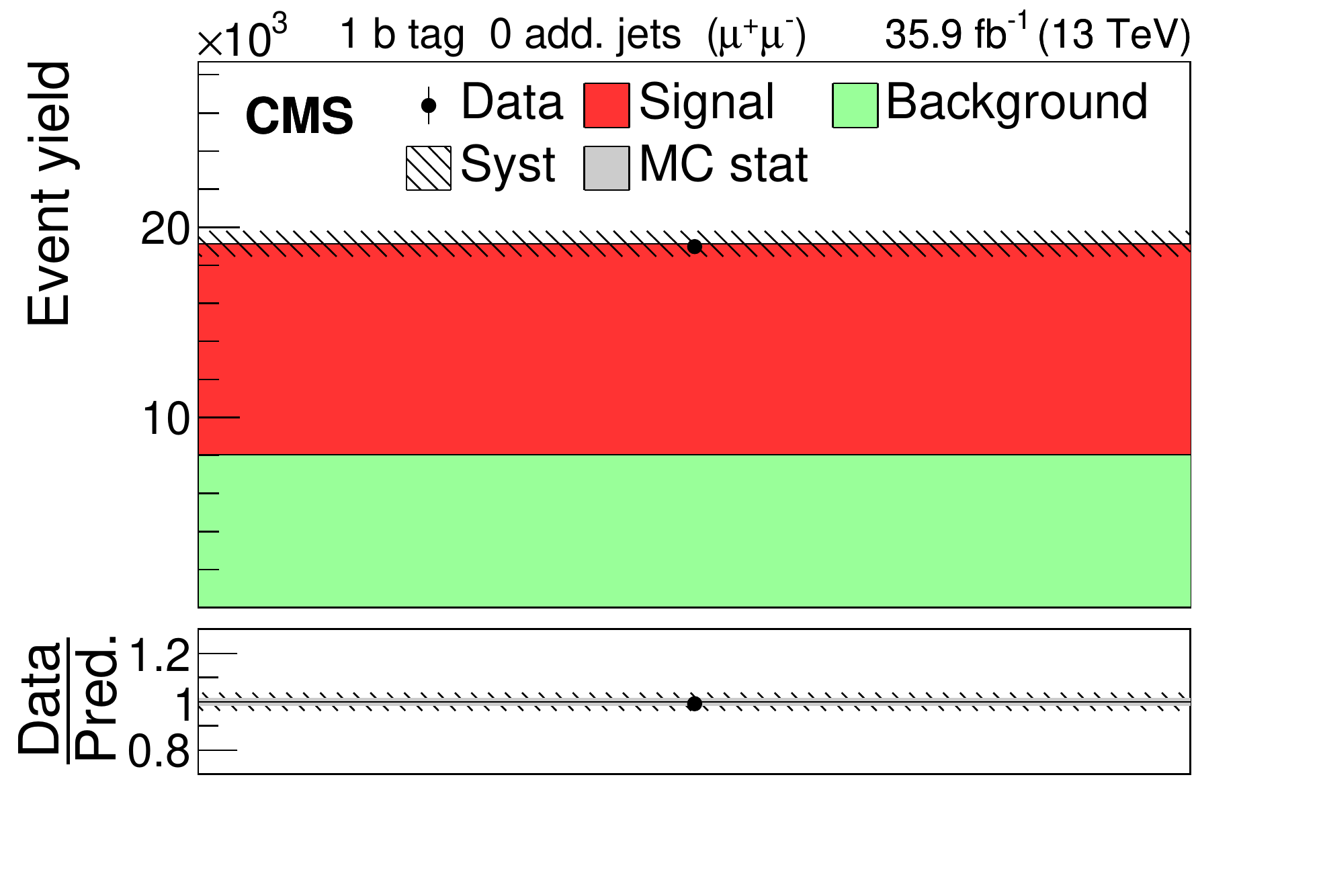}
    \includegraphics[width=0.325\textwidth]{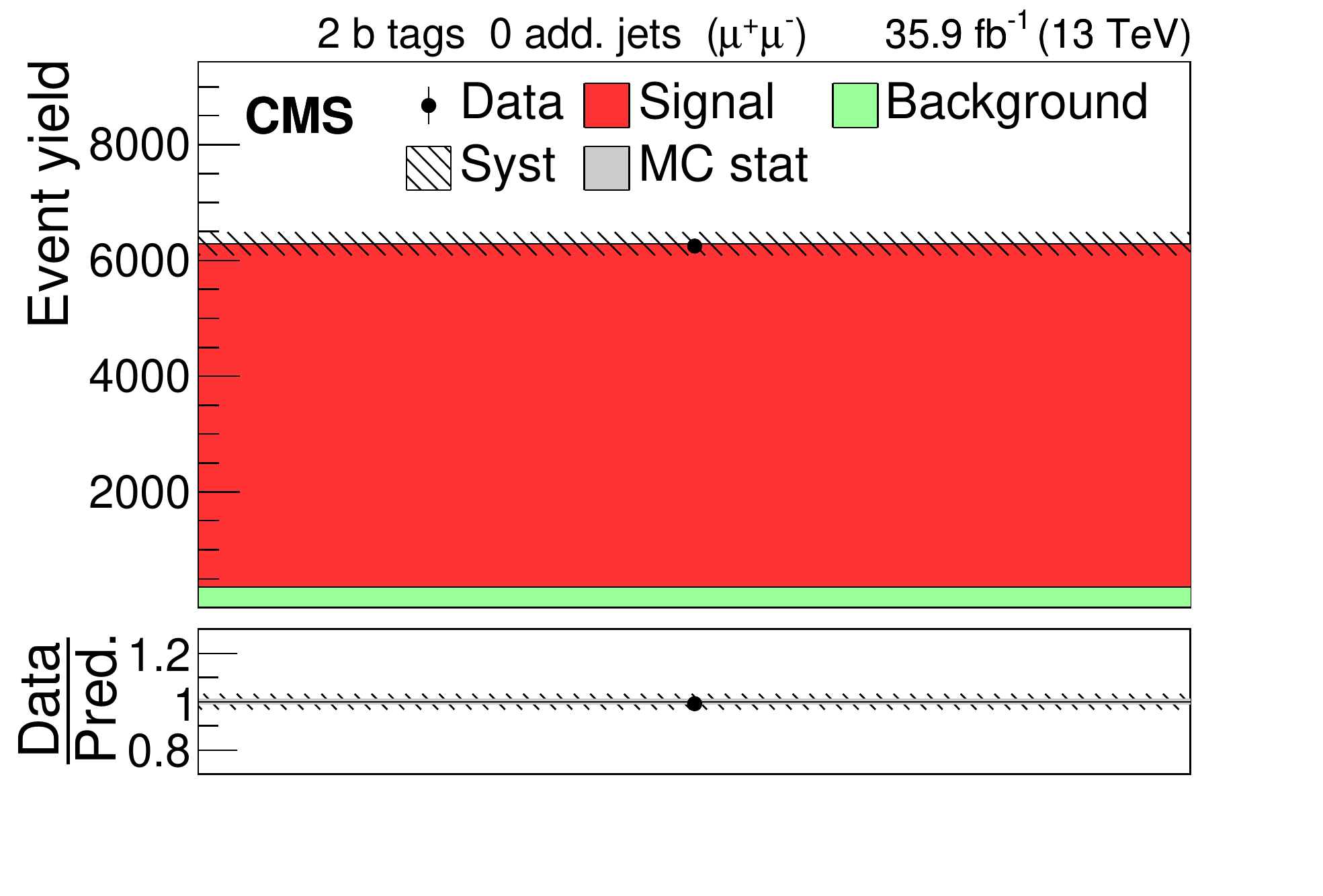}\\

    \includegraphics[width=0.325\textwidth]{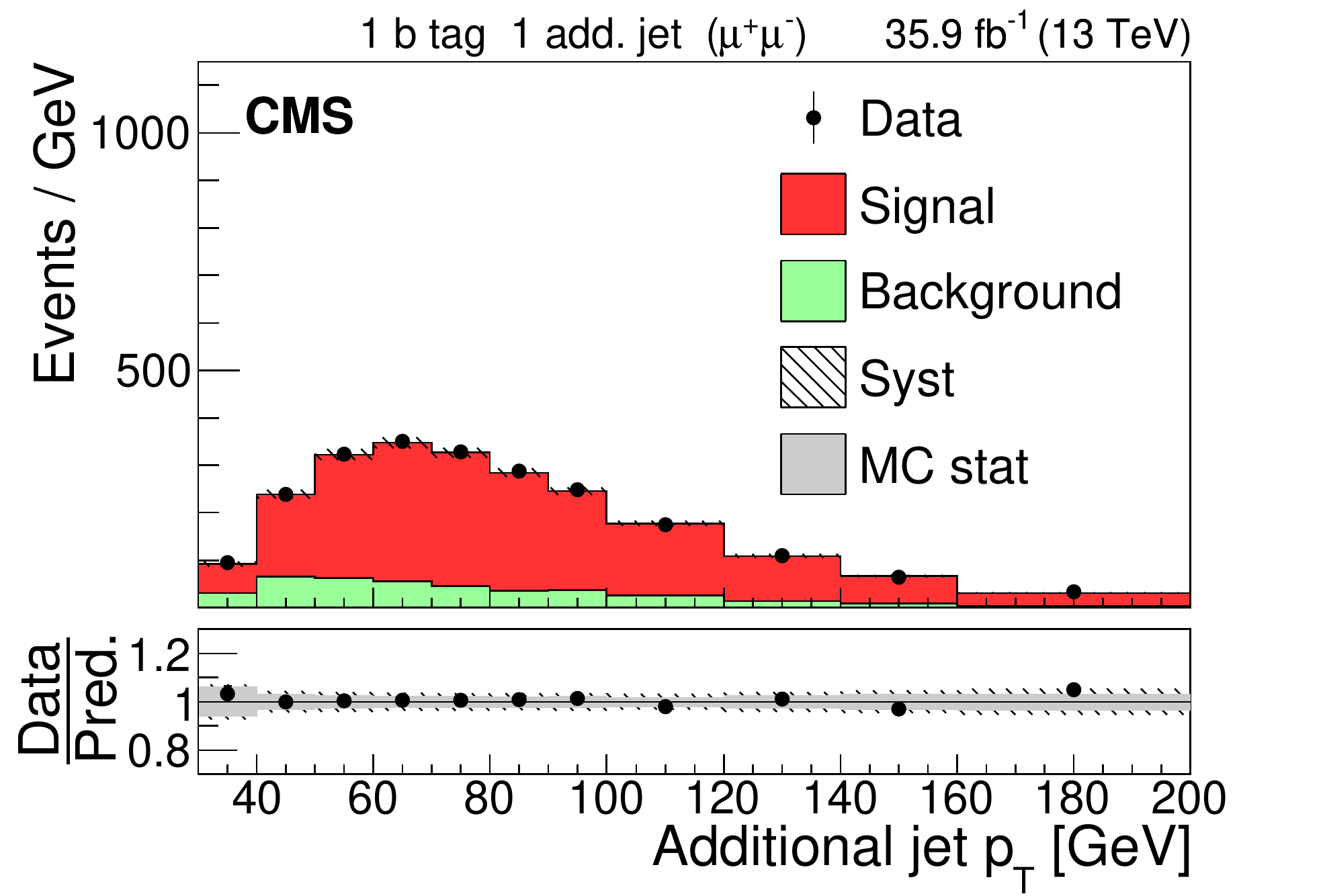}
    \includegraphics[width=0.325\textwidth]{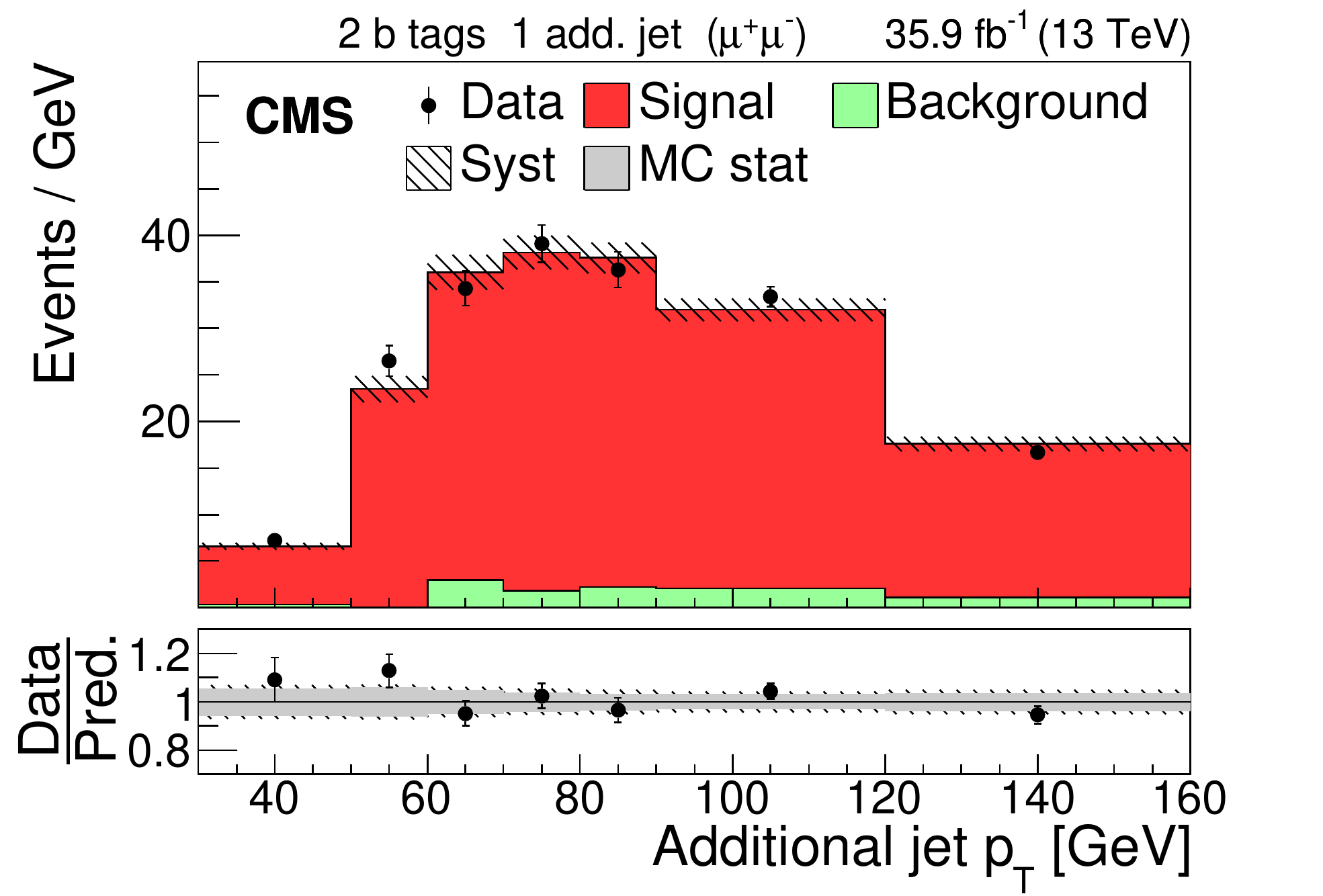}\\

    \includegraphics[width=0.325\textwidth]{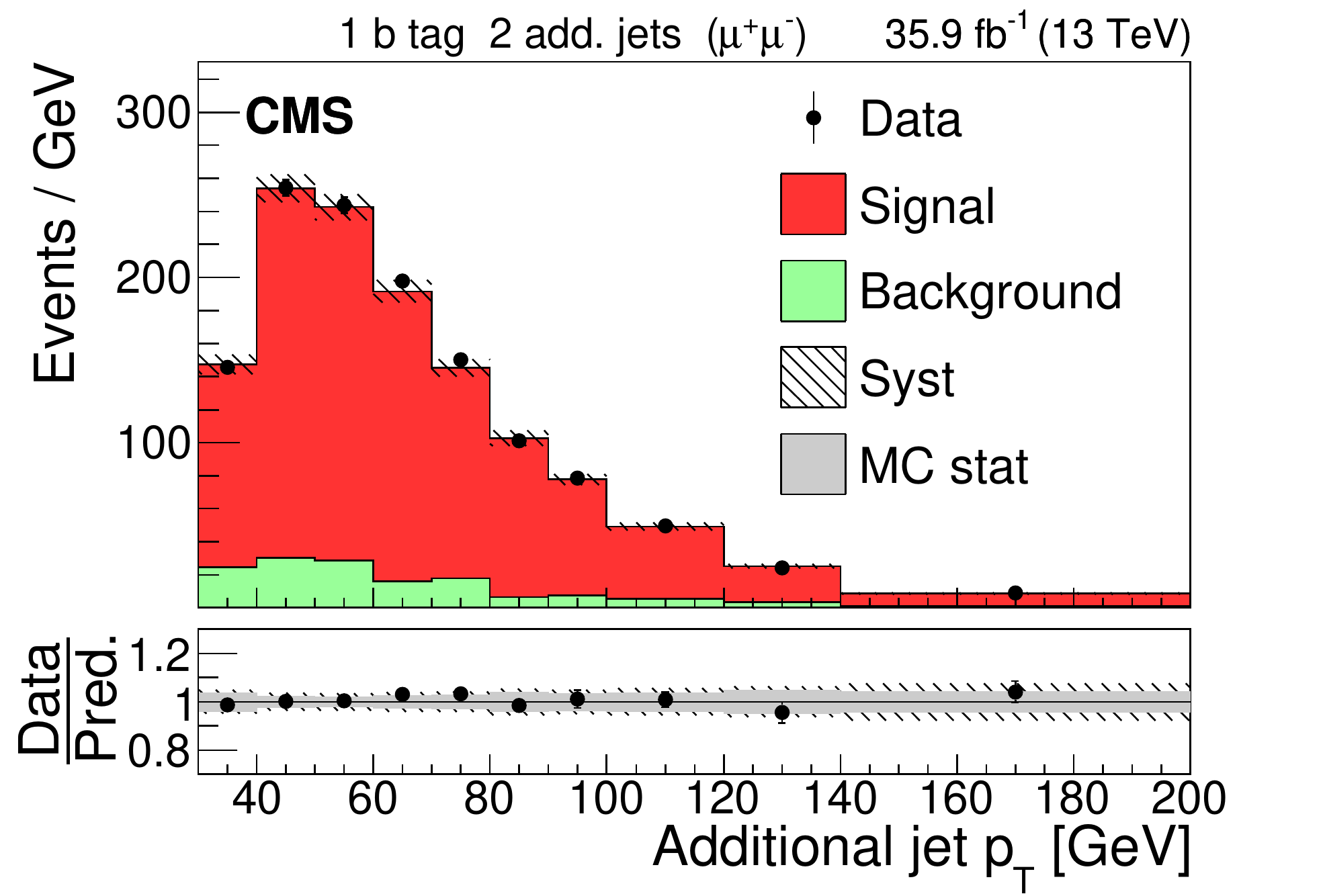}
    \includegraphics[width=0.325\textwidth]{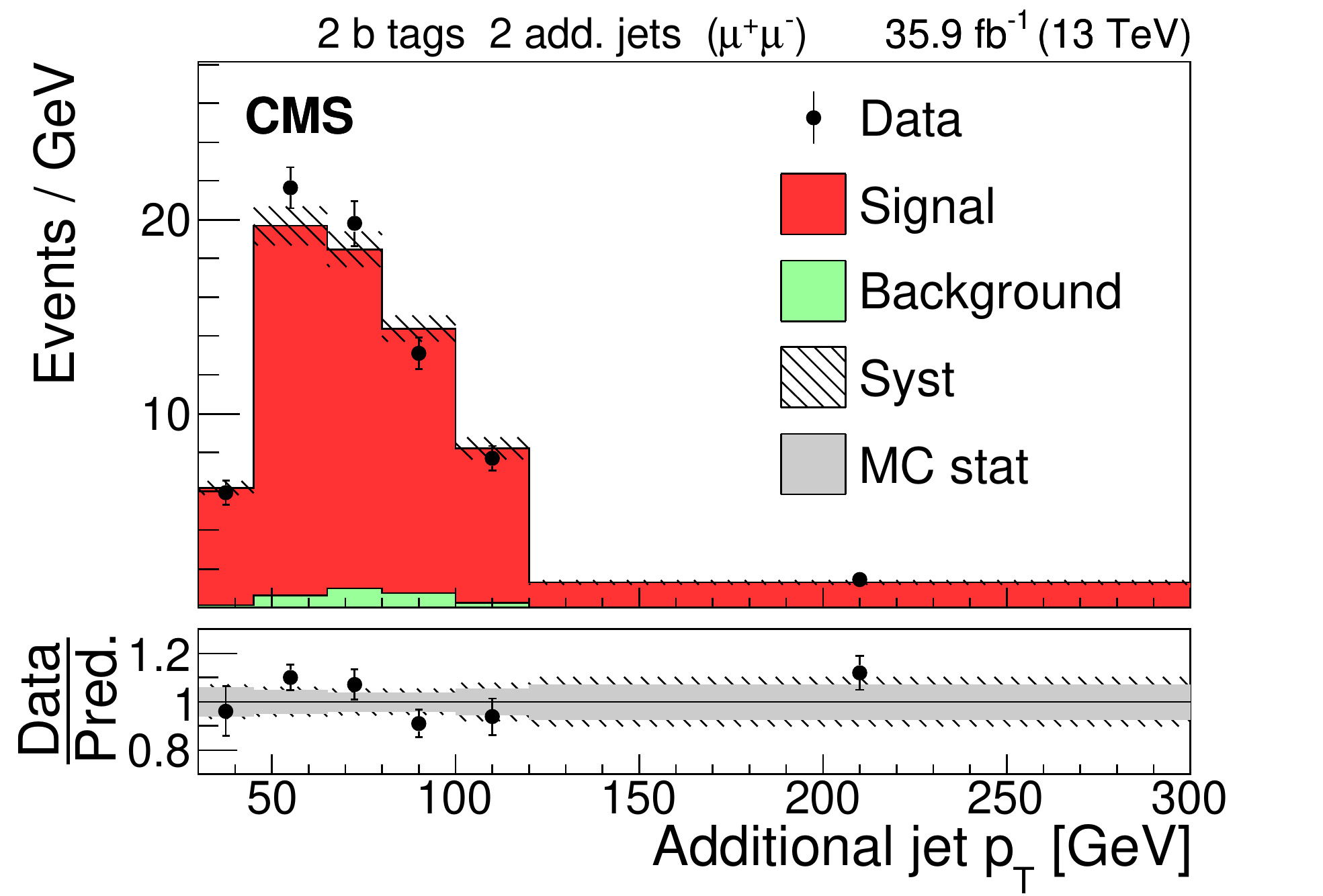}\\

    \includegraphics[width=0.325\textwidth]{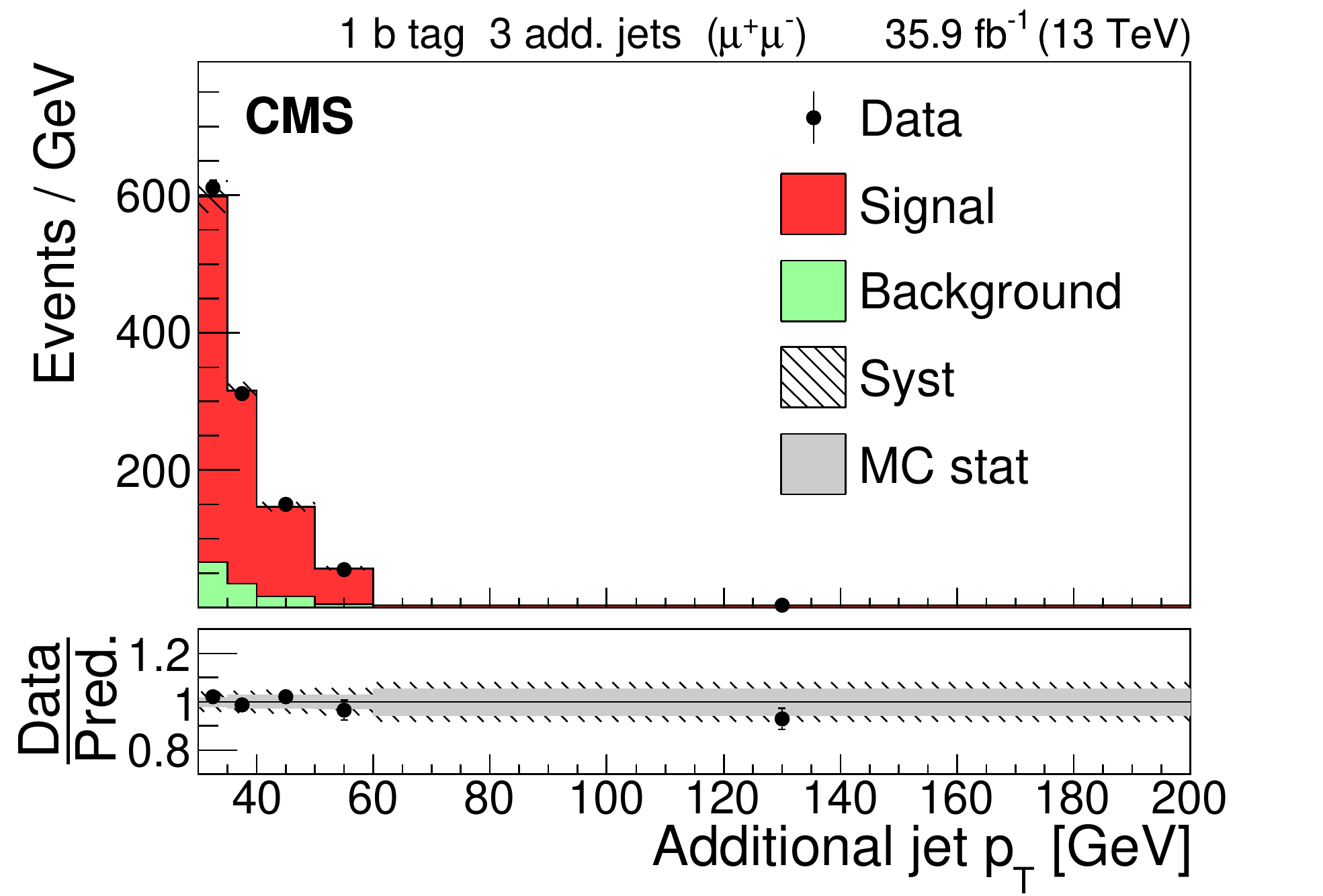}
    \includegraphics[width=0.325\textwidth]{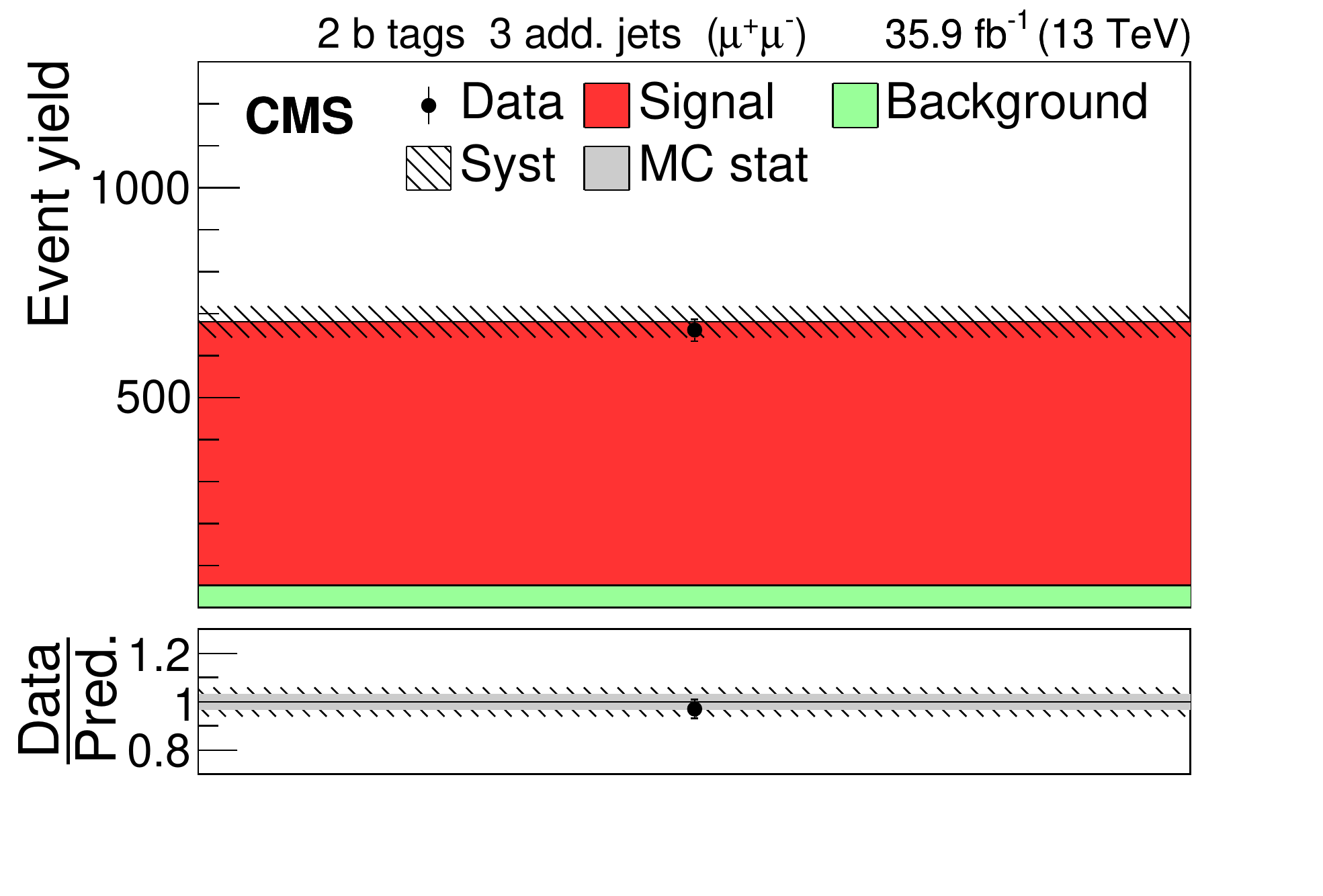}

\caption{
Distributions in the \mumu channel after the fit to the data. The left (right)
column shows events with exactly one (two) \cPqb-tagged jets.
Events with zero, one, two, or three or more additional non-\cPqb-tagged jets are shown in the first, second, third,
and fourth row, respectively.
The hatched bands correspond to the total uncertainty in the sum of the predicted yields
including all correlations. The ratios of the data to the sum of the simulated
yields after the fit are shown in the lower panel of each figure. Here, the solid gray
band
represents the contribution of the statistical uncertainty in the MC simulation.
  \label{fig:lh_mumu_postfitdistr8}}
\end{figure*}

\begin{figure*}[htbp!]
  \centering

    \includegraphics[width=0.325\textwidth]{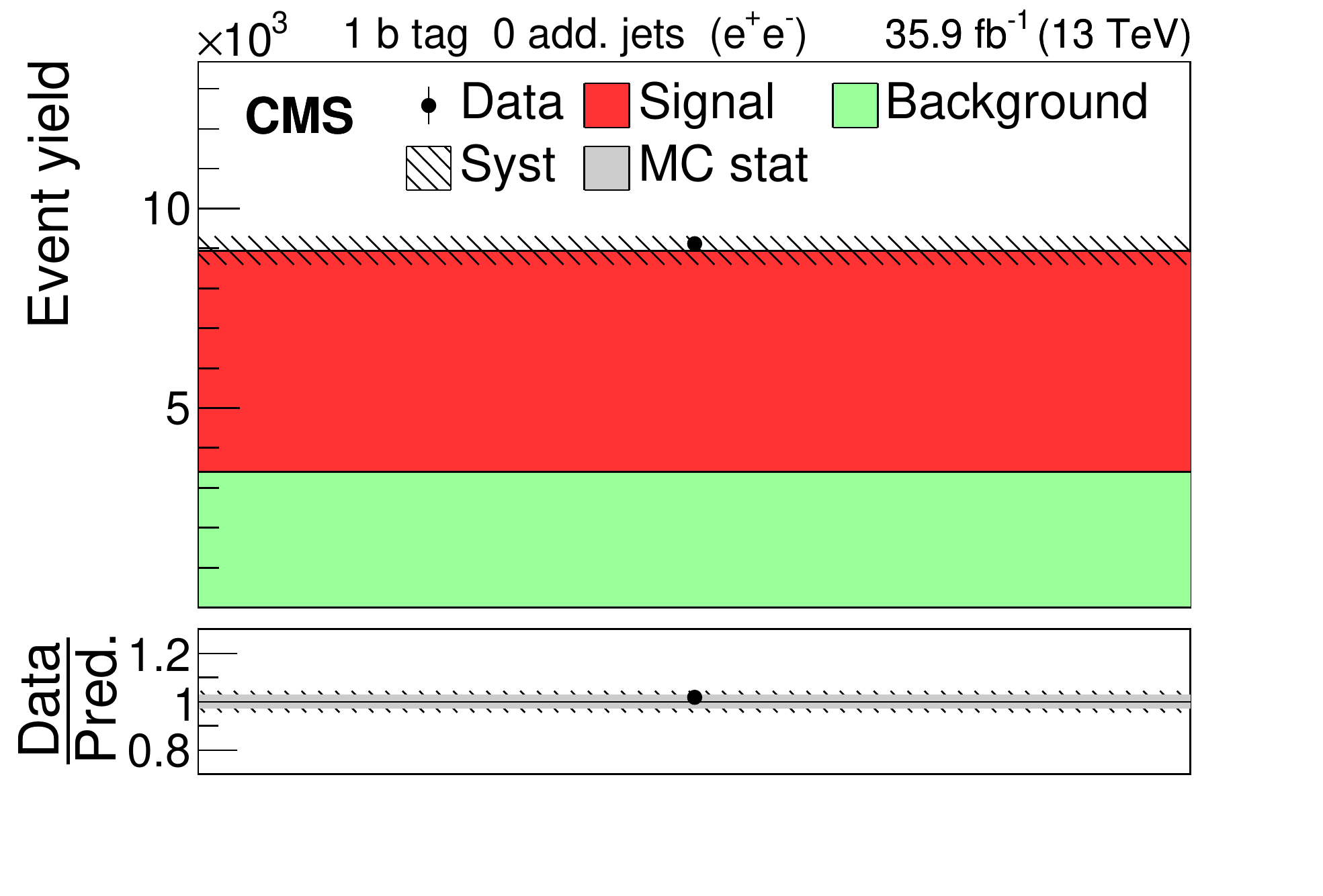}
    \includegraphics[width=0.325\textwidth]{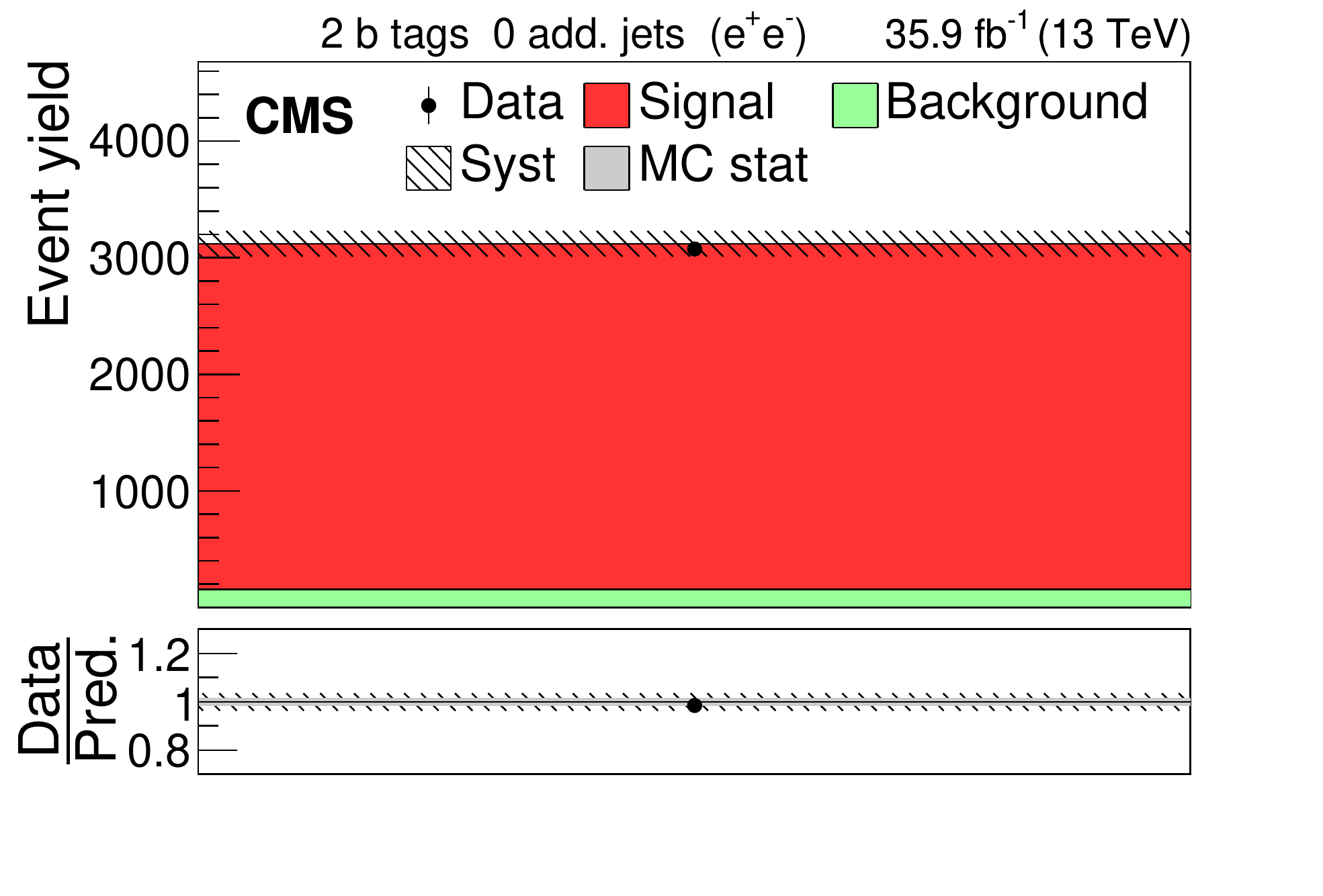}\\

    \includegraphics[width=0.325\textwidth]{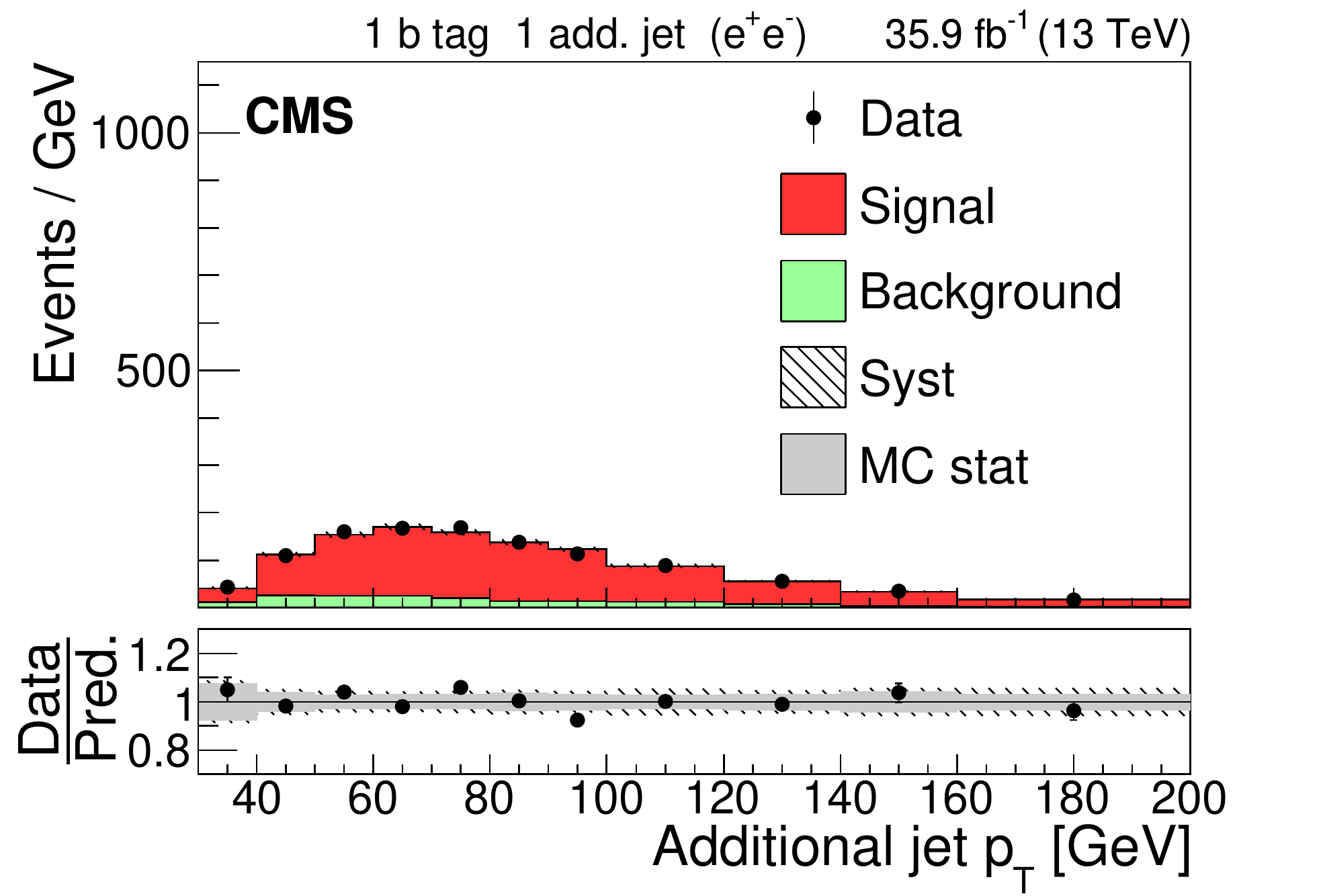}
    \includegraphics[width=0.325\textwidth]{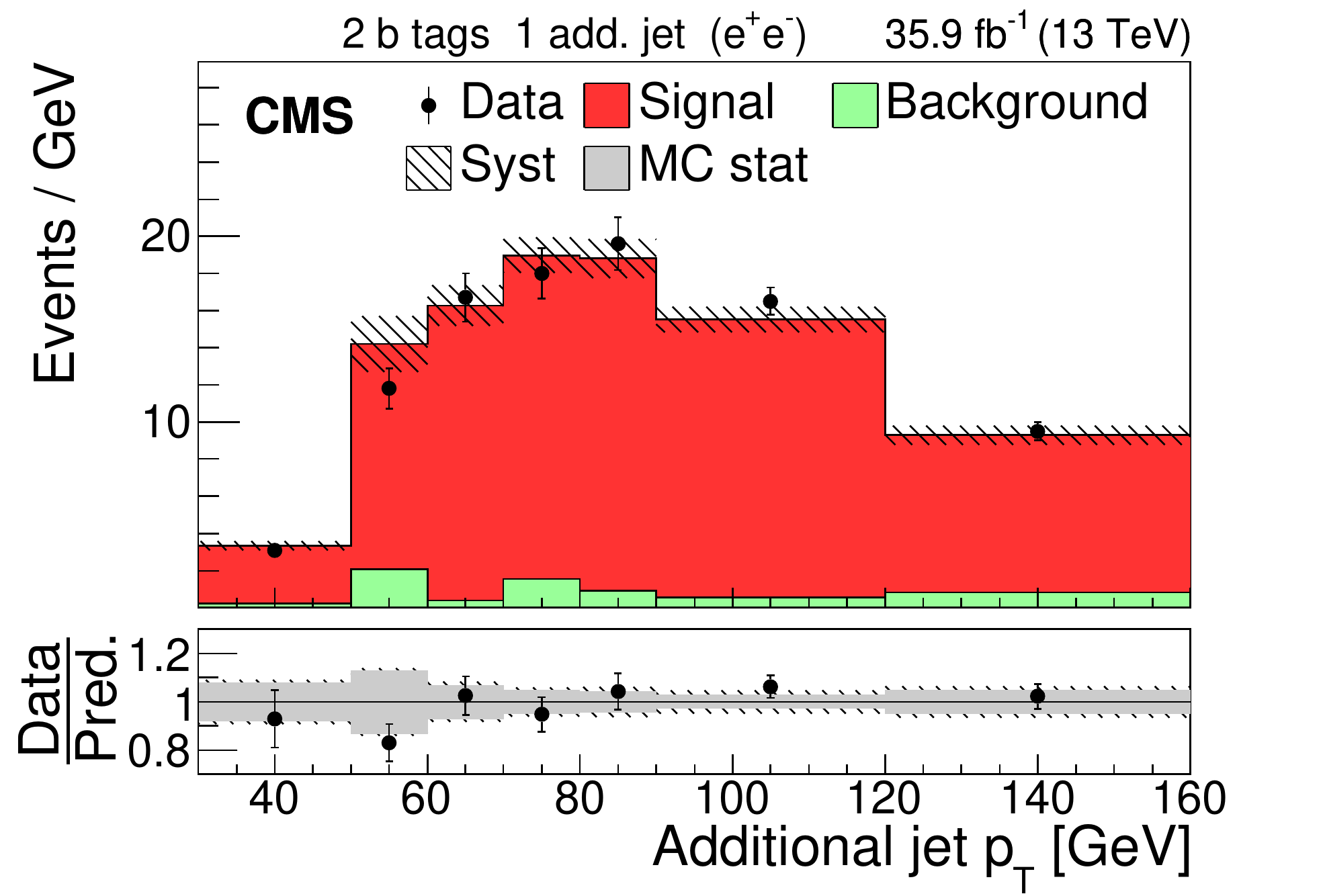}\\

    \includegraphics[width=0.325\textwidth]{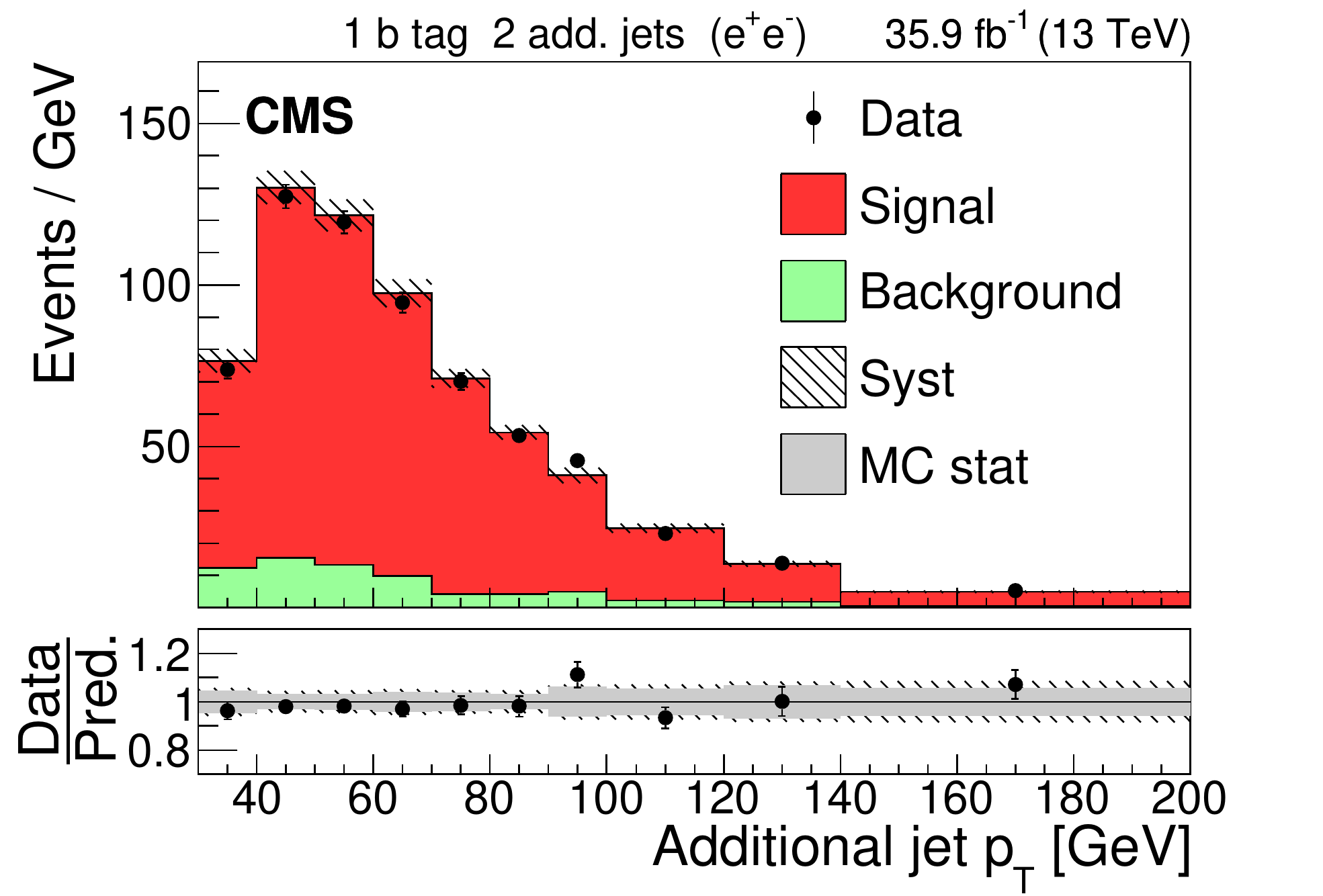}
    \includegraphics[width=0.325\textwidth]{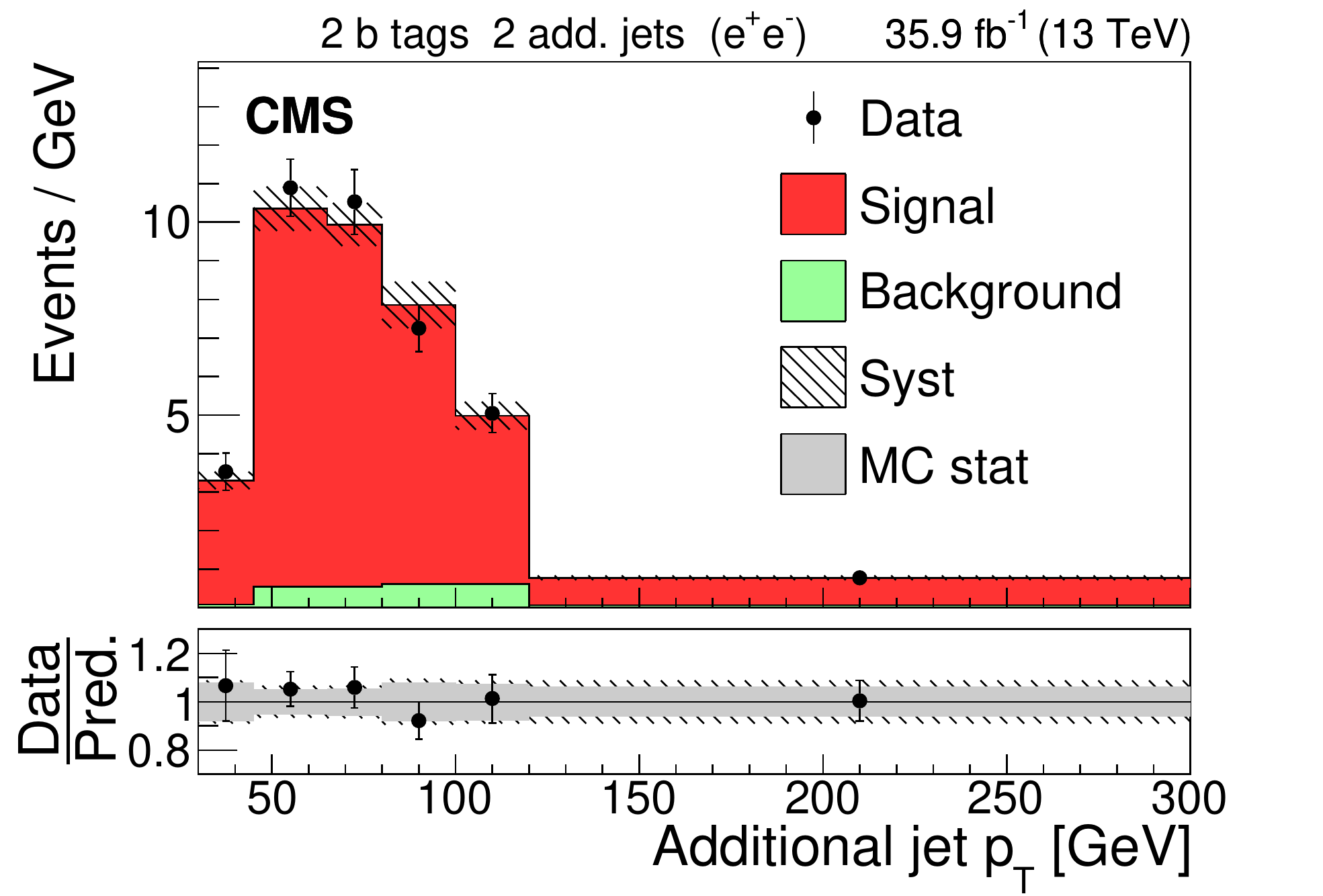}\\

    \includegraphics[width=0.325\textwidth]{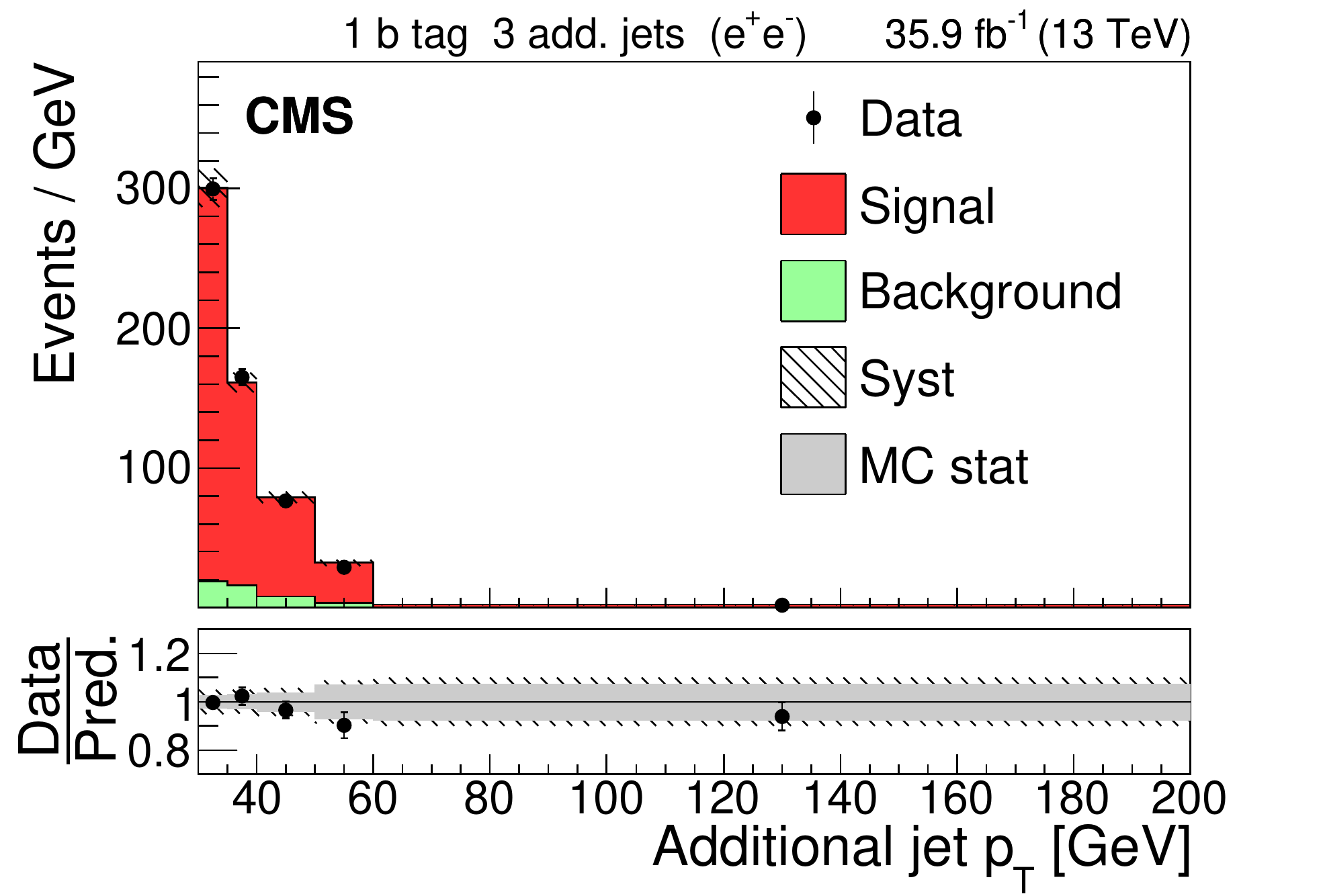}
    \includegraphics[width=0.325\textwidth]{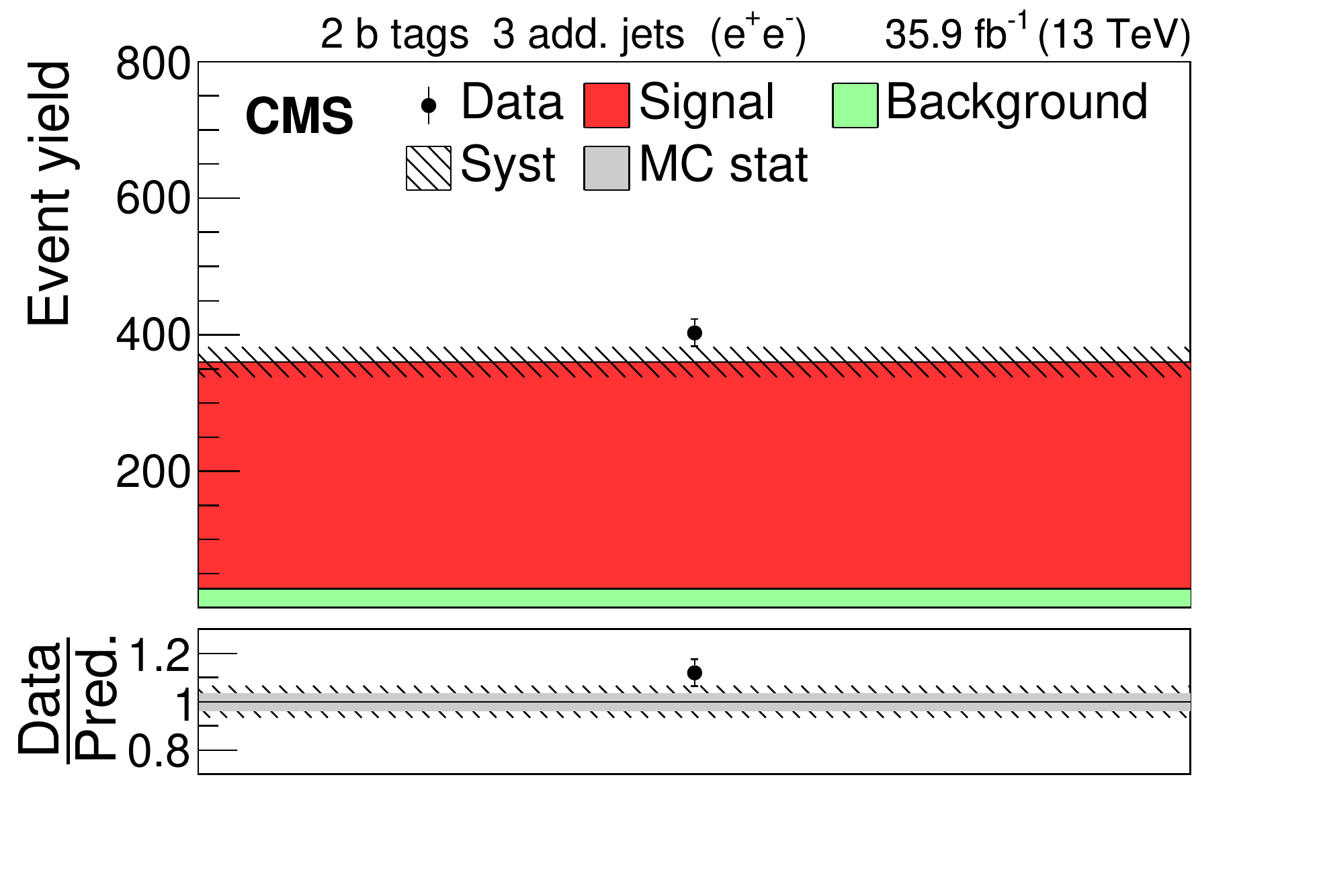}

\caption{Same distributions as in Fig.~\ref{fig:lh_mumu_postfitdistr8}, but in the \ee channel.
       \label{fig:lh_ee_postfitdistr8}}
\end{figure*}

The input distributions to the fit are shown in Figs.~\ref{fig:lh_emu_postfitdistr8}--\ref{fig:lh_ee_postfitdistr8},
where the data are compared to the signal and background distributions resulting from the fit to the data. In the top row, the number of events without additional non-\cPqb-tagged jets is displayed. For events with at least one additional non-\cPqb-tagged jet, the \pt distributions of the non-\cPqb-tagged jet with the smallest \pt in the respective category is considered, except for the category corresponding to events with 2 \cPqb-tagged jets and at least three additional non-\cPqb-tagged jets, where the statistical uncertainty of the simulation is high.
This distribution is chosen in order to constrain the jet energy scale at lower jet \pt, where the corresponding systematic uncertainty is larger~\cite{Khachatryan:2016kdb}.
Good agreement is found between the data and the simulation.

\section{Systematic uncertainties}
\label{sec:systematics}

The contributions from each source of systematic uncertainty are represented by nuisance parameters (see Section~\ref{sec:fit}). For each uncertainty, the simulation is used to construct template
histograms that describe the expected signal and background distributions for a given nuisance parameter variation. In the fit of the templates to the data, the best values for \sttvis (and \mtmc in the case of the combined fit) and all nuisance parameters are determined, as described in Section~\ref{sec:fit}. The prior probability
density functions for the nuisance parameters have a Gaussian shape. Table~\ref{tab:lh_syst_sum} shows the value of the contributions of the uncertainties after the fit.

\begin{table}[htbp!]
\centering \topcaption{\label{tab:lh_syst_sum}
The relative uncertainties in \sttvis and \stt and their sources, as obtained from the template fit. The uncertainty in the integrated luminosity and the MC statistical uncertainty are determined separately. The individual uncertainties are given without their correlations, which are however accounted for in the total uncertainties. Extrapolation uncertainties only affect \stt. For these uncertainties, the $\pm$ notation is used if a positive variation produces an increase in \stt, while the $\mp$ notation is used otherwise.}
\begin{tabular}{ l  c }
Source & Uncertainty [\%] \\ \hline
Trigger & ${0.3}$ \\
Lepton ident./isolation & ${2.0}$ \\
Muon momentum scale & ${0.1}$ \\
Electron momentum scale & ${0.1}$ \\
Jet energy scale & ${0.4}$ \\
Jet energy resolution & ${0.4}$ \\
\cPqb~tagging & ${0.4}$ \\
Pileup & ${0.1}$ \\
\ttbar ME scale & ${0.2}$ \\
\tW ME scale & ${0.2}$ \\
DY ME scale & ${0.1}$ \\
PDF & ${1.1}$ \\
Top quark \pt & ${0.5}$ \\
ME/PS matching & ${0.2}$ \\
UE tune & ${0.3}$ \\
\ttbar ISR scale & ${0.4}$ \\
\tW ISR scale & ${0.1}$ \\
\ttbar FSR scale & ${0.8}$ \\
\tW FSR scale & ${0.1}$ \\
\cPqb~quark fragmentation & ${0.7}$ \\
\cPqb~hadron BF & ${0.1}$ \\
Colour reconnection & ${0.3}$ \\
DY background & ${0.9}$ \\
\tW background & ${1.1}$ \\
Diboson background & ${0.2}$ \\
\Wjets background & ${0.2}$ \\
\ttbar background & ${0.2}$ \\
Statistical & ${0.2}$ \\
Integrated luminosity & ${2.5}$ \\
MC statistical  & ${1.1}$ \\
\rule{0pt}{2.6ex}Total \sttvis uncertainty & ${3.8}$ \\[0.15cm]
\multicolumn{2}{l}{Extrapolation uncertainties} \\ [0.05cm] \hline
\rule{0pt}{2.3ex}\ttbar ME scale  & $\mp^{0.3}_{0.1}$ \\
\rule{0pt}{2.3ex}PDF  & $\pm^{0.8}_{0.6}$ \\
\rule{0pt}{2.3ex}Top quark \pt  & $\mp^{0.5}_{<0.1}$ \\
\rule{0pt}{2.3ex}\ttbar ISR scale  & $\mp^{0.1}_{<0.1}$ \\
\rule{0pt}{2.3ex}\ttbar FSR scale  & $\pm^{0.1}_{<0.1}$ \\
\rule{0pt}{2.3ex}UE tune  & $<$0.1\rule[-1.2ex]{0pt}{0pt}\\
\rule{0pt}{2.5ex}Total \stt uncertainty & $4.0 $ \\
\end{tabular}
\end{table}

Most of the experimental uncertainties are determined from ancillary measurements in which data and simulation are compared and small corrections to the simulation, referred to as scale factors (SFs), are determined. To assess the impact of the uncertainty in these corrections, the SFs are varied within their uncertainty and the analysis is repeated.

The trigger efficiencies are determined using multiple independent methods, which show agreement
within 0.3\%. An additional statistical uncertainty arises because the SFs are determined from the data in intervals of \pt and $\eta$.

The uncertainty in the SFs of the lepton identification efficiency is typically 1.5\% for electrons and 1.2\% for muons, with a small dependency on the lepton \pt and $\eta$. The uncertainties in the calibration of the muon and electron momentum scales are included as nuisance parameters for each lepton separately. Their impact on the measurement is negligible.

The impact of the jet energy scale (JES) uncertainties is estimated by varying the jet momenta within
the JES uncertainties, split into 18 contributions~\cite{Khachatryan:2016kdb}. To account for the jet energy resolution (JER), the SFs are varied within their $\abs{\eta}$-dependent uncertainties~\cite{CMS-PAS-JME-16-003}.

The uncertainties associated with the \cPqb~tagging efficiency are determined by varying the related corrections for the simulation of \cPqb~jets and light-flavour jets, split into 16 orthogonal contributions for \cPqb~jets. These uncertainties depend on the \pt of each jet and amount to approximately 1.5\% for \cPqb~jets in \ttbar signal events~\cite{Sirunyan:2017ezt}.

The uncertainty in the modelling of the number of pileup events is obtained by changing the inelastic \pp cross section, which is used to model the pileup in simulation, by $\pm$4.6\%~\cite{Aaboud:2016mmw}.

The integrated luminosity uncertainty is not included in the fit as a nuisance parameter, but treated as an external uncertainty. It is estimated to be 2.5\%~\cite{CMS-PAS-LUM-17-001}.

The ME scale uncertainties for the simulation of the \ttbar and DY are assessed by varying the renormalization and factorization scale choices in \POWHEG by factors of two up and down independently~\cite{Cacciari:2003fi,Catani:2003zt}, avoiding cases where $\muf/\mur= 1/4$ or 4.

To estimate the uncertainty due to the NLO generator, the \POWHEG \ttbar signal sample is replaced by a \ttbar sample generated using the \MGvATNLO program with FxFx matching~\cite{Frederix:2012ps}. This uncertainty is only included in the  combined measurement of \stt and \mtmc (Section~\ref{sec:simultaneous_fit}) in order to compare with the latest direct top quark mass measurement from CMS in the lepton+jets channel~\cite{Sirunyan:2018gqx}.

The PDF uncertainty is estimated using the 28 orthogonal Hessian eigenvectors of the CT14~\cite{Dulat:2015mca} PDF, which are used as independent inputs to the fit.

Differential measurements of \stt at $\sqrts = 13 \TeV$ have demonstrated that the \pt distribution of the top quark is softer than predicted by the \POWHEG simulation~\cite{Sirunyan:2018ucr,Sirunyan:2017mzl,Khachatryan:2016mnb}. An additional uncertainty, referred to as ``Top quark \pt'', is estimated by reweighting the simulation. This nuisance parameter has a one-sided prior distribution.

The uncertainty due to the matching of the ME to the PS in simulation is estimated by varying the $h_{\text{damp}}$ parameter in \POWHEG, as described in Ref.~\cite{CMS-PAS-TOP-16-021}. The uncertainty due to the assumptions in the UE tune is estimated by varying the tuning parameters~\cite{CMS-PAS-TOP-16-021}. The impact of the PS scale uncertainty is estimated by varying the initial-state radiation (ISR) and the final-state radiation (FSR) scales by a factor of two up and down~\cite{Skands:2014pea}, similar to the case of renormalization and factorization scales.

The uncertainties due to the assumed \cPqb~hadron branching fraction (BF) and fragmentation are taken into account following the procedures described in Ref.~\cite{Sirunyan:2018gqx}. For the fragmentation, variations of the Bowler--Lund fragmentation function~\cite{Bowler:1981sb} and the comparison to the Peterson fragmentation function~\cite{Peterson:1982ak} are considered.

The effects of colour reconnection (CR) processes on the top quark final state are estimated by enabling early resonance decays (ERD) in \PYTHIA. In the nominal sample, ERD are turned off. Alternative colour reconnection models are considered, such as ``gluon move''~\cite{Argyropoulos:2014zoa} and ``QCD inspired''~\cite{Christiansen:2015yqa}, since they were found to potentially have relevant effects for the measurement of the top quark mass~\cite{Sirunyan:2018gqx}.

For the uncertainties related to the background contributions, prior normalization uncertainties of 30\% are assumed~\cite{Khachatryan:2015uqb}. The contributions of these uncertainties are small and/or strongly constrained in the fit. For the DY background, separate nuisance parameters are used for each \cPqb-tagged jet category in order to remove the dependence of the fit result on the prediction of the \cPqb-tagged jet multiplicity distribution by the DY MC simulation. Similarly, the DY background is given an additional uncertainty of 5, 10, 30, and 50\% for events with exactly 0, 1, 2, and 3 or more jets, respectively. The first three numbers are estimated by performing scale variations in \Wjets predictions with NLO precision, whereas the last one is assigned conservatively.

\begin{figure*}[htb]
  \centering
    \includegraphics[width=0.85\textwidth]{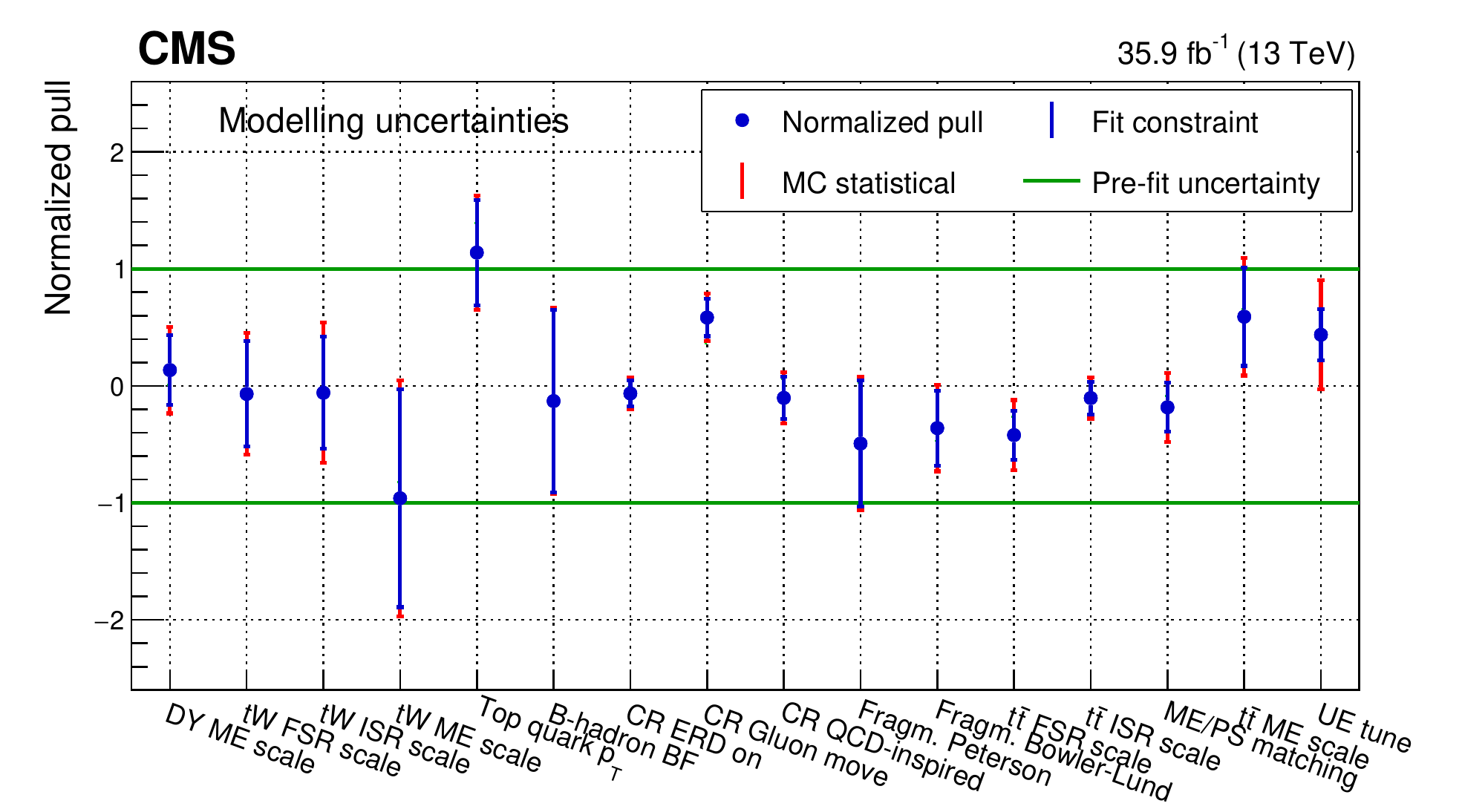}
\caption{Normalized pulls and constraints of the nuisance parameters related to the modelling uncertainties for the cross section fit. The markers denote the fitted values, while the inner vertical bars represent the constraint and the outer vertical bars denote the additional uncertainty as determined from pseudo-experiments. The constraint is defined as the ratio of the post-fit uncertainty to the pre-fit uncertainty of a given nuisance parameter, while the normalized pull is the difference between the post-fit and the pre-fit values of the nuisance parameter normalized to its pre-fit uncertainty. The horizontal lines at $\pm 1$ represent the pre-fit uncertainty.}
       \label{fig:topxsec_pulls}
\end{figure*}

In total, 103 uncertainty sources are used in the fit. In Fig.~\ref{fig:topxsec_pulls}, the normalized pulls and constraints for the nuisance parameters related to the modelling uncertainties are shown. For each nuisance parameter, the normalized pull is defined as the difference between the best-fit and the input values, normalized to the pre-fit uncertainty, and the constraint is defined as the ratio of the post-fit to the pre-fit uncertainty.
The vast majority of the nuisance parameters lie within one standard deviation of their priors, reflecting the good agreement of the nominal simulation with the data.
Most \ttbar\ signal uncertainties show significant constraints with respect to their prior uncertainty, illustrating the strength of the analysis ansatz.
The nuisance parameter for the \pt distribution of the top quarks is pulled by one standard deviation. This is expected since it is known that the observed \pt distribution of the top quark is softer than predicted by the simulation~\cite{Sirunyan:2017mzl,Khachatryan:2016mnb}.

\section{Cross section measurement}
\label{sec:crosssection}

The visible cross section is defined for \ttbar events in the fiducial region with two oppositely charged leptons (electron or muon). Contributions from leptonically decaying \Pgt\xspace leptons are included. The leading lepton is required to have $\pt > 25\GeV$, and the subleading lepton must have $\pt>20\GeV$. Both leptons have to be in the range $\abs{\eta} < 2.4$.
From the likelihood fit, described in Section~\ref{sec:fit}, the visible cross section is measured to be
\begin{eqnarray*}
\sttvis & = & \resultxsecvismain.
\end{eqnarray*}
Here, the uncertainties denote the statistical uncertainty, the systematic uncertainty, and that coming from the uncertainty in the integrated luminosity. The full list of uncertainties is presented in Table~\ref{tab:lh_syst_sum}.

The total cross section \stt is obtained by extrapolating the measured visible cross section to the full phase space. As explained in Section~\ref{sec:fit}, the extrapolation is described by a
multiplicative acceptance correction factor $A_{\ell\ell}$ (see Eq.~(\ref{eq:extrapol})). The extrapolation uncertainty is determined for each relevant model systematic source $j$ as described in the following:
all
nuisance parameters except the one under study are fixed to their post-fit values; the nuisance parameter $\lambda_j$ is set to values $+1$ and $-1$, and the variations of $A_{\ell\ell}$ are recorded.
The resulting variations of \stt with respect to the nominal value, obtained with the post-fit value of $\lambda_j$, are taken as the additional extrapolation uncertainties. The individual uncertainties in \stt from these sources are summed in quadrature to estimate the total systematic uncertainty, as summarized in Table~\ref{tab:lh_syst_sum}.
A fixed value of  $\mtmc = 172.5 \GeV$ is chosen in the simulation, and no uncertainty
is assigned.

The total cross section \stt is measured to be
\begin{eqnarray*}
\sigma_{\ttbar} & = & \resultxsecmain.
\end{eqnarray*}
{\tolerance=800
As shown in Table~\ref{tab:lh_syst_sum}, in comparison to the fiducial cross section, the relative systematic uncertainty in the total cross section is marginally increased.
The result is in good agreement with the theoretical calculation at NNLO+NNLL, which predicts a \ttbar
cross section of \xsectheo, as described in Section~\ref{sec:detector}. \par
}

An independent cross section measurement is performed using a simple event-counting method and a more restrictive event selection, following closely the analysis of Ref.~\cite{Khachatryan:2016kzg}. The analysis uses events in the \emu channel with at least two jets, at least one of which is \cPqb~tagged.
The cross section is measured to be $ \stt =  \resultxseccheck $, in good agreement with the main result.

\section{Simultaneous measurement of \texorpdfstring{\stt and \mtmc}{cross section and mass}}
\label{sec:simultaneous_fit}

The analysis is designed such that the dependence of the measured \ttbar cross section on \mtmc is small. However, because of the impact of the top quark mass on the simulated detector efficiency and acceptance, the measurement is expected to have a residual dependence on the chosen value of \mtmc. In previous measurements, this dependence was determined by repeating the analysis with varied mass values.

Here, the approach proposed in Refs.~\cite{Kieseler:2015jzh,Sirunyan:2017uhy} is followed. The value of \mtmc is introduced in the fit as an additional free parameter. In the simultaneous fit, \stt and \mtmc are directly constrained from the data. The resulting \stt and its uncertainty therefore account for the dependence on \mtmc and can be used, \eg for the extraction of \mt and \as using fixed-order calculations. The value of \mtmc, in turn, can be compared to the results of direct measurements using, \eg kinematic fits~\cite{Sirunyan:2018gqx}.

\begin{figure*}[htbp!]

  \centering

    \includegraphics[width=0.325\textwidth]{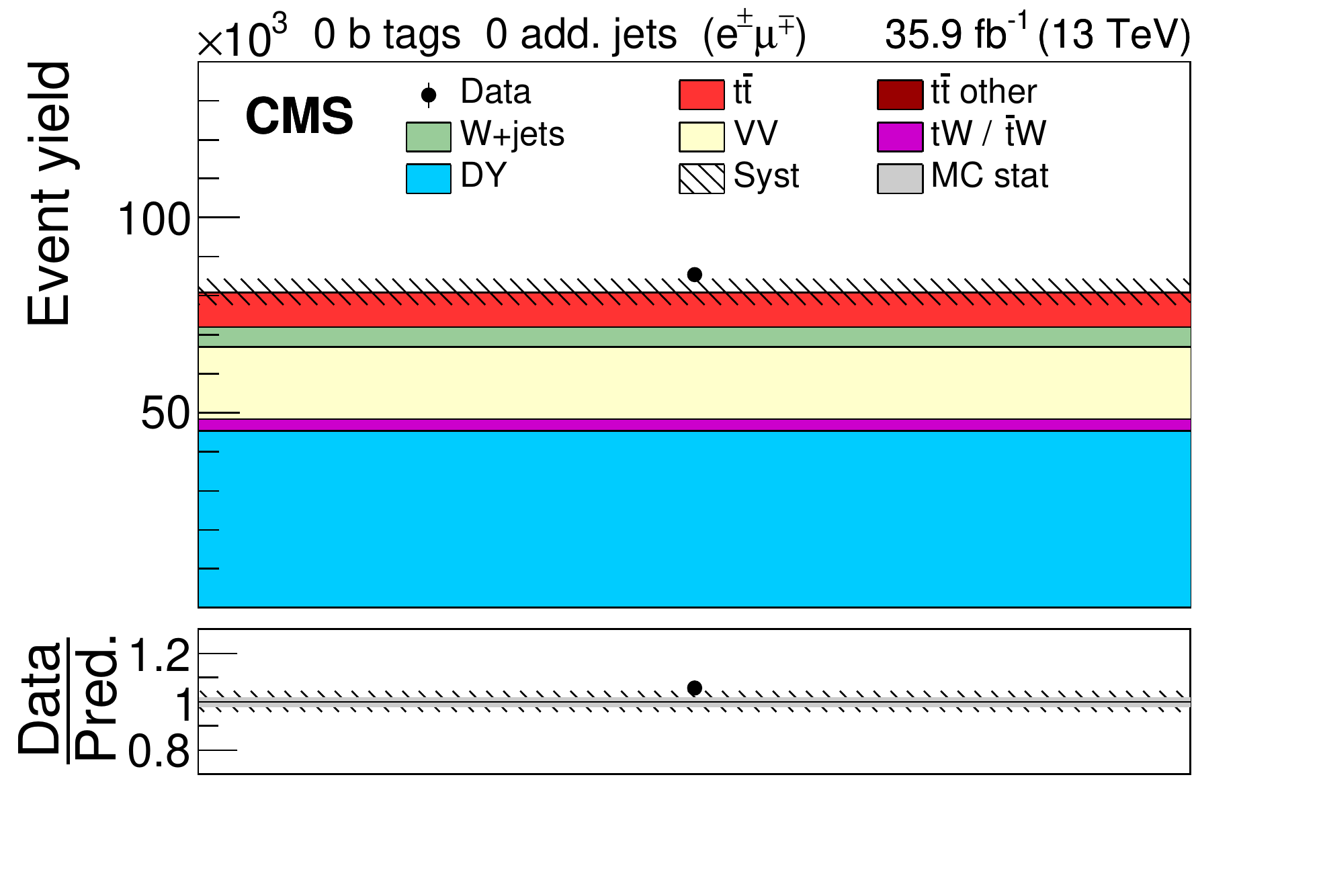}
    \includegraphics[width=0.325\textwidth]{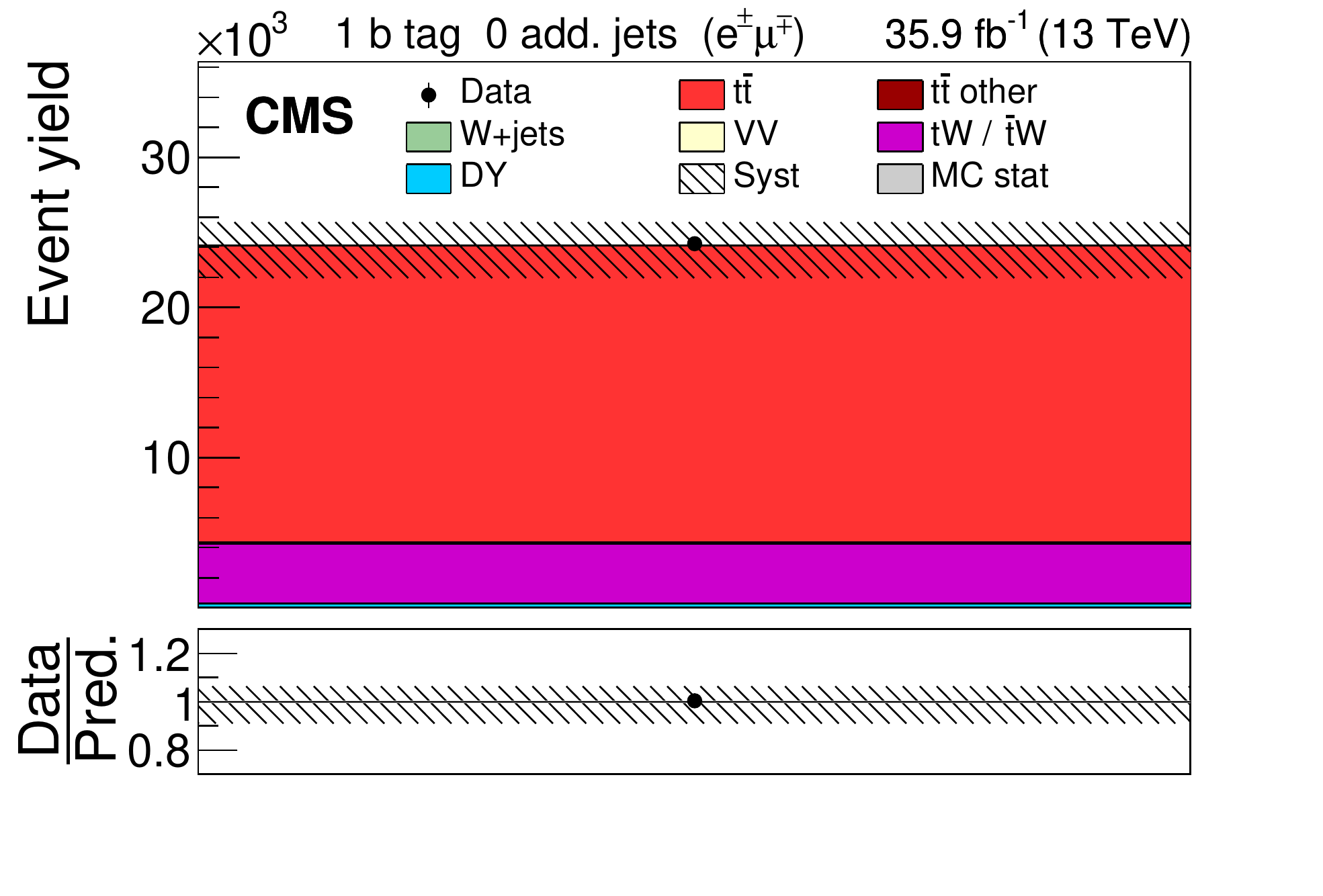}
    \includegraphics[width=0.325\textwidth]{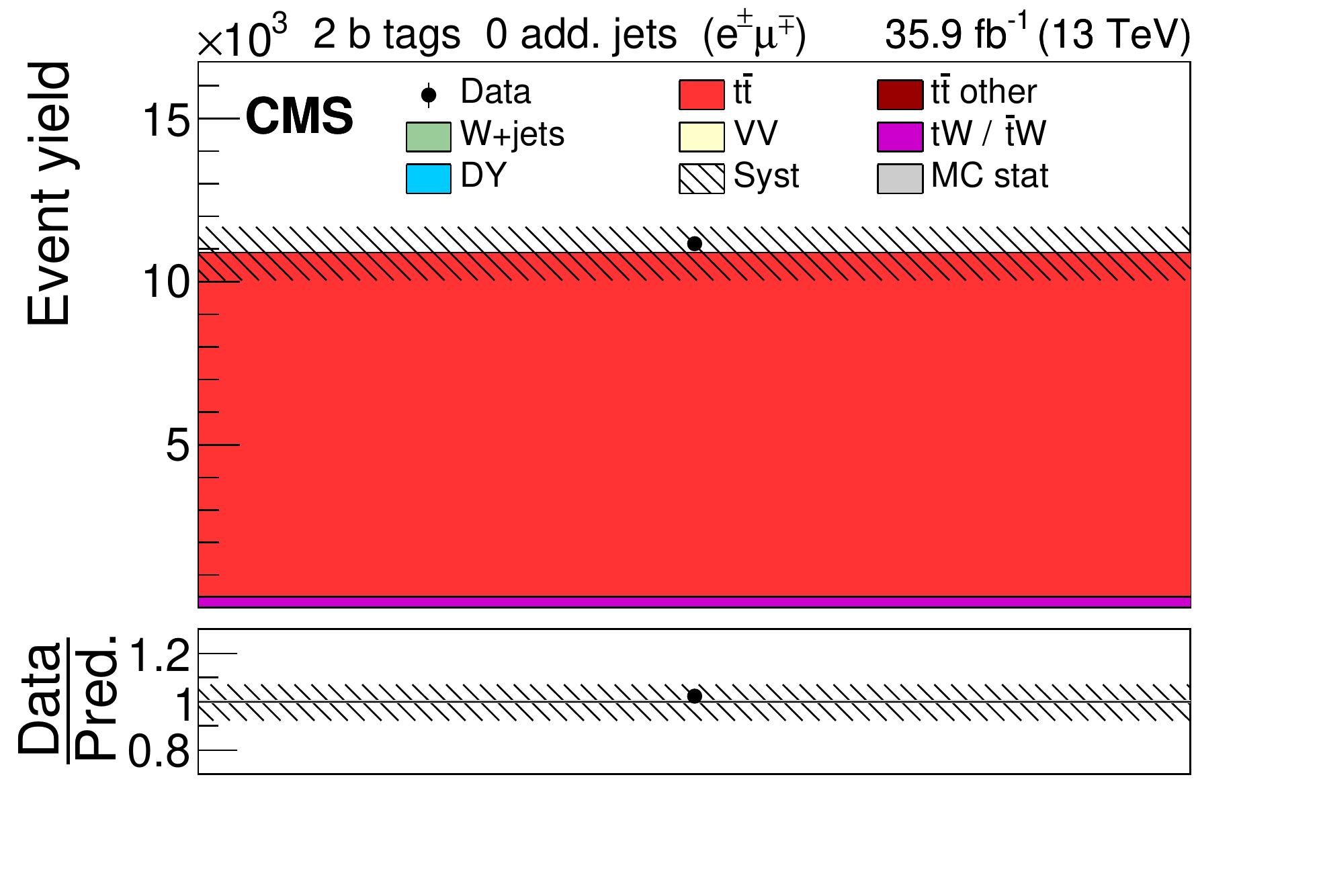}

    \includegraphics[width=0.325\textwidth]{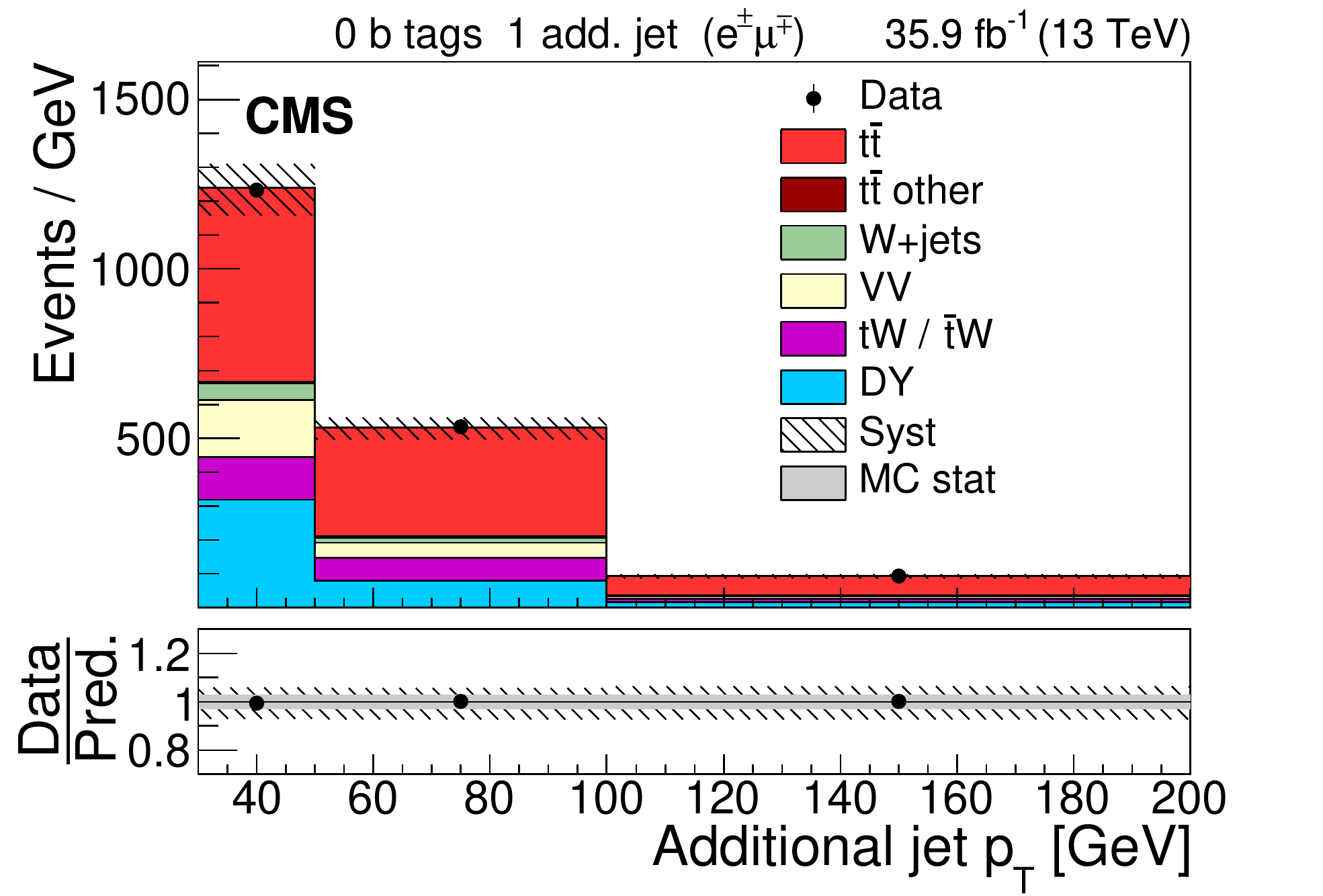}
    \includegraphics[width=0.325\textwidth]{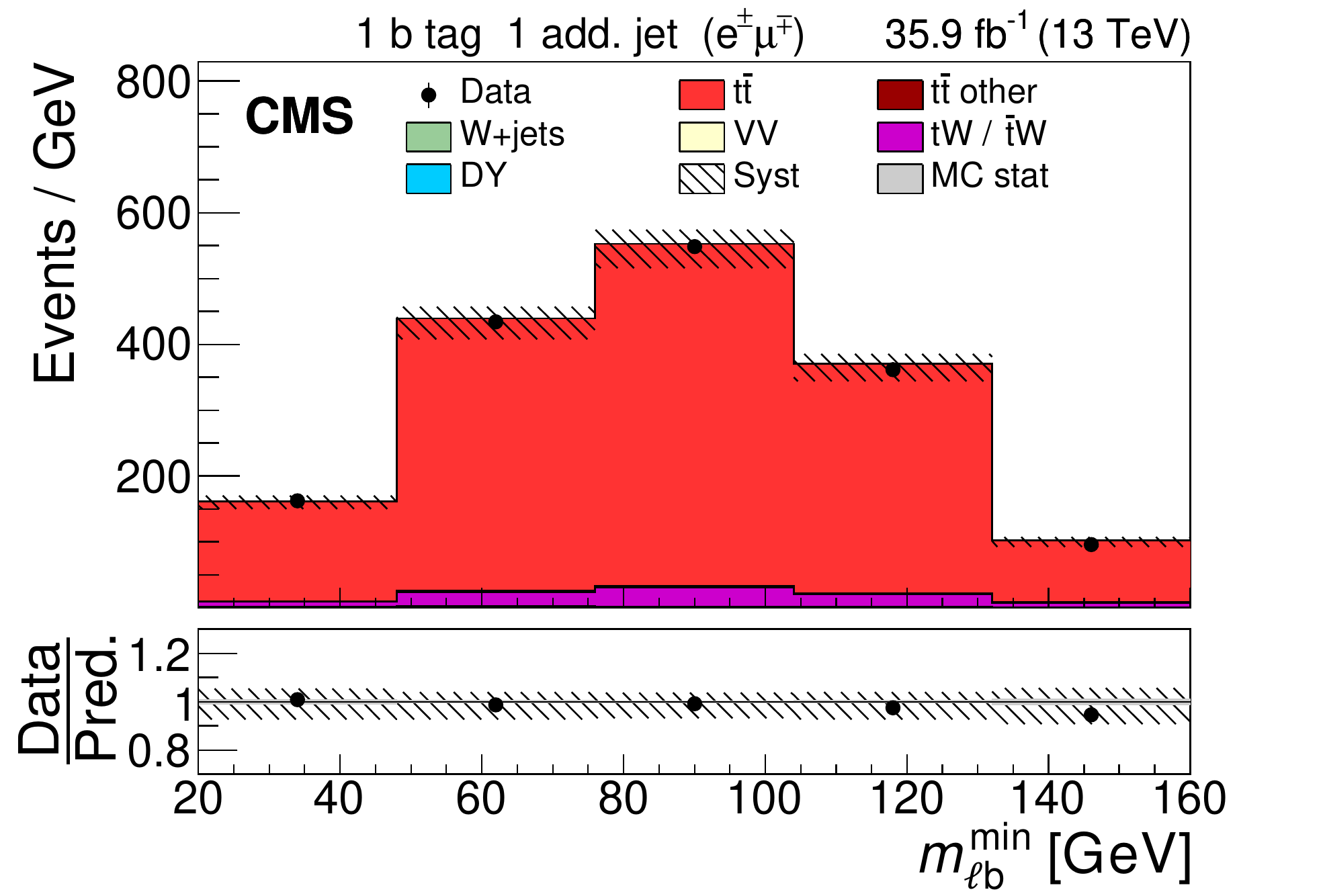}
    \includegraphics[width=0.325\textwidth]{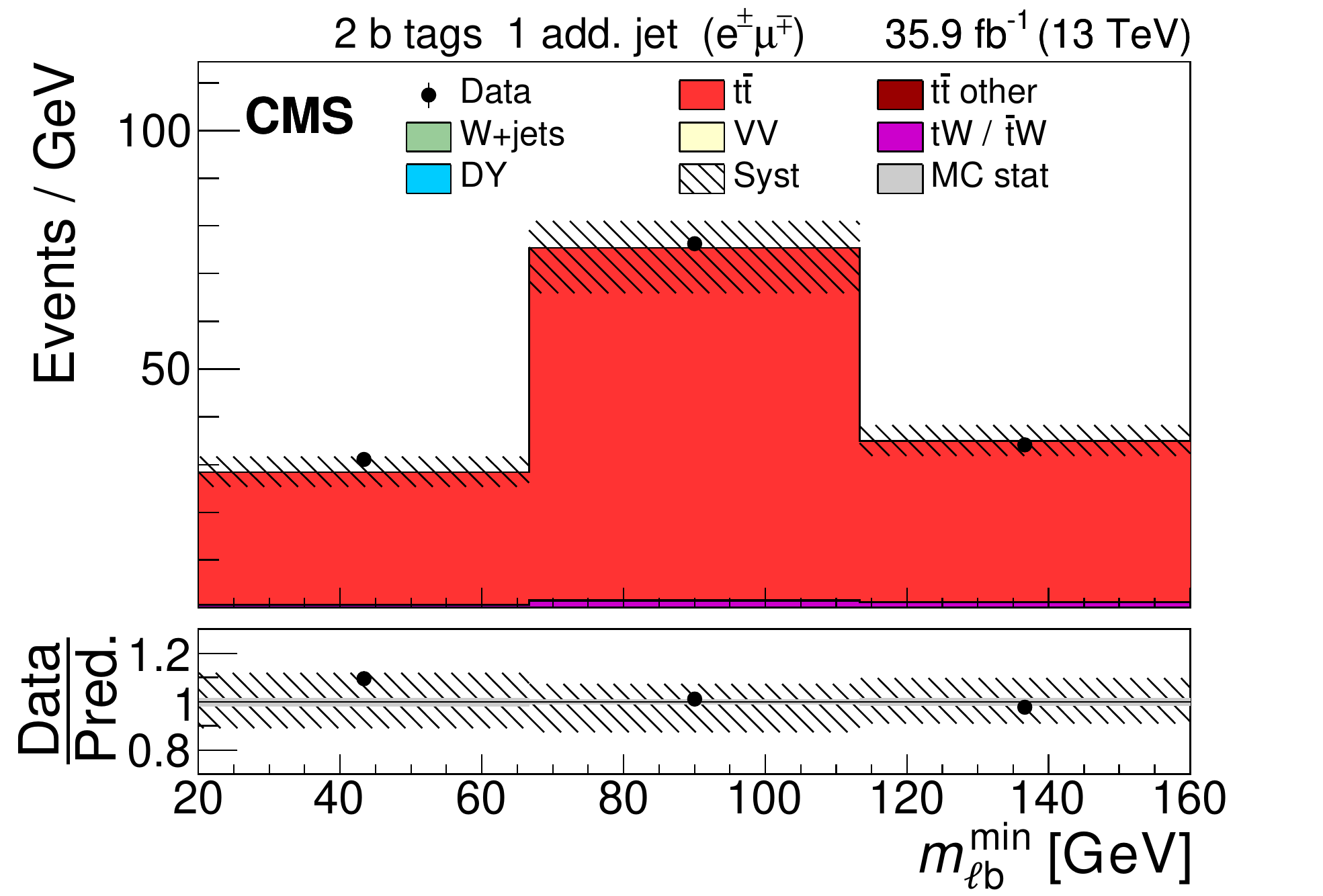}

    \includegraphics[width=0.325\textwidth]{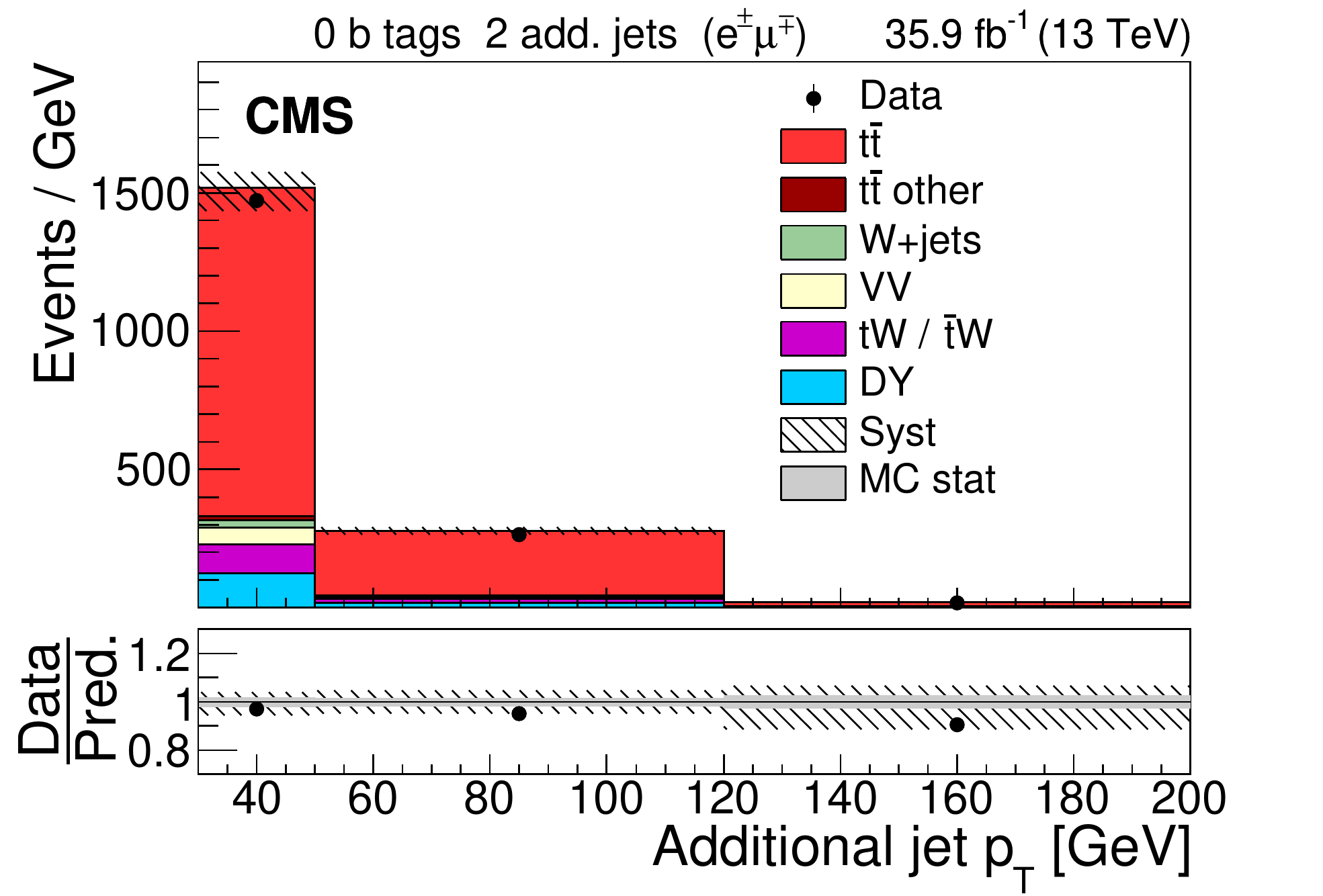}
    \includegraphics[width=0.325\textwidth]{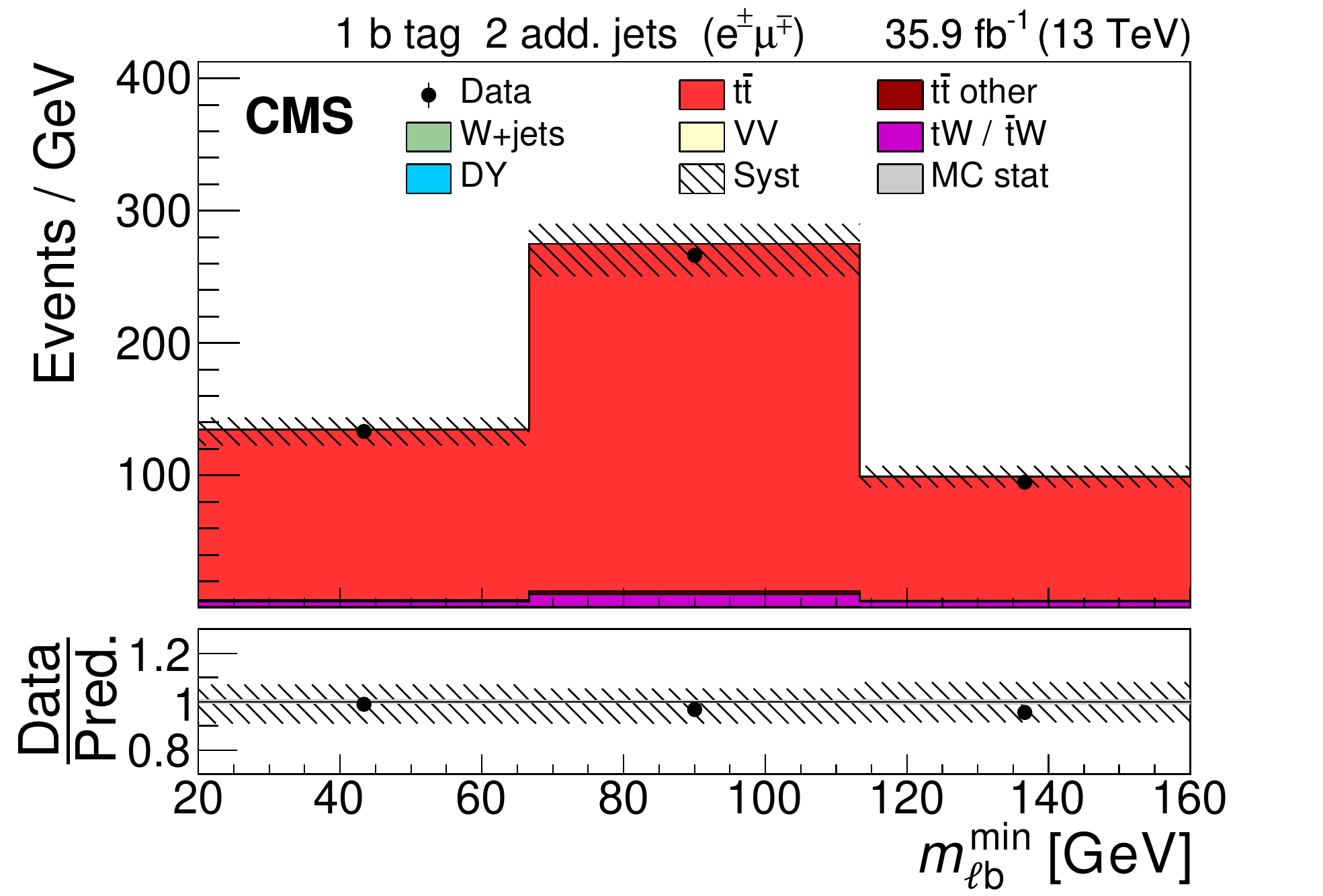}
    \includegraphics[width=0.325\textwidth]{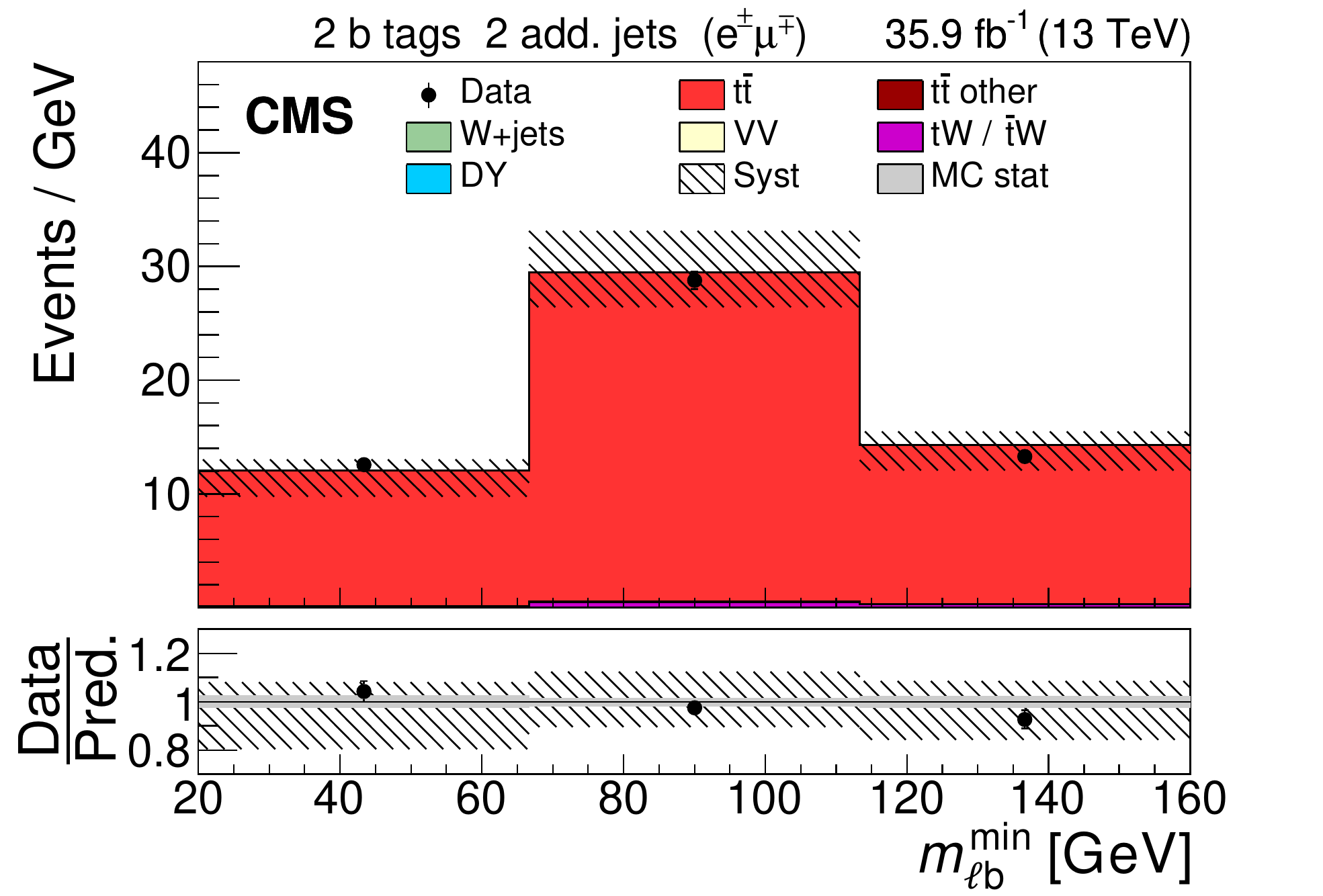}

    \includegraphics[width=0.325\textwidth]{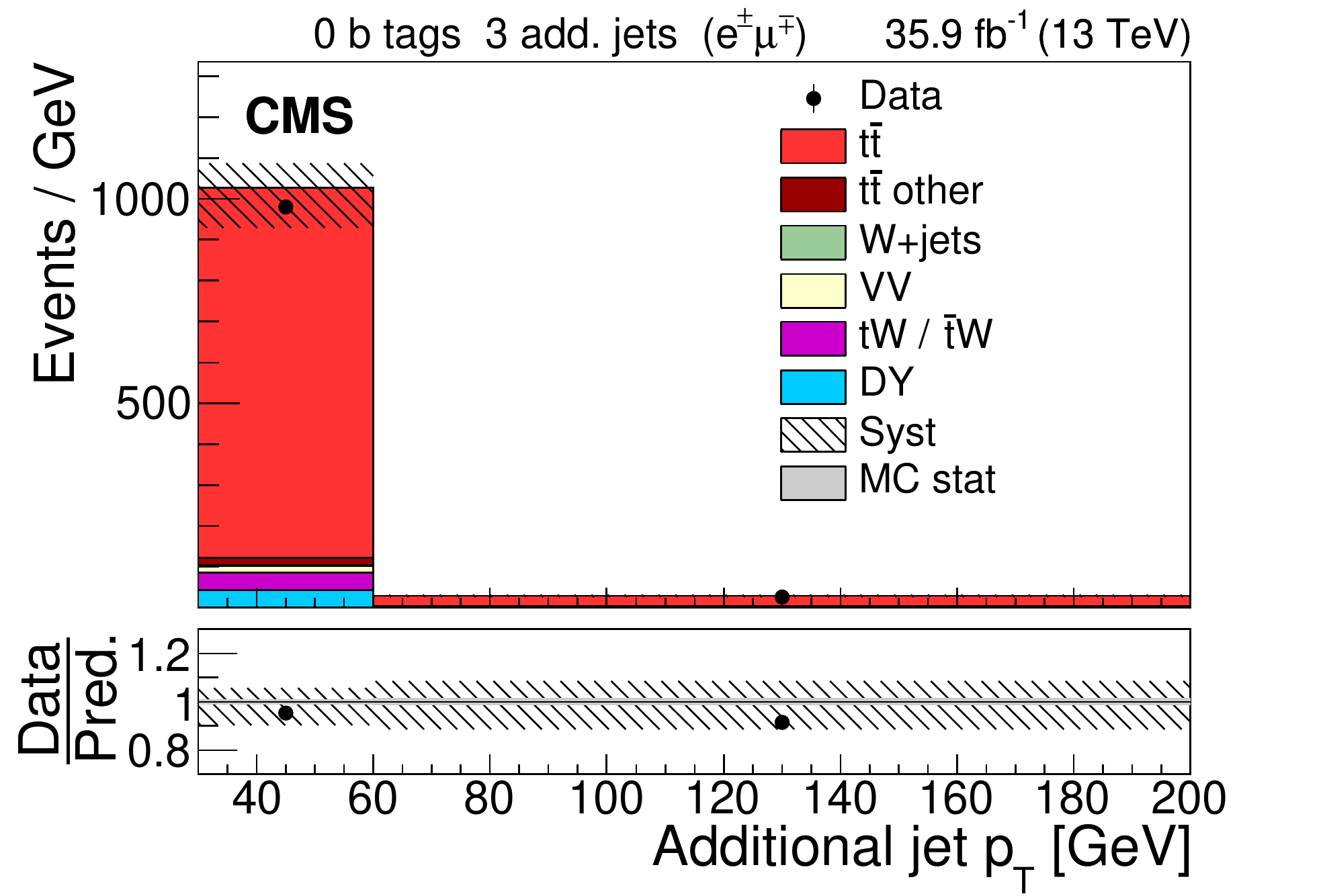}
    \includegraphics[width=0.325\textwidth]{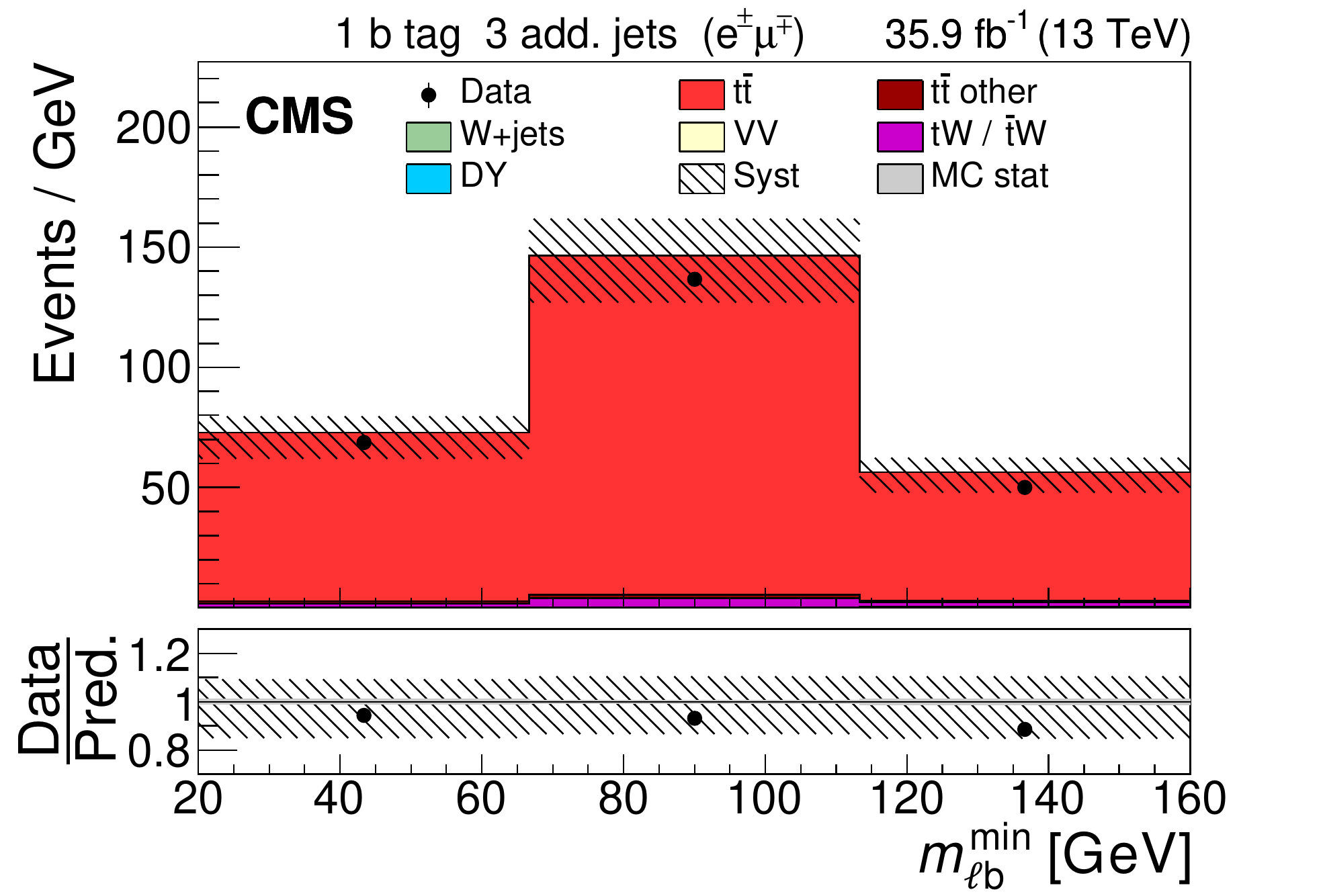}
    \includegraphics[width=0.325\textwidth]{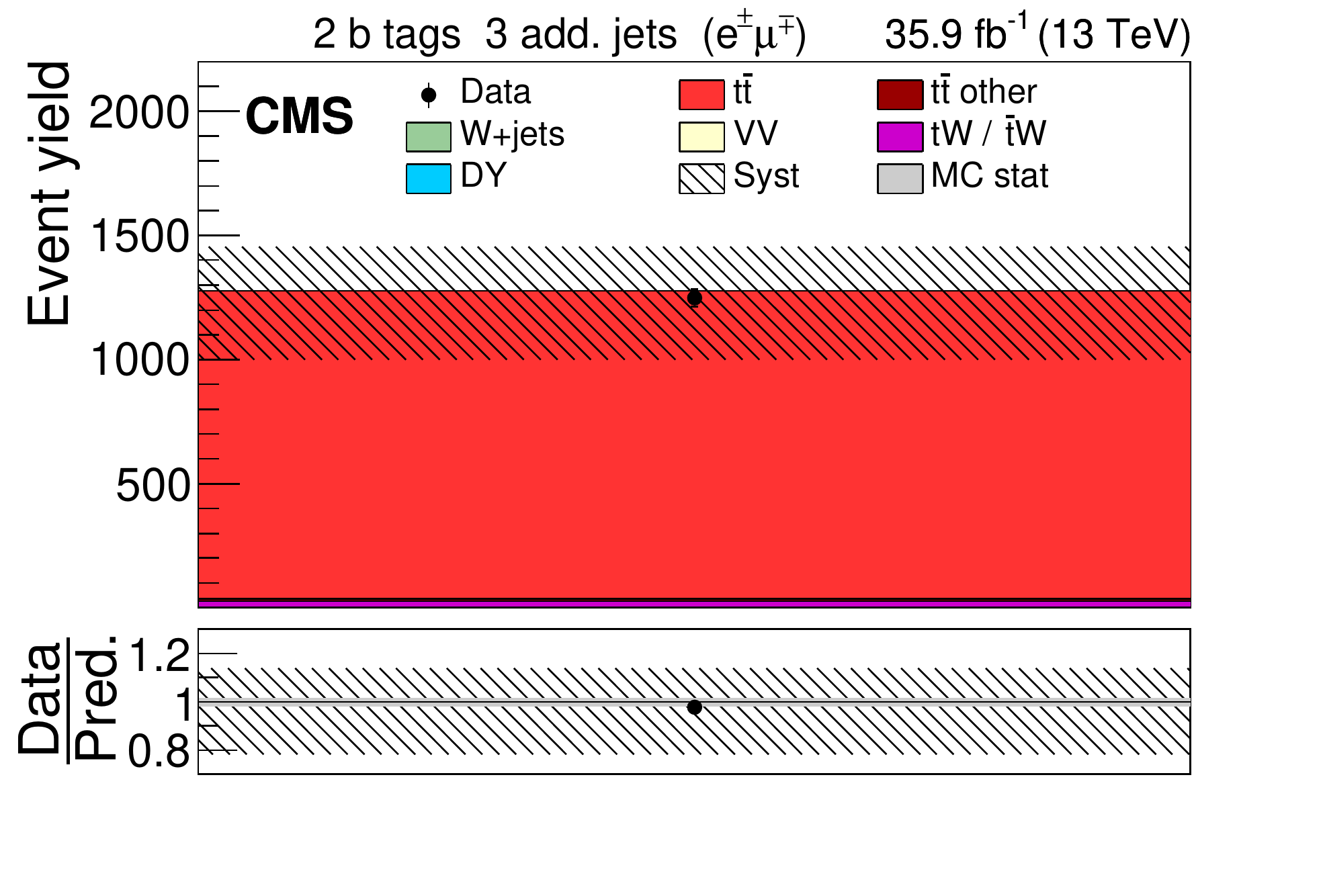}

\caption{Comparison of data (points) and pre-fit distributions of the
expected signal and backgrounds from simulation (shaded
histograms) used in the simultaneous fit of \stt and \mtmc in the \emu channel.
In the left column events with zero or three or more \cPqb-tagged jets are shown. The middle (right) column shows events with exactly one (two) \cPqb-tagged jets. Events with zero, one, two, or three or more additional non-\cPqb-tagged jets are shown in the first, second, third, and fourth row, respectively.
The hatched bands correspond to the total uncertainty in the sum of the predicted yields. The ratios of data to the sum of the predicted yields are shown in the lower panel of each figure. Here, the solid gray band represents the contribution of the statistical uncertainty.}
	\label{fig:lh_emu_inputdistr8_topmass}
\end{figure*}

In contrast to the \stt measurement presented in Section~\ref{sec:crosssection}, the sensitivity of the simultaneous fit to \mtmc is maximized by introducing a new observable: the minimum invariant mass
\mlb, which is defined as the smallest invariant mass found
when combining the charged leptons with the \cPqb~jets in an event.
To minimize the impact from background, only the \emu sample
is used.
The simultaneous fit of \stt and \mtmc is performed in 12 mutually exclusive categories, according to the number of \cPqb-tagged jets and of additional non-\cPqb-tagged jets in the event. The same observables as in Fig.~\ref{fig:lh_emu_postfitdistr8} are used as input to the fit, where the jet \pt spectrum is replaced by the \mlb distribution in categories with at least one \cPqb-tagged jet, as shown in Fig.~\ref{fig:lh_emu_inputdistr8_topmass}.

To construct the templates describing the dependence of the final-state distributions on \mtmc, separate MC simulation samples of \ttbar and \tW production are used in which \mtmc is varied in the range $\mtmc = 172.5 \pm 3 \GeV$.
The data and MC samples, the event selection, the modelling of the systematic uncertainties, and the fit procedure are identical to those described in Section~\ref{sec:fit}. In the simultaneous fit, the same systematic uncertainties are included as in a previous CMS measurement~\cite{Sirunyan:2018gqx} of the \mtmc. The results of the two measurements are thus directly comparable.

\begin{figure*}[htbp!]
  \centering
    \includegraphics[width=0.325\textwidth]{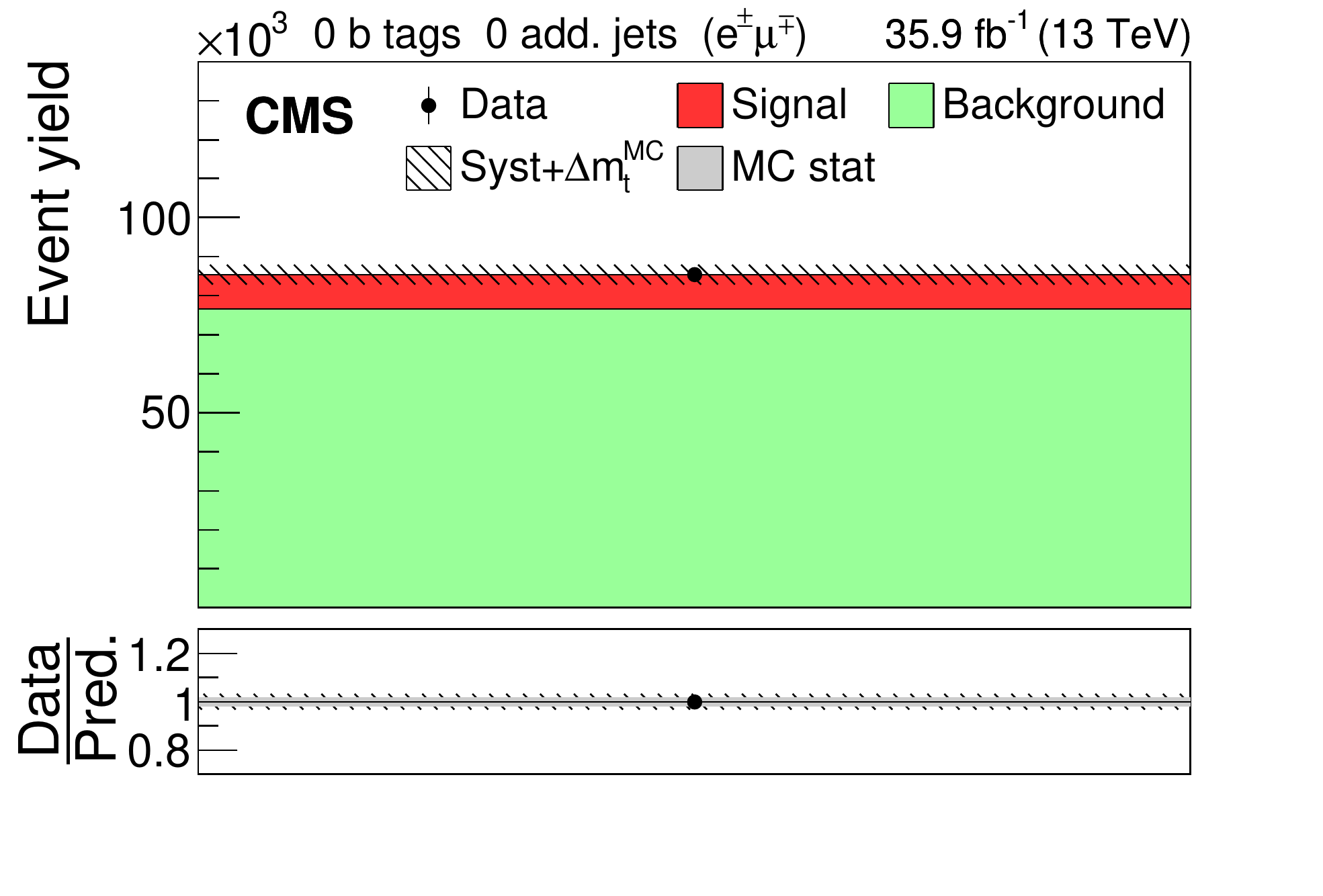}
    \includegraphics[width=0.325\textwidth]{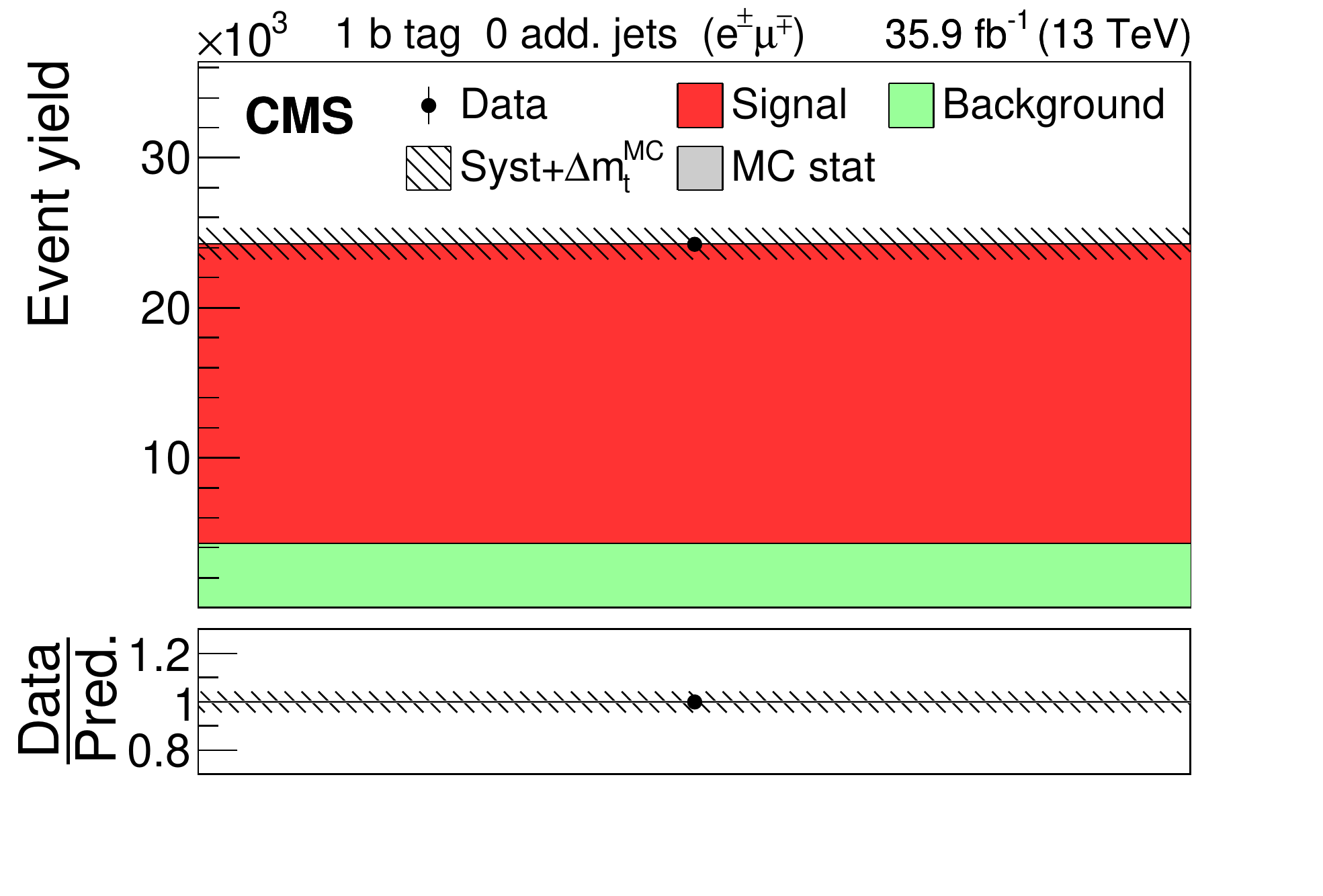}
    \includegraphics[width=0.325\textwidth]{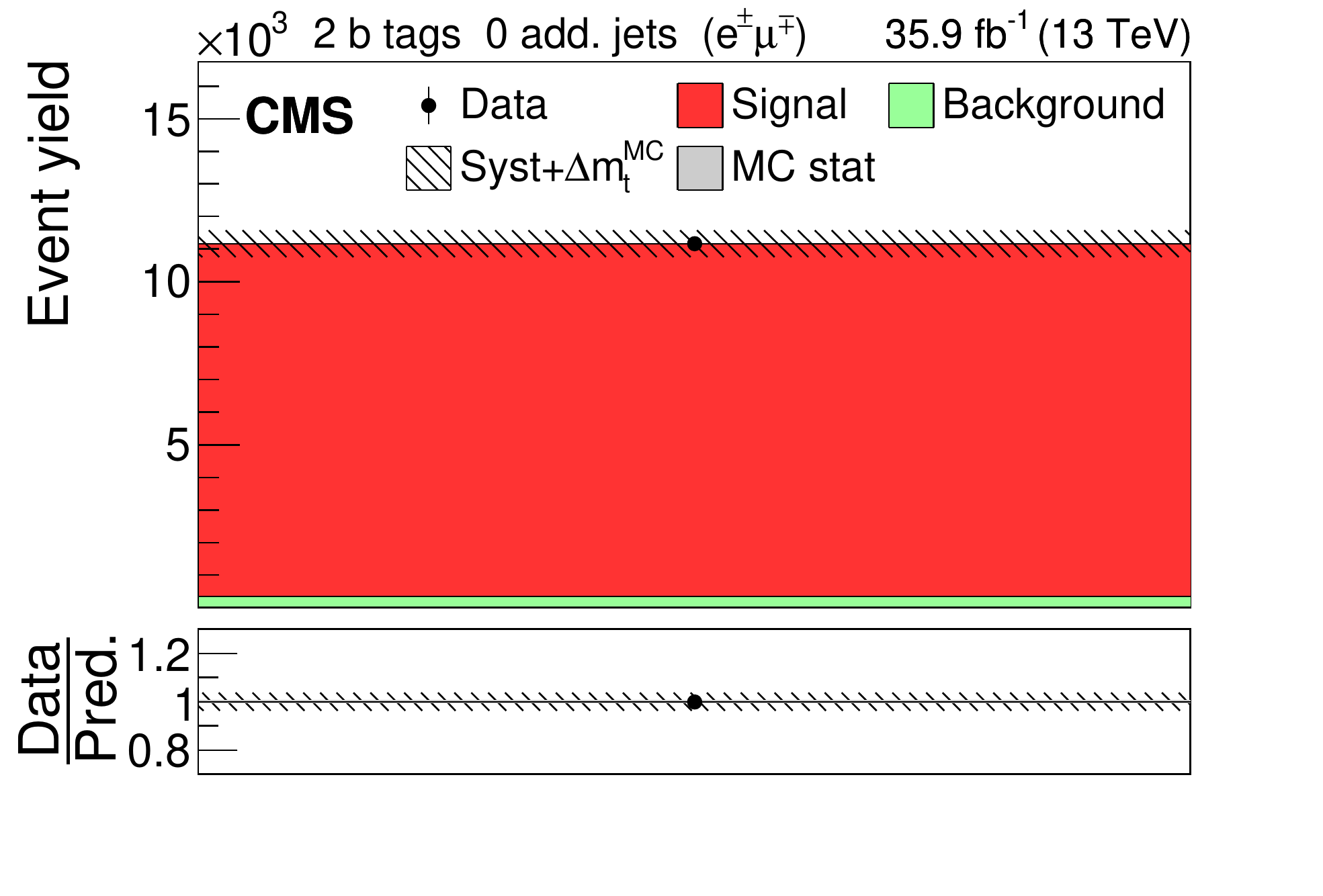}

    \includegraphics[width=0.325\textwidth]{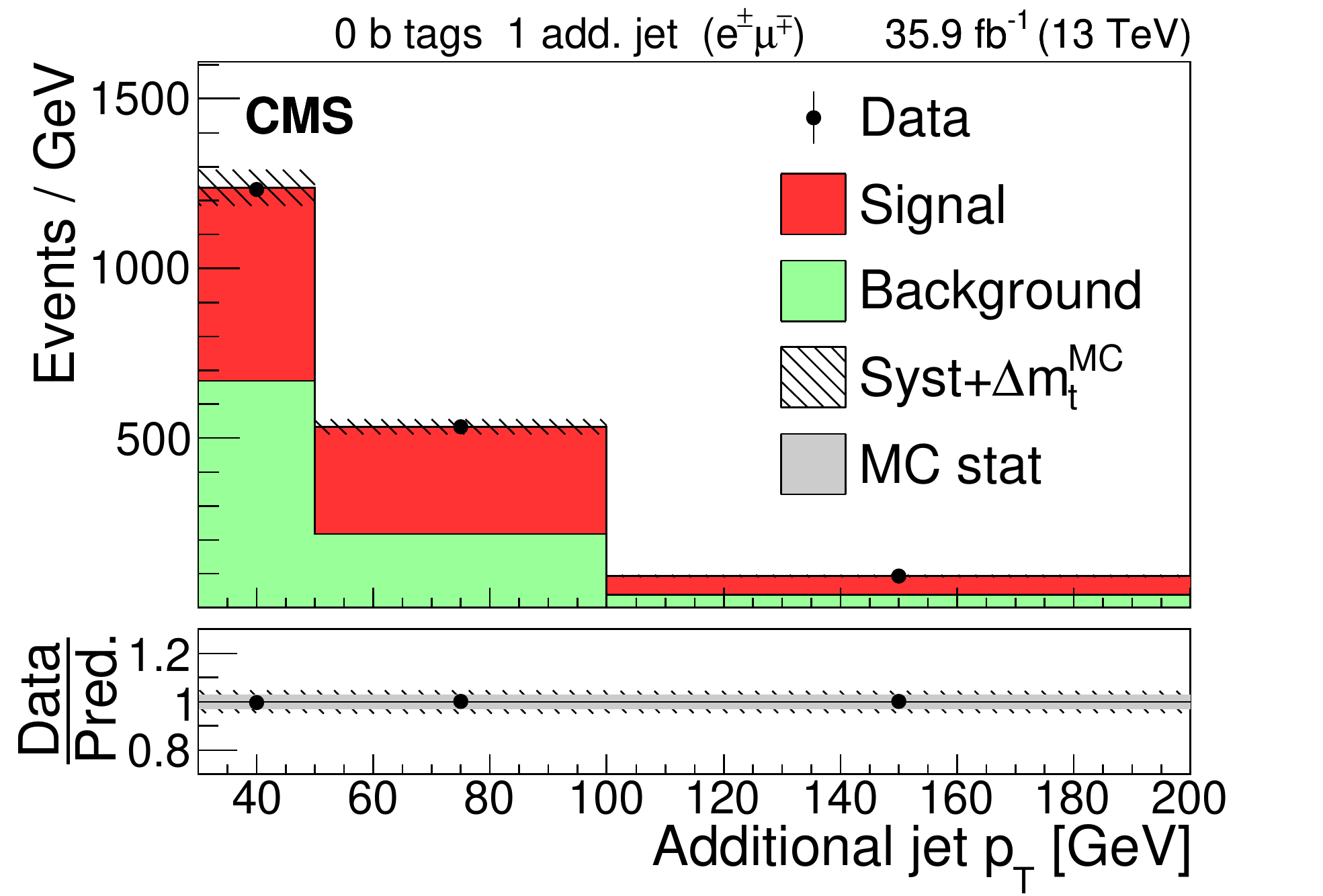}
    \includegraphics[width=0.325\textwidth]{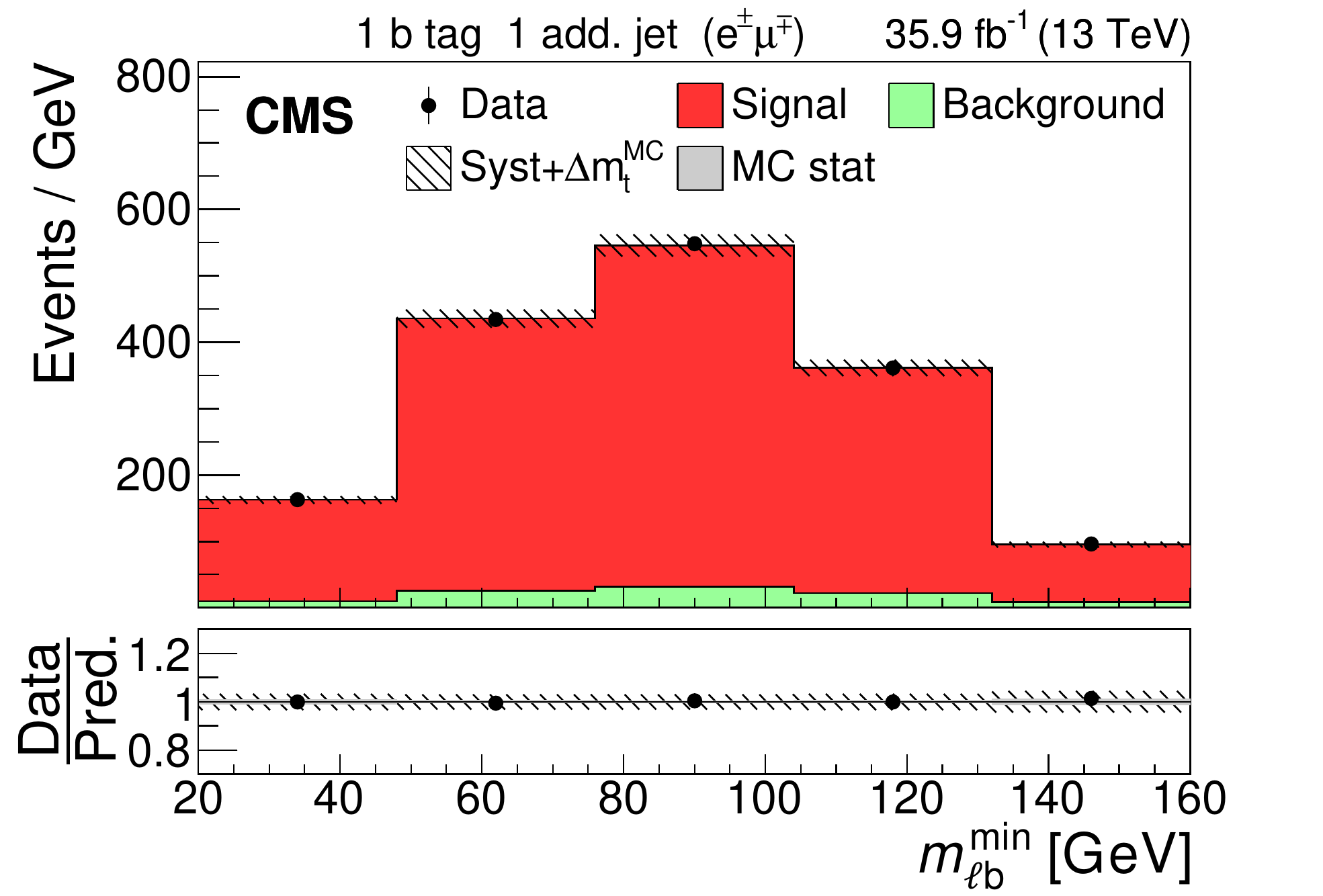}
    \includegraphics[width=0.325\textwidth]{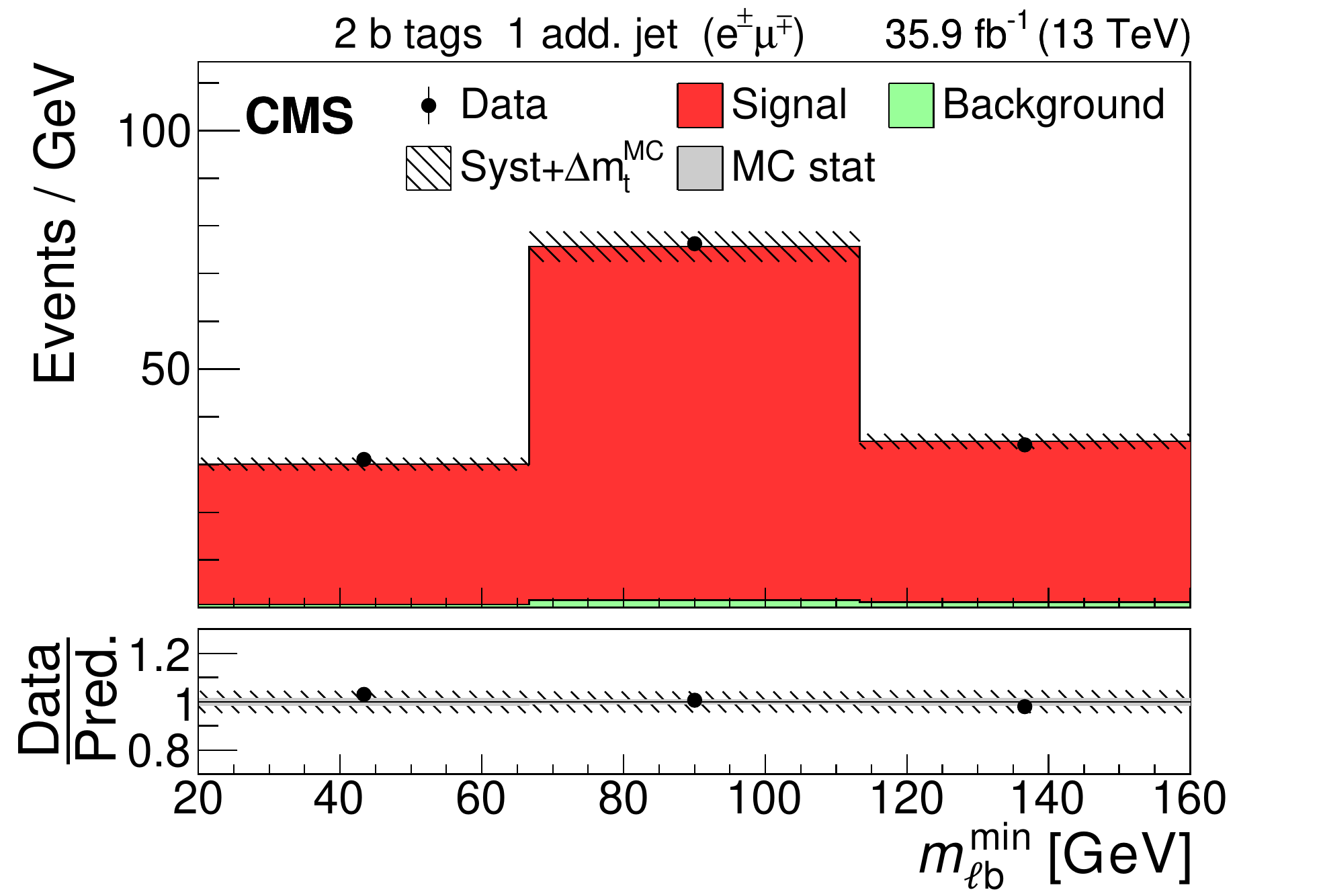}

    \includegraphics[width=0.325\textwidth]{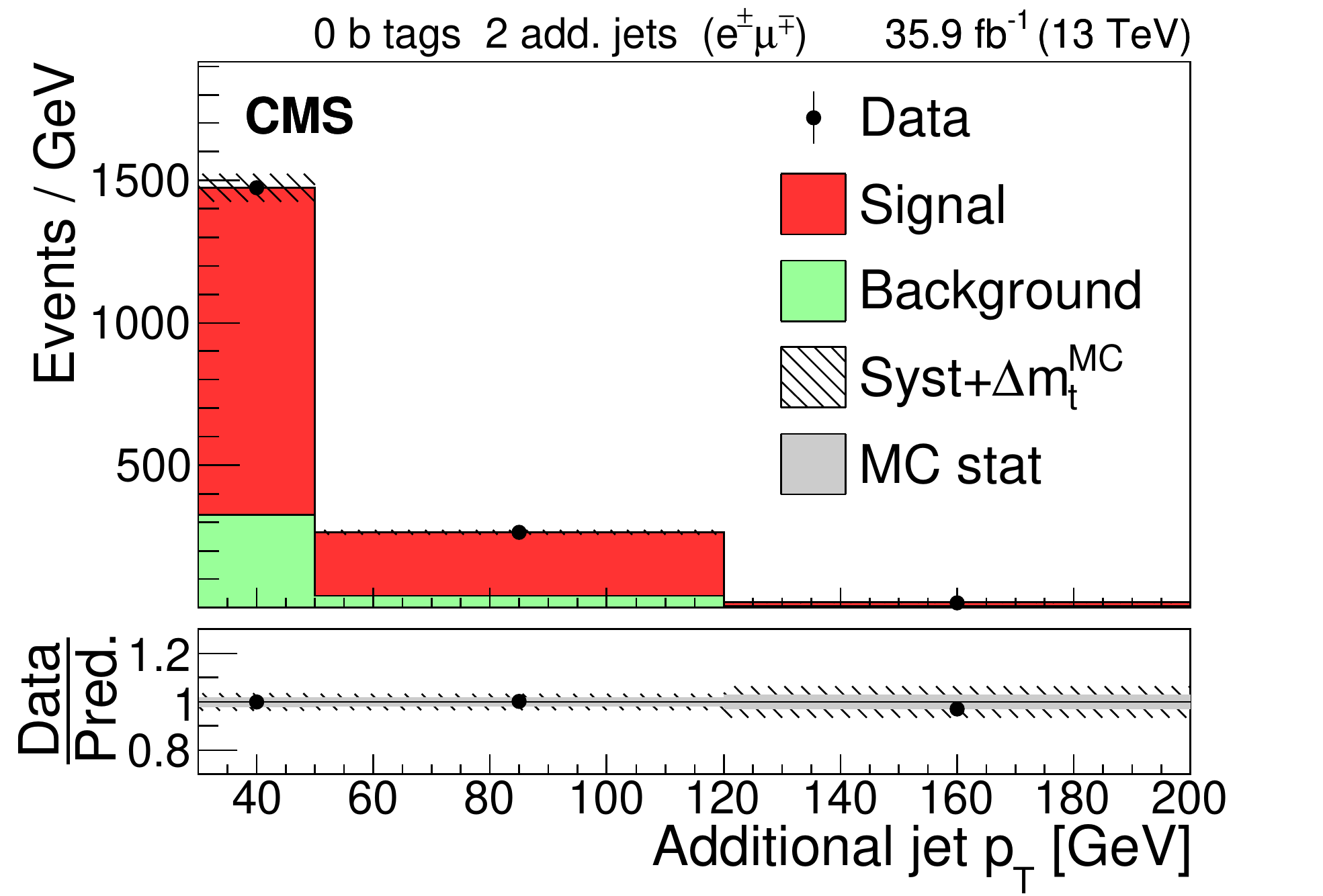}
    \includegraphics[width=0.325\textwidth]{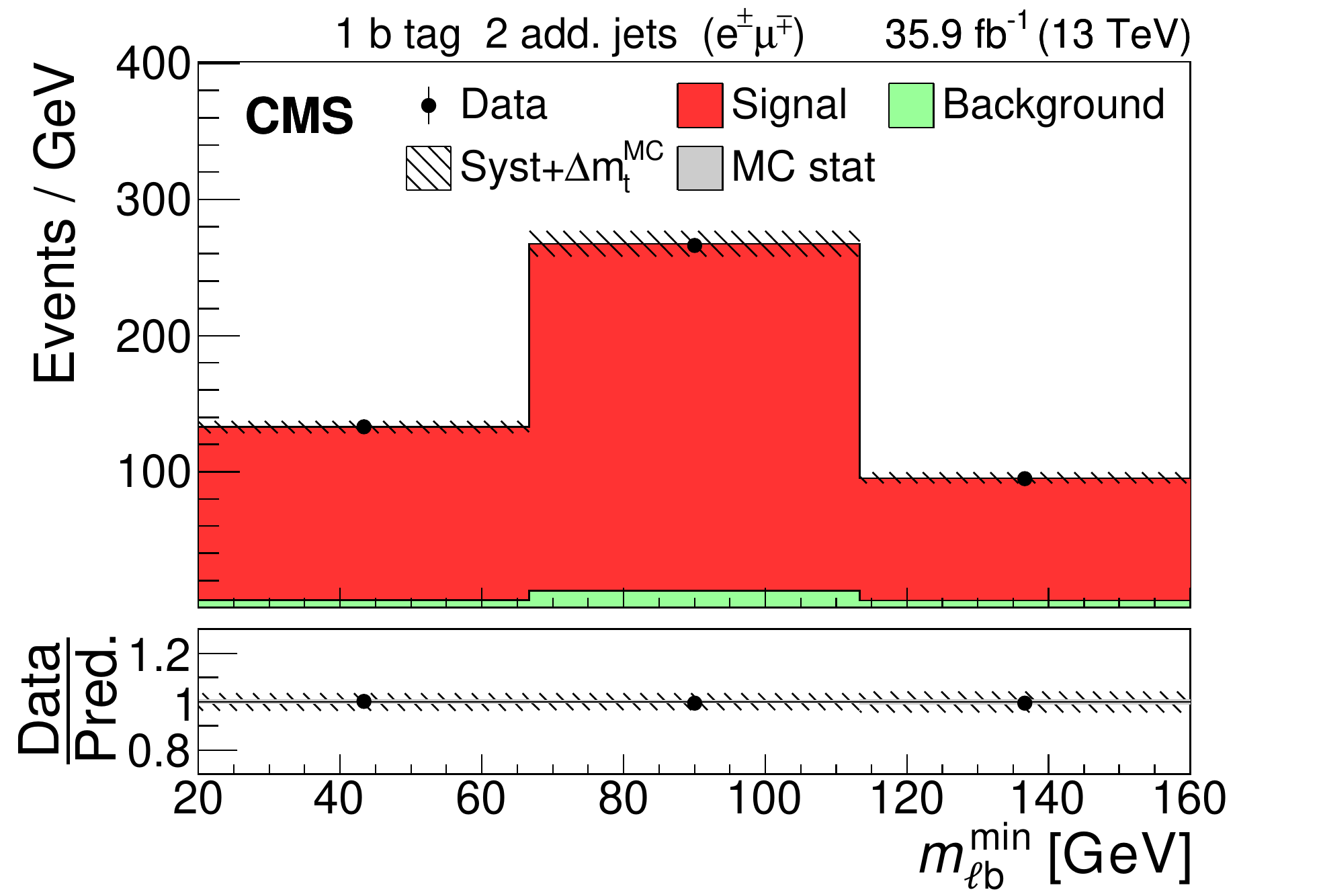}
    \includegraphics[width=0.325\textwidth]{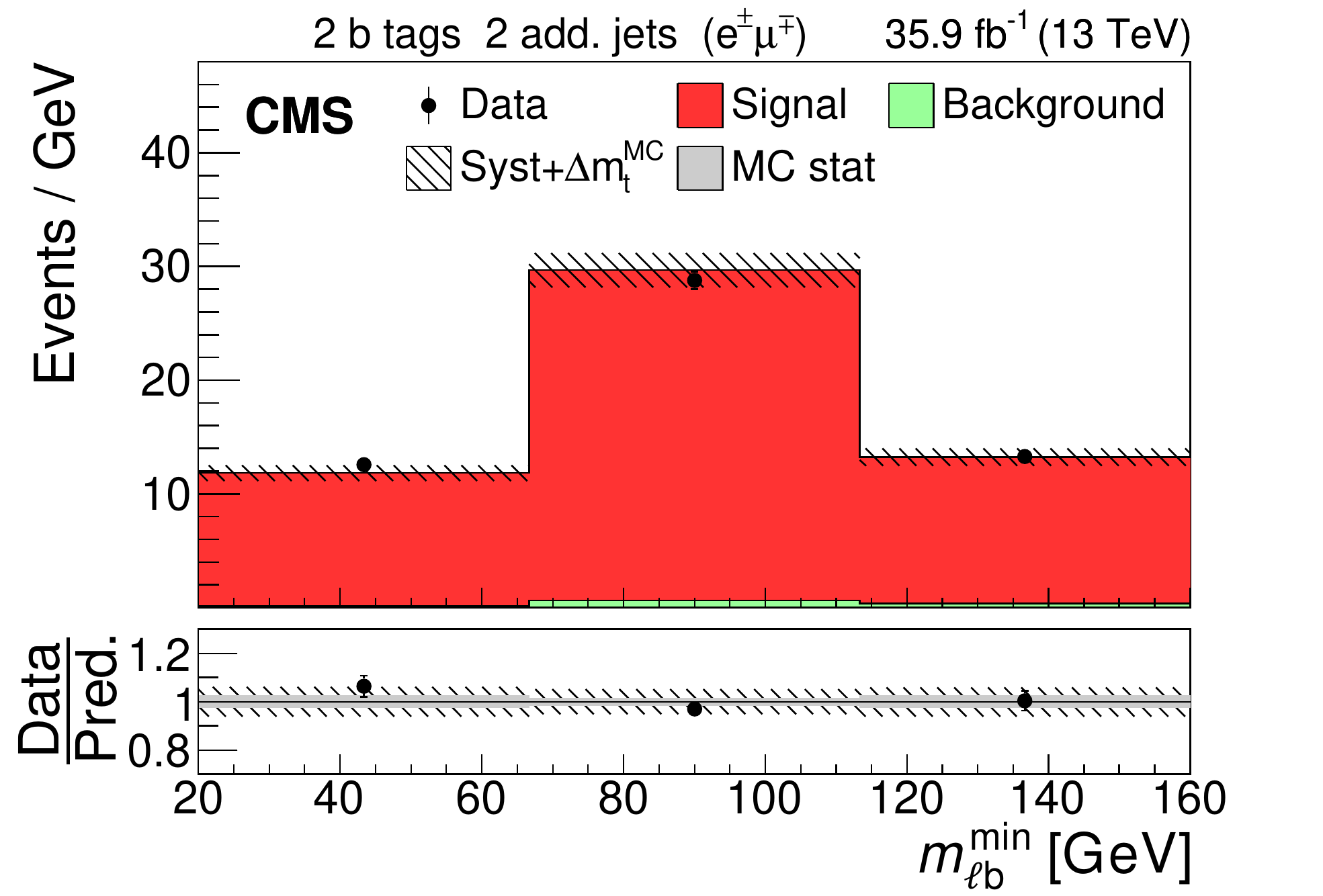}

    \includegraphics[width=0.325\textwidth]{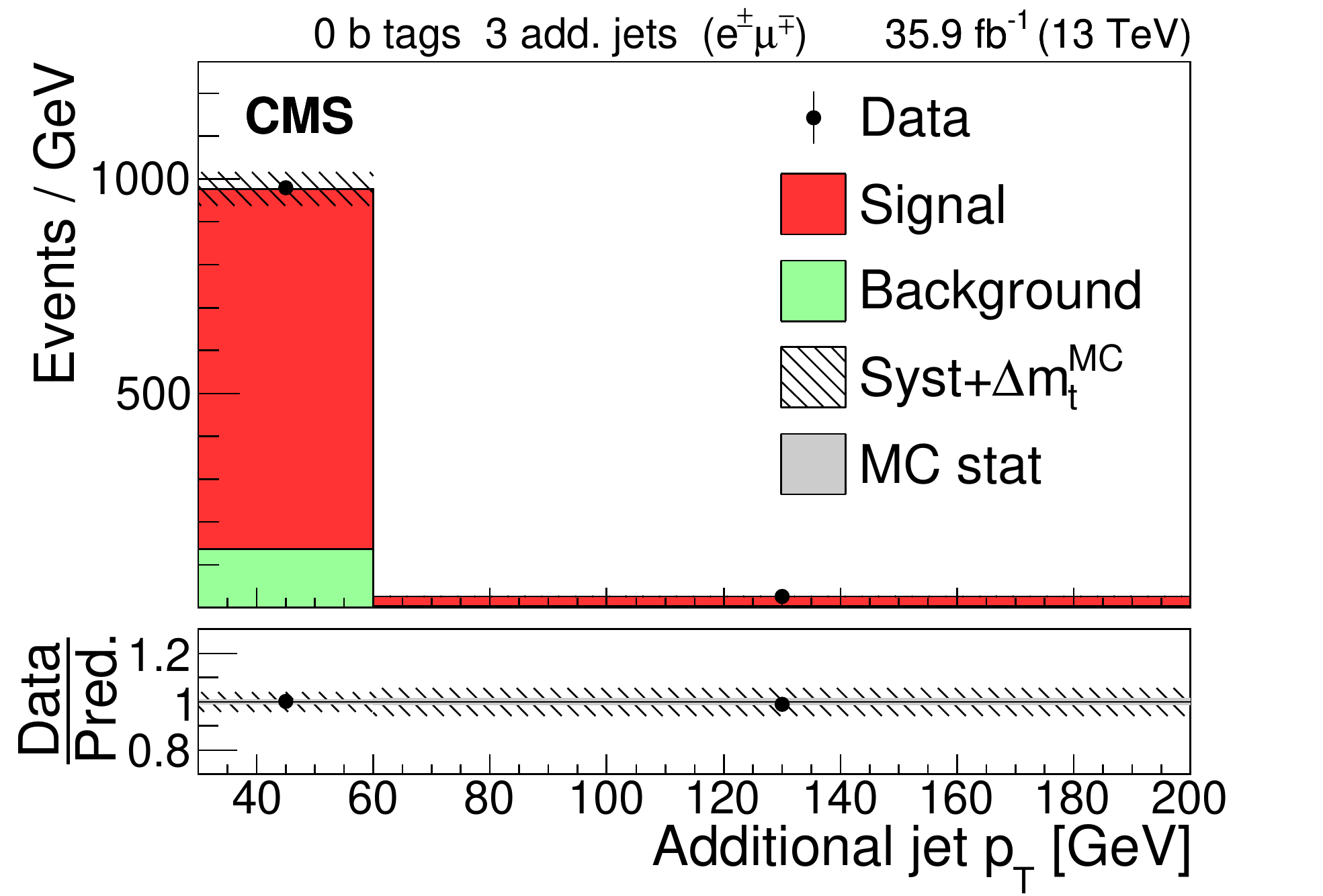}
    \includegraphics[width=0.325\textwidth]{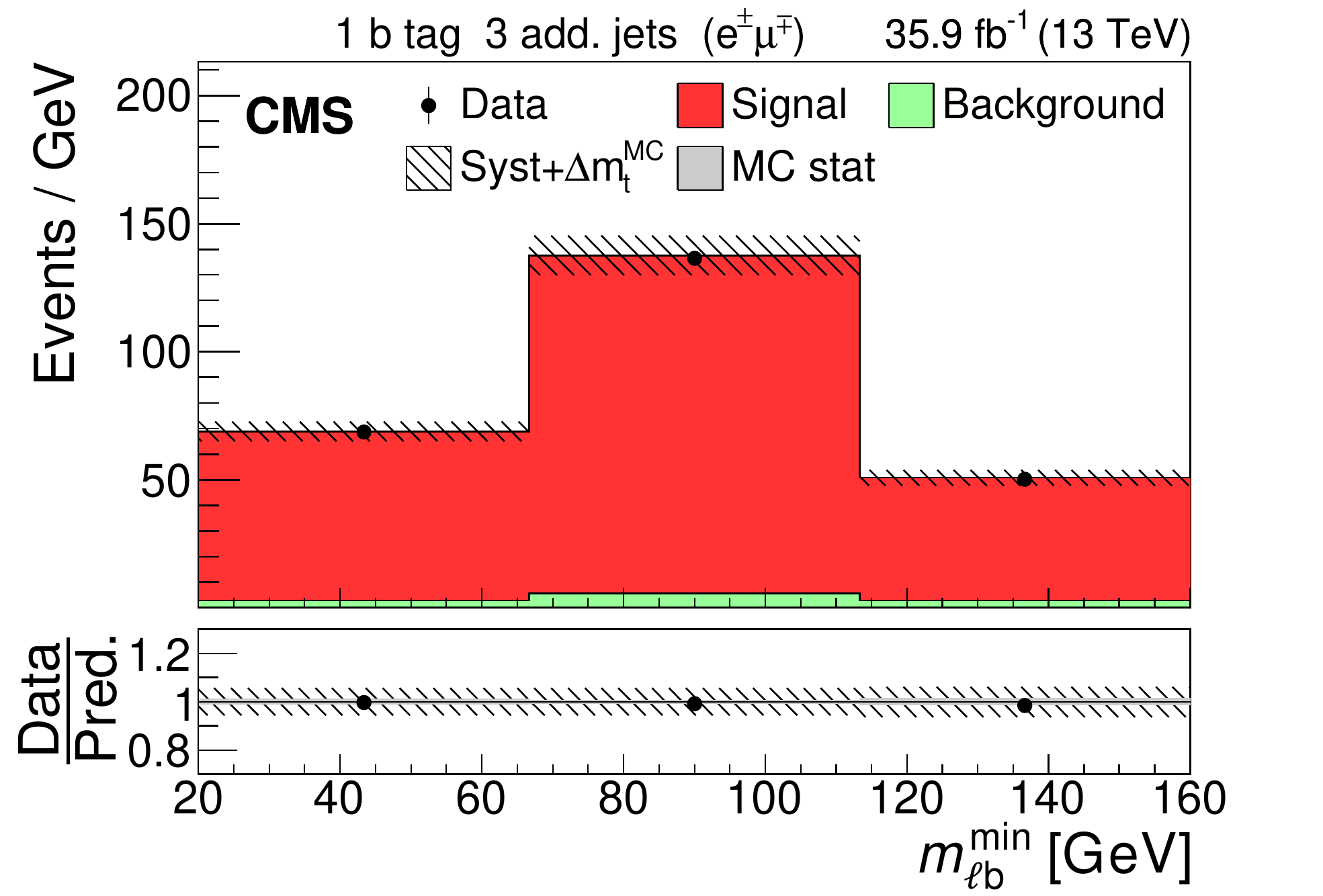}
    \includegraphics[width=0.325\textwidth]{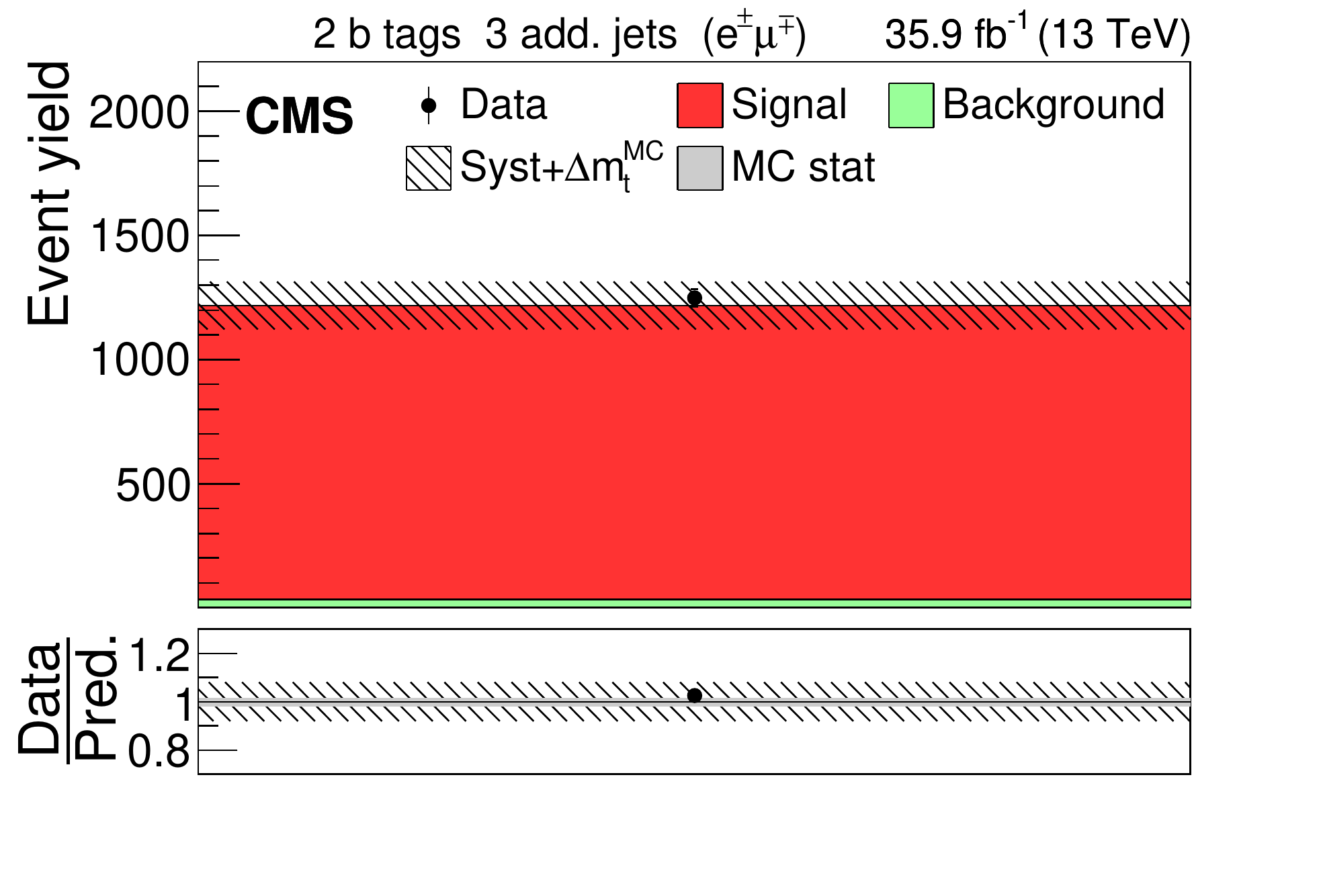}

\caption{Comparison of data (points) and post-fit distributions of the
expected signal and backgrounds from simulation (shaded
histograms) used in the simultaneous fit of \stt and \mtmc in the \emu channel.
In the left column events with zero or three or more \cPqb-tagged jets are shown. The middle (right) column shows events with exactly one (two) \cPqb-tagged jets. Events with zero, one, two, or three or more additional non-\cPqb-tagged jets are shown in the first, second, third, and fourth row, respectively.
The hatched bands correspond to the total uncertainty in the sum of the predicted yields and include the contribution from the top quark mass ($\Delta\mtmc$). The ratios of data to the sum of the predicted yields are shown in the lower panel of each figure. Here, the solid gray band represents the contribution of the statistical uncertainty.}

       \label{fig:lh_emu_outputdistr8_topmass}

\end{figure*}

Comparisons of the data and the prediction from the MC simulation before and after the fit are presented in Figs.~\ref{fig:lh_emu_inputdistr8_topmass} and~\ref{fig:lh_emu_outputdistr8_topmass}, respectively. Good agreement is found in both cases.

The result of the fit is found to be stable against the choice of the fit distributions, and the introduction of the \mlb distribution was confirmed not to alter the final result on \stt or the behaviour with respect to the nuisance parameters.
The procedure is calibrated by performing fits where data is replaced by simulations with different \mtmc hypotheses: full closure of the method is obtained and no additional correction is applied.
The effect of the statistical uncertainty in the simulation on the fit results is estimated as explained in Section~\ref{sec:fit} and is considered as an additional uncertainty. The results for \stt and \mtmc are
\begin{align*}
\stt & =  \resultxsectopmass, \\
\mtmc & =  \resulttopmassMC.
\end{align*}
The value for the cross section is in good agreement with the result obtained for a fixed value of $\mtmc = 172.5 \GeV$, reported in Section~\ref{sec:crosssection}. The correlation between the two parameters  is found to be \correlationMassXsec.

\begin{figure*}[htbp!]
  \centering
    \includegraphics[width=0.85\textwidth]{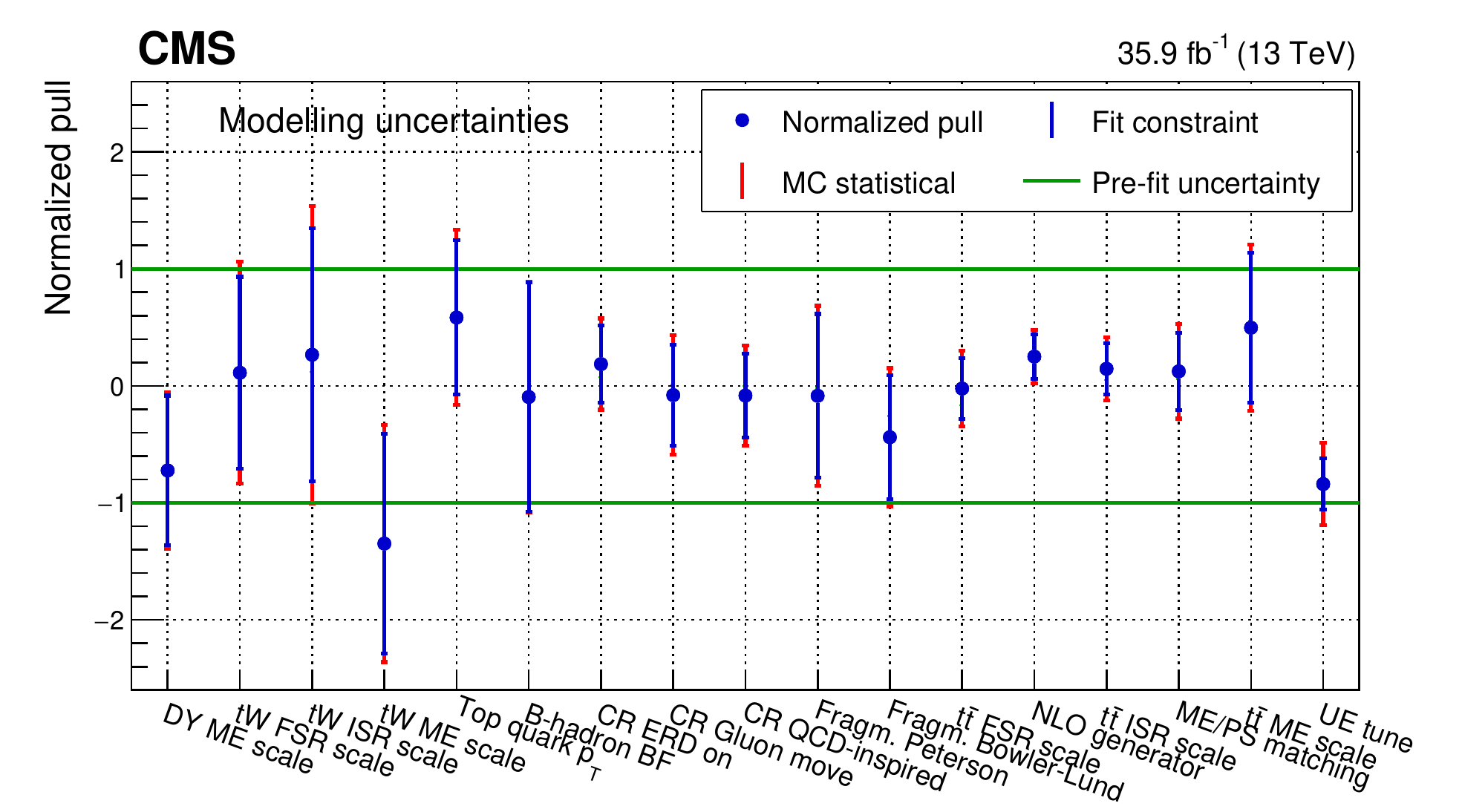}
\caption{Normalized pulls and constraints of the nuisance parameters related to the modelling uncertainties for the simultaneous fit of \stt and \mtmc. The markers denote the fitted value, while the inner vertical bars represent the constraint and the outer vertical bars denote the additional uncertainty as determined from pseudo-experiments. The constraint is defined as the ratio of the post-fit uncertainty to the pre-fit uncertainty of a given nuisance parameter, while the normalized pull is the difference between the post-fit and the pre-fit values of the nuisance parameter normalized to its pre-fit uncertainty. The horizontal lines at $\pm 1$ represent the pre-fit uncertainty.}
       \label{fig:topmass_pulls}
\end{figure*}

The results of the simultaneous fit to \stt and \mtmc are summarized in Tables~\ref{tab:fit_results_topmass_stt} and~\ref{tab:fit_results_topmass_mtmc}, respectively, together with the contribution of each systematic uncertainty to the total uncertainty.
Normalized pulls and constraints of the nuisance parameters related to modelling uncertainties are shown in Fig.~\ref{fig:topmass_pulls}.
The nuisance parameters displayed in this figure show similar trends to those in Fig.~\ref{fig:topxsec_pulls}, described above.
Here, the constraints on the nuisance parameters tend to be less stringent because only data in the \emu channel are used to determine the two parameters of interest, using mostly the \mlb spectra in place of the jet \pt distributions within the jet and \cPqb-tagged jet categories.

As a cross-check, a measurement of \mtmc is performed by fitting a single \mlb distribution containing all events with at least one \cPqb-tagged jet. The resulting value is \mtmc = \resulttopmassMCxcheck. Since the uncorrelated uncertainty with respect to the main result is estimated to be at least \uncorrelatedSyst, which is larger than the difference between the two measurements, the two results are in good agreement.

\begin{table}[hbtp!]
\centering \topcaption{\label{tab:fit_results_topmass_stt}
The same as Table~\ref{tab:lh_syst_sum}, but for the simultaneous fit of \stt and \mtmc.
}
\begin{tabular}{ l  c }
 Source & Uncertainty [\%] \\ \hline
 Trigger & ${0.4}$ \\
 Lepton ident./isolation & ${2.2}$ \\
 Muon momentum scale & ${0.2}$ \\
 Electron momentum scale & ${0.2}$ \\
 Jet energy scale & ${0.7}$ \\
 Jet energy resolution & ${0.5}$ \\
 \cPqb~tagging & ${0.3}$ \\
 Pileup & ${0.3}$ \\
 \ttbar ME scale & ${0.5}$ \\
 \tW ME scale & ${0.7}$ \\
 DY ME scale & ${0.2}$ \\
 NLO generator & ${1.2}$ \\
 PDF & ${1.1}$ \\
 \mtmc & ${0.4}$ \\
 Top quark \pt & ${0.5}$ \\
 ME/PS matching & ${0.2}$ \\
 UE tune & ${0.3}$ \\
 \ttbar ISR scale & ${0.4}$ \\
 \tW ISR scale & ${0.4}$ \\
 \ttbar FSR scale & ${1.1}$ \\
 \tW FSR scale & ${0.2}$ \\
 \cPqb~quark fragmentation & ${1.0}$ \\
 \cPqb~hadron BF & ${0.2}$ \\
 Colour reconnection & ${0.4}$ \\
 DY background & ${0.8}$ \\
 \tW background & ${1.1}$ \\
 Diboson background & ${0.3}$ \\
 \Wjets background & ${0.3}$ \\
 \ttbar background & ${0.2}$ \\
 Statistical & ${0.2}$ \\
 Integrated luminosity & ${2.5}$ \\
 MC statistical & ${1.2}$ \\
 \rule{0pt}{2.6ex}Total \sttvis uncertainty & ${4.2}$ \\[0.15cm]
 \multicolumn{2}{l}{Extrapolation uncertainties} \\ [0.05cm] \hline
  \rule{0pt}{2.3ex}\ttbar ME scale  & $\mp^{0.4}_{<0.1}$ \\
 \rule{0pt}{2.3ex}PDF  & $\pm^{0.8}_{0.6}$ \\
 \rule{0pt}{2.3ex}Top quark \pt  & $\pm^{0.2}_{0.3}$ \\
 \rule{0pt}{2.3ex}\ttbar ISR scale  & $\mp^{0.2}_{<0.1}$ \\
 \rule{0pt}{2.3ex}\ttbar FSR scale  & $\pm$0.1 \\
 UE tune  & $<$0.1 \\
 \rule{0pt}{2.3ex}\mtmc  & $\mp^{0.2}_{0.3}$\rule[-1.2ex]{0pt}{0pt}\\
 \rule{0pt}{2.5ex}Total \stt uncertainty & $^{+4.3}_{-4.2}$\rule[-1.2ex]{0pt}{0pt}\\
 \end{tabular}
 \end{table}

\begin{table}[hbtp!]
\centering \topcaption{\label{tab:fit_results_topmass_mtmc}
The absolute uncertainties in \mtmc and their sources, from the simultaneous fit of \stt and \mtmc.
The MC statistical uncertainty is determined separately. The individual uncertainties are given without their correlations, which are however accounted for in the total uncertainties.
}
\begin{tabular}{ l  c }
 Source & Uncertainty [{\GeVns}] \\ \hline
 Trigger & ${0.02}$ \\
 Lepton ident./isolation & ${0.02}$ \\
 Muon momentum scale & ${0.03}$ \\
 Electron momentum scale & ${0.10}$ \\
 Jet energy scale & ${0.57}$ \\
 Jet energy resolution & ${0.09}$ \\
 \cPqb~tagging & ${0.12}$ \\
 Pileup & ${0.09}$ \\
 \ttbar ME scale & ${0.18}$ \\
 \tW ME scale & ${0.02}$ \\
 DY ME scale & ${0.06}$ \\
 NLO generator & ${0.14}$ \\
 PDF & ${0.05}$ \\
 $\sigma_{\ttbar}$ & ${0.09}$ \\
 Top quark \pt & ${0.04}$ \\
 ME/PS matching & ${0.16}$ \\
 UE tune & ${0.03}$ \\
 \ttbar ISR scale & ${0.16}$ \\
 \tW ISR scale & ${0.02}$ \\
 \ttbar FSR scale & ${0.07}$ \\
 \tW FSR scale & ${0.02}$ \\
 \cPqb~quark fragmentation & ${0.11}$ \\
 \cPqb~hadron BF & ${0.07}$ \\
 Colour reconnection & ${0.17}$ \\
 DY background & ${0.24}$ \\
 \tW background & ${0.13}$ \\
 Diboson background & ${0.02}$ \\
 \Wjets background & ${0.04}$ \\
 \ttbar background & ${0.02}$ \\
 Statistical & ${0.14}$ \\
 MC statistical & ${0.36}$ \\
 \rule{0pt}{2.6ex}Total \mtmc uncertainty & $^{+0.68}_{-0.73}$\rule[-1.2ex]{0pt}{0pt}\\
 \end{tabular}
 \end{table}

\section{\texorpdfstring{Extraction of \mt and \asmz in the \msbar scheme}{Extraction of m(top) and alpha(s) in the MS bar scheme}}
\label{sec:alphas}

The cross section value obtained in the simultaneous fit to \stt and \mtmc is used to extract \asmz and \mt in the \msbar renormalization scheme. For this purpose, the measured and the predicted cross sections are compared via a \chisq minimization. The \chisq fit is performed using the open-source QCD analysis framework \xFitter~\cite{Alekhin:2014irh} and a \chisq definition from Ref.~\cite{Abramowicz:2015mha}. The method to determine \mt and \asmz is very similar to the one used in earlier CMS analyses to extract \asmz using jet cross section measurements, \eg in Ref.~\cite{Khachatryan:2016mlc}.

It is assumed that the measured \stt is not affected by non-SM physics. The SM theoretical prediction for \stt at NNLO~\cite{PhysRevLett.109.132001,Czakon:2012zr,Czakon:2012pz,Czakon:2013goa} is calculated using the \Hathor~2.0~\cite{Aliev:2010zk} program, interfaced with \xFitter. This is the only available calculation to date that provides the \mt definition in the \msbar scheme. The top quark mass in the \msbar scheme is denoted by \mtmt, following the convention of presenting the value of a running coupling at a fixed value. In the calculation, the renormalization and factorization scales, \mur and \muf, are set to \mtmt. These are varied by a factor of two up and down, independently, avoiding cases where \muf/\mur= 1/4 or 4, in order to estimate the uncertainty due to the missing higher-order corrections (referred to in the following as the scale variation uncertainty).

The values of \asmz and \mt cannot be determined simultaneously, since both parameters alter the predicted \stt in such a way that any variation of one parameter can be compensated by a variation of the other. In the presented analysis, the values of \mt and \asmz are therefore determined at fixed values of \asmz and \mt, respectively.

{\tolerance=3000
The four most recent PDF sets available~\cite{Buckley:2014ana} at NNLO are used: ABMP16nnlo~\cite{Alekhin:2017kpj}, CT14nnlo~\cite{Dulat:2015mca}, MMHT14nnlo~\cite{Harland-Lang:2014zoa}, and NNPDF3.1nnlo~\cite{Ball:2017nwa}. While CT14nnlo does not use any \ttbar data as input, the PDF sets ABMP16nnlo and MMHT14nnlo use measurements of inclusive \ttbar cross sections at the Tevatron and LHC, and NNPDF3.1nnlo makes use of all available inclusive and differential \ttbar cross section measurements. Using the currently available \ttbar measurements has only a marginal effect on a global PDF and \asmz fit~\cite{Dulat:2015mca, Alekhin:2017kpj}. The details of the PDFs relevant for this analysis are summarized in Table~\ref{tab:pdf_input}.
In the MMHT14nnlo, CT14nnlo, and NNPDF3.1nnlo PDFs, the value of \asmz is assumed to be 0.118. In ABMP16nnlo, \asmz is fitted simultaneously with the PDFs. The ABMP16nnlo PDF employs the \msbar scheme for the heavy-quark mass treatment in its determination. Similar to the value of \asmz, the value of \mtmt in the ABMP16nnlo set is obtained in a simultaneous fit with the PDFs. For the other PDFs, the values of \mtp are assumed, as listed in Table~\ref{tab:pdf_input}. Since the analysis is performed in the \msbar scheme, the assumed \mtp of each PDF is converted into \mtmt using the \RunDec~\cite{Chetyrkin:2000yt,Schmidt:2012az} code, according to the prescription by the corresponding PDF group.  \par}

\begin{table*}[htbp!]
\centering
\topcaption{\label{tab:pdf_input}
Values of the top quark pole mass \mtp and
strong coupling constant \asmz used in the different
PDF sets.  Also shown are the corresponding \mtmt values
obtained using the \RunDec~\cite{Chetyrkin:2000yt,Schmidt:2012az} conversion, the number of
loops in the conversion, and the \as range used to estimate
the PDF uncertainties.
}
  \begin{tabular}{  l  l  l  l  l }
    & ABMP16 & NNPDF3.1 & CT14 & MMHT14 \\ \cline{2-5}
    \mtp [{\GeVns}] & 170.37  & 172.5  &173.3  & 174.2  \\
    \RunDec loops & 3  & 2  & 2  & 3  \\
    \mtmt [{\GeVns}] & 160.86  & 162.56  &163.30  & 163.47  \\
    \asmz & 0.116 & 0.118 & 0.118 & 0.118\\
    \as range & 0.112$-$0.120 & 0.108$-$0.124 & 0.111$-$0.123& 0.108$-$0.128 \\
\end{tabular}

\end{table*}

For each used PDF set, a series of \asmz values is provided. The PDF uncertainties for all sets correspond to a 68\% confidence level (\CL), whereby the uncertainties in the CT14nnlo PDF set are scaled down from 95\% \CL.

Because of the strong correlation between \as and \mt in the prediction of \stt, for the \mt extraction, the value of \asmz in the theoretical prediction is set to that of the particular PDF set. Similarly, in the theoretical prediction of \stt used for the \asmz determination, the value of \mt is the one used in the PDF evaluation. The correlation of the values of \mtmt, \asmz, and the proton PDFs in the prediction of \stt is also studied.

\begin{figure*}[!htb]
\centering
\includegraphics[width=0.49\textwidth]{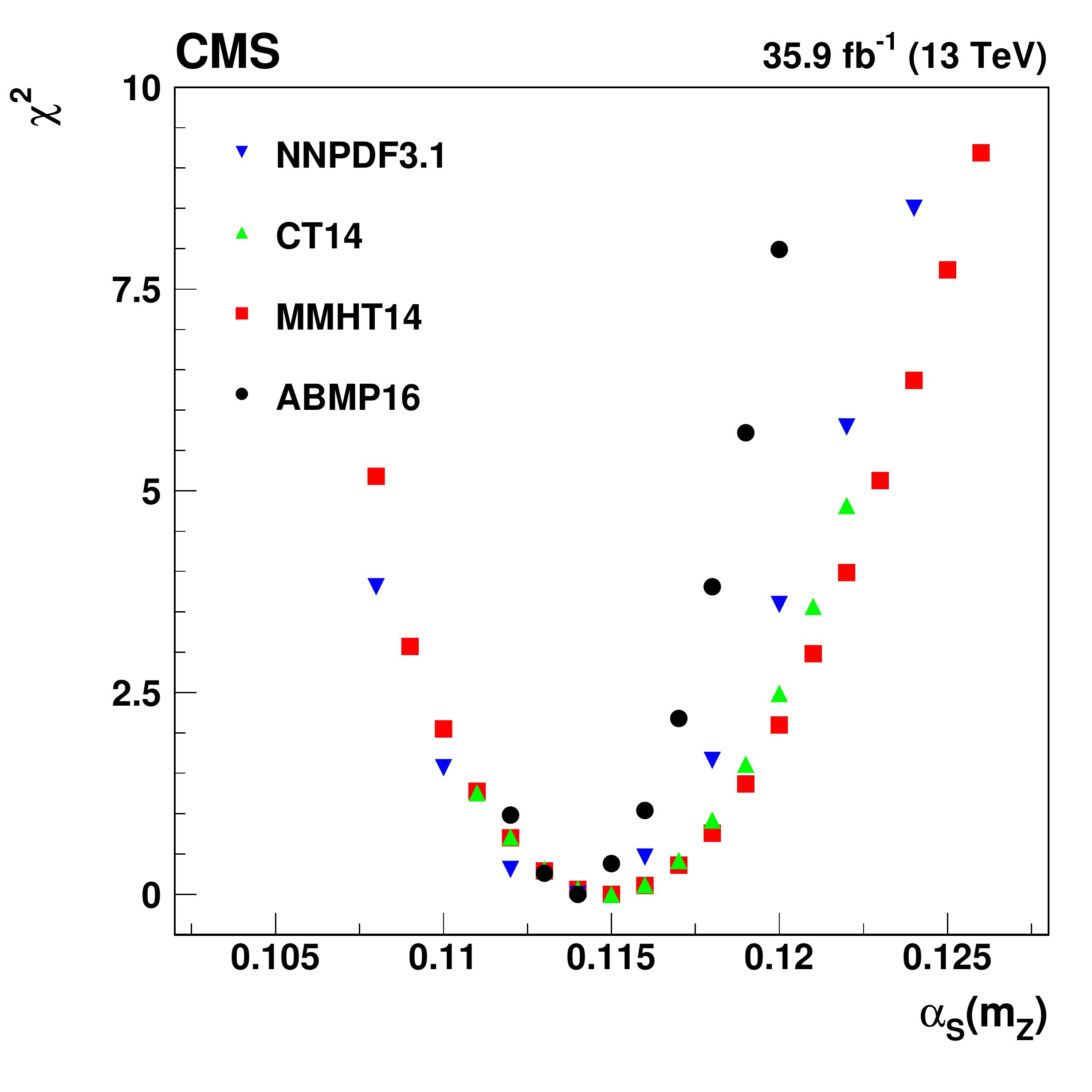}
\includegraphics[width=0.49\textwidth]{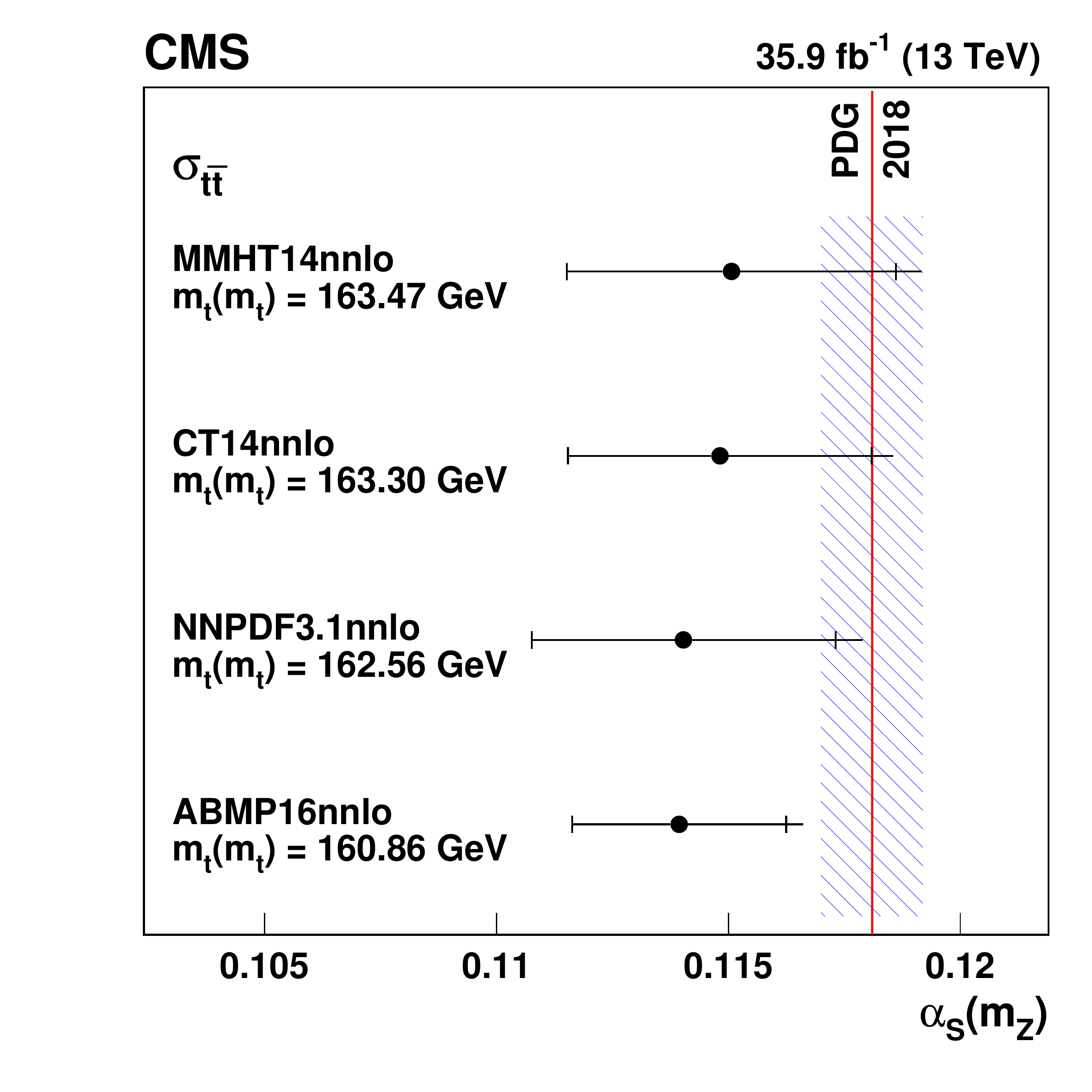}
\caption{
Left: \chisq versus \as obtained from the
comparison of the measured \stt value to the NNLO prediction in the \msbar scheme using different
PDFs (symbols of different styles).
Right: \asmz obtained from the comparison of the
measured \stt value to the theoretical prediction using
different PDF sets in the \msbar scheme. The corresponding
value of \mtmt is given for each PDF set. The inner horizontal
bars on the points represent the experimental and PDF
uncertainties added in quadrature. The outer horizontal bars show
the total uncertainties. The vertical line displays the
world-average \asmz value~\cite{PDG2018}, with the hatched band representing its uncertainty.}

\label{fig:as_scan}
\end{figure*}

{\tolerance=800
To extract the value of \asmz from \stt, the measured cross section is compared to the theoretical prediction, and for each \asmz member of each PDF set, the \chisq is evaluated. In the case of ABMP16nnlo and NNPDF3.1nnlo, the complete set of PDF uncertainties is provided for each member of the \asmz series and is accounted for in the analysis. The uncertainties in the CT14nnlo and MMHT14nnlo PDFs are evaluated only for the central \asmz value of 0.118 and are used for each \asmz variant in the fit. The optimal value of \asmz is subsequently determined from a parabolic fit of the form
\begin{equation}
\chisq(\as)=\chisq_\text{min}+\left(\frac{\as - \asmin}{\delta(\asmin)}\right)^2
\end{equation}
to the $\chisq (\as)$ values. Here, $\chisq_\text{min}$ is the \chisq value at $\as = \asmin$ and $\delta(\asmin)$ is
the fitted experimental uncertainty in \asmin, which also accounts for the PDF uncertainty. The $\chisq (\as) $ scan is illustrated in Fig.~\ref{fig:as_scan} for the PDF sets used,
demonstrating a clear parabolic behaviour.
To estimate the scale variation uncertainties, this procedure is repeated with \mur and \muf being varied, and the largest deviations of the
resulting values of \asmin from that of the central scale choice are considered as the corresponding uncertainties.
The values of the \asmz obtained using different PDFs are listed in Table~\ref{tab:as_pdf_ms} and shown in Fig.~\ref{fig:as_scan}. The uncertainties in the measured \stt and the PDF contribute about equally to the resulting \asmz uncertainty. \par}

\begin{table}[htbp!]
\centering
\topcaption{\label{tab:as_pdf_ms} Values of \asmz with their uncertainties
obtained from a comparison of the measured
\stt value to the NNLO prediction in the \msbar scheme
using different PDF sets.  The first uncertainty is the
combination of the experimental and PDF uncertainties, and the
second is from the variation of the renormalization and
factorization scales.}

  \begin{tabular}{  l  c}
\rule{0pt}{2ex}PDF set & \asmz \\ \hline
    \rule{0pt}{2.3ex}ABMP16   & 0.1139  $\pm$ 0.0023 (fit + PDF) $^{+0.0014}_{-0.0001}$ (scale) \\
    \rule{0pt}{2.3ex}NNPDF3.1 & 0.1140  $\pm$ 0.0033 (fit + PDF) $^{+0.0021}_{-0.0002}$ (scale) \\
    \rule{0pt}{2.3ex}CT14     & 0.1148  $\pm$ 0.0032 (fit + PDF) $^{+0.0018}_{-0.0002}$ (scale) \\
    \rule{0pt}{2.3ex}MMHT14   & 0.1151  $\pm$ 0.0035 (fit + PDF) $^{+0.0020}_{-0.0002}$ (scale)\rule[-1.2ex]{0pt}{0pt}\\
\end{tabular}
\end{table}

The values of \asmz obtained using different PDF sets are consistent among each other and are in agreement with the world-average value~\cite{PDG2018} within the uncertainties, although suggesting
a smaller value of \asmz. The value of \asmz is also in good agreement with the recent result of the analysis in Ref.~\cite{Andreev:2017vxu} of jet production in
deep-inelastic scattering using the NNLO calculation by the H1 experiment, and is of comparable precision.

The same procedure is used to extract \mtmt by fixing \asmz to the nominal value at which the used PDF is evaluated. The fit is performed by varying
\mtmt in a 5-\GeVns range around the central value used in each PDF.
The uncertainties related to the variation of \asmz in the PDF are estimated by repeating the fit using
the PDF eigenvectors with \asmz varied within its uncertainty, as provided by NNPDF3.1nnlo, MMHT2014nnlo, and CT14nnlo.
In the case of ABMP16nnlo, the  value of \asmz is a free parameter in the PDF fit and its uncertainty is implicitly included
in the ABMP16nnlo PDF uncertainty eigenvectors. The resulting \mtmt values are summarized in Table~\ref{tab:topmass_extraction_MS},
where the fit uncertainty corresponds to the precision of the \stt measurement.
The results obtained with different PDF sets are in agreement, although the ABMP16nnlo PDF set yields a systematically lower value. This difference
is expected and has its origin in a larger value of $\asmz = 0.118$ assumed in the NNPDF3.1, MMHT2014, and CT14 PDFs.

\begin{table}[htbp!]
\centering \topcaption{\label{tab:topmass_extraction_MS}Values of \mtmt obtained from the comparison of
the \stt measurement with the NNLO predictions using
different PDF sets.  The first uncertainty shown comes from the
experimental, PDF, and \asmz uncertainties, and the second from the variation in the
renormalization and factorization scales.}
\begin{tabular} {l l}
\rule{0pt}{2ex}PDF set & \multicolumn{1}{c}{\mtmt~[{\GeVns}]} \\ \hline
  \rule{0pt}{2.3ex}ABMP16   & 161.6 $\pm$ 1.6 (fit + PDF + \as) $^{+0.1}_{-1.0}$ (scale)\\
  \rule{0pt}{2.3ex}NNPDF3.1 & 164.5 $\pm$ 1.6 (fit + PDF + \as) $^{+0.1}_{-1.0}$ (scale) \\
  \rule{0pt}{2.3ex}CT14     & 165.0 $\pm$ 1.8 (fit + PDF + \as) $^{+0.1}_{-1.0}$ (scale) \\
  \rule{0pt}{2.3ex}MMHT14   & 164.9 $\pm$ 1.8 (fit + PDF + \as) $^{+0.1}_{-1.1}$ (scale)\rule[-1.2ex]{0pt}{0pt}\\
\end{tabular}
\end{table}

\begin{figure}[htbp!]
  \centering
    \includegraphics[width=0.49\textwidth]{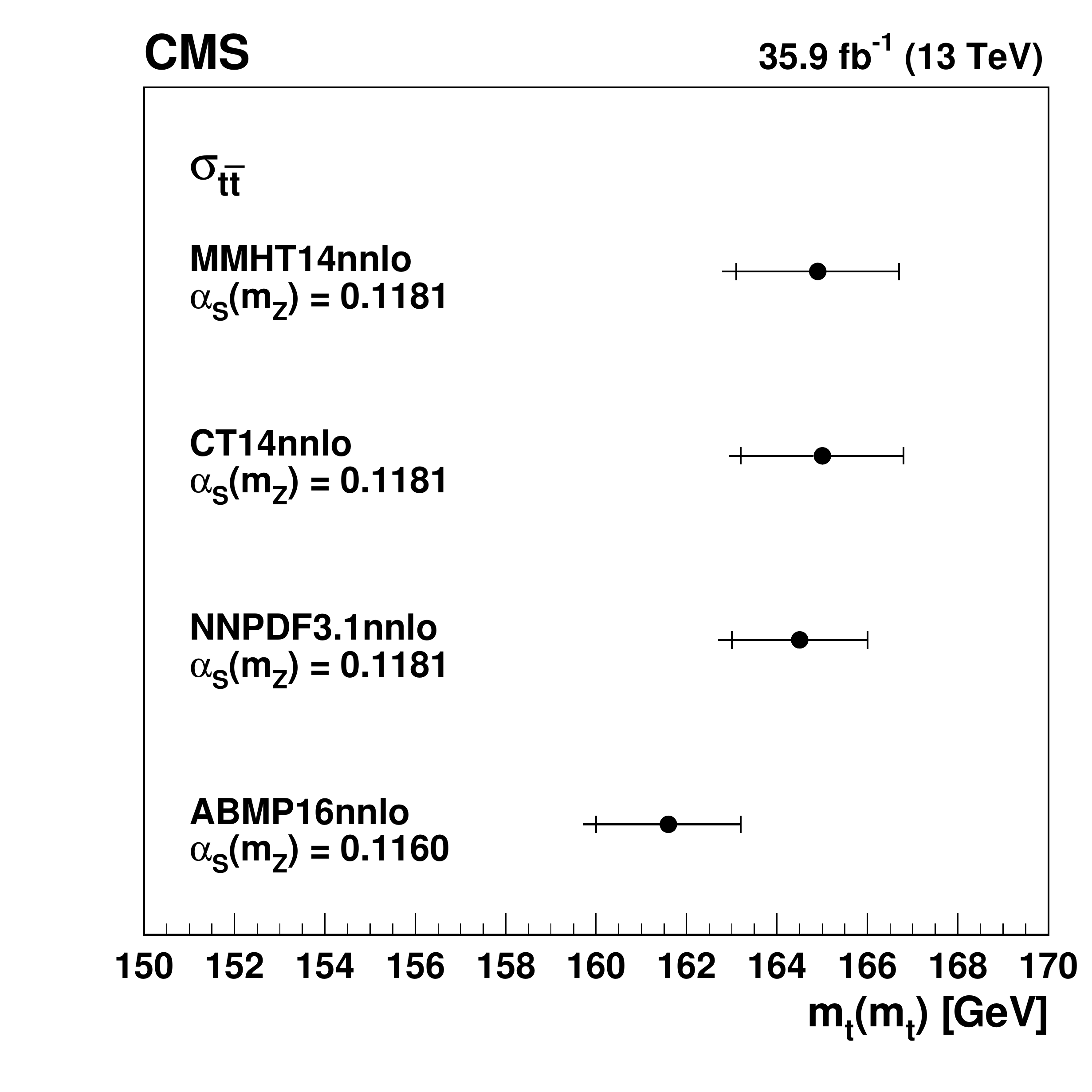}
\caption{Values of \mtmt obtained from comparing the
\stt measurement to the theoretical NNLO predictions
using different PDF sets.  The inner horizontal bars on the
points represent the quadratic sum of the experimental, PDF, and
\asmz uncertainties, while the outer
horizontal bars give the total uncertainties.
       \label{fig:mass_sum}}
\end{figure}

The values of \mtmt are in agreement with those originally used in the evaluation of each PDF set. The results are shown in Fig.~\ref{fig:mass_sum} for the four
different PDFs used.

\begin{figure}[htbp!]
\centering
\includegraphics[width=0.49\textwidth]{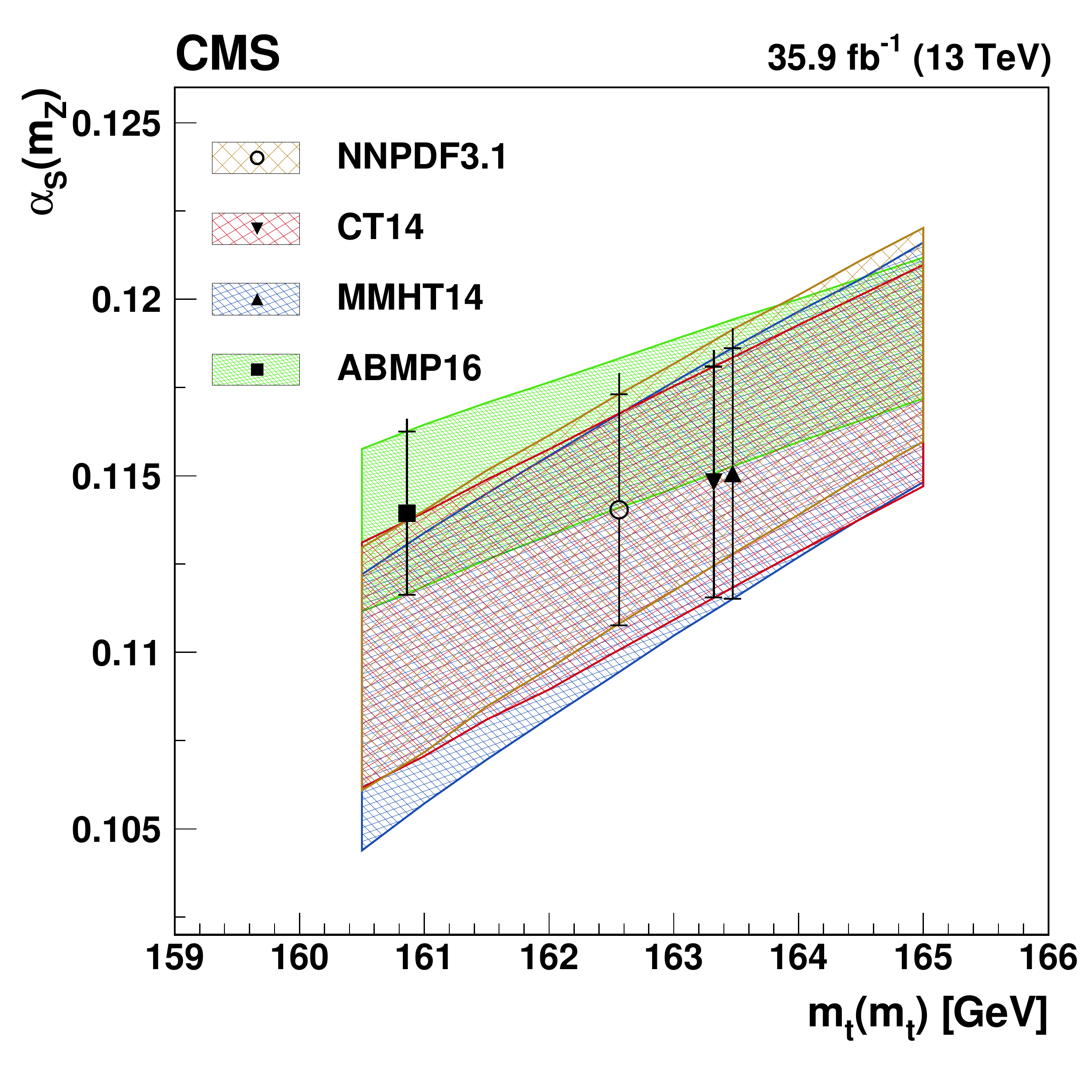}
\caption{Values of \asmz obtained
in the comparison of the \stt measurement to the NNLO
prediction using different PDFs, as a function of the \mtmt
value used in the theoretical calculation. The results from
using the different PDFs are shown by the bands with different
shadings, with the band width corresponding to the quadratic sum
of the experimental and PDF uncertainties in \asmz.  The
resulting measured values of \asmz are shown by the
different style points at the \mtmt values used for each PDF.
The inner vertical bars on the points represent the quadratic sum
of the experimental and PDF uncertainties in \asmz, while
the outer vertical bars show the total uncertainties.
}
\label{fig:mass_scan}
\end{figure}

The dependence of the \asmz result on the assumption on \mtmt is investigated for each PDF by
performing the $\chisq (\as)$ scan for ten values of \mtmt varying from 160.5 to 165.0\GeV. A linear dependence is observed, as shown in Fig.~\ref{fig:mass_scan}.

\section{Extraction of \texorpdfstring{\mt}{m top} in the pole mass scheme}
\label{sec:mtp}
The extraction of \mt is repeated in the pole mass scheme using the \Toppp~2.0 program~\cite{Czakon:2011xx}, which employs the calculation of \stt at NNLO, improved by the NNLL soft-gluon resummation. The results are summarized in Table~\ref{tab:topmass_extraction_pole}. The scale variation uncertainties are estimated in the same way as in the case of the \mtmt extraction. These uncertainties are larger than those determined in the \msbar scheme. This is because of the better convergence of the perturbative series when using the \msbar renormalization scheme in the calculation of \stt.

\begin{table}[htbp!]
\centering \topcaption{\label{tab:topmass_extraction_pole}
Values of \mtp obtained by comparing the
\stt measurement with predictions at NNLO+NNLL using
different PDF sets.
}
\begin{tabular} {l l}
\rule{0pt}{2ex}PDF set & \multicolumn{1}{c}{\mtp~[{\GeVns}]} \\ \hline
  \rule{0pt}{2.3ex}ABMP16   & 169.9 $\pm$ 1.8~(fit + PDF + \as) $^{+0.8}_{-1.2}$ (scale)\\
  \rule{0pt}{2.3ex}NNPDF3.1 & 173.2 $\pm$ 1.9~(fit + PDF + \as) $^{+0.9}_{-1.3}$ (scale) \\
  \rule{0pt}{2.3ex}CT14     & 173.7 $\pm$ 2.0~(fit + PDF + \as) $^{+0.9}_{-1.4}$ (scale) \\
  \rule{0pt}{2.3ex}MMHT14   & 173.6 $\pm$ 1.9~(fit + PDF + \as) $^{+0.9}_{-1.4}$ (scale)\rule[-1.2ex]{0pt}{0pt}\\
\end{tabular}
\end{table}

\section{Summary}
\label{sec:conclusions}

A measurement of the top quark-antiquark pair production cross section \stt by the CMS Collaboration in proton-proton collisions at a centre-of-mass energy of 13\TeV is presented, corresponding to an integrated luminosity of \lumiv. Assuming a top quark mass in the simulation of $\mtmc = 172.5 \GeV$, a visible cross section is measured in the fiducial region using dilepton events (\emu, \mumu, \ee) and then extrapolated to the full phase space.
The total \ttbar production cross section is found to be $\stt = \resultxsecmain$. The measurement is in good agreement with the theoretical prediction calculated to next-to-next-to-leading order in perturbative QCD, including soft-gluon resummation to next-to-next-to-leading logarithm.

{\tolerance=900
The measurement is repeated including the top quark mass in the \POWHEG simulation as an additional free parameter in the fit. The sensitivity to \mtmc is maximized by fitting the minimum invariant mass found when combining the charged leptons with the \cPqb~jets in an event. This yields a cross section  of $\stt = \resultxsectopmass$ and a value of $\mtmc = \resulttopmassMC$, in good agreement with previous measurements. The value of \stt obtained in the simultaneous fit is further used to extract the values of the top quark mass and the strong coupling constant at next-to-next-to-leading order in the minimal subtraction renormalization scheme, as well as the value of the top quark pole mass for different sets of parton distribution functions. \par
}

\begin{acknowledgments}
We congratulate our colleagues in the CERN accelerator departments for the excellent performance of the LHC and thank the technical and administrative staffs at CERN and at other CMS institutes for their contributions to the success of the CMS effort. In addition, we gratefully acknowledge the computing centres and personnel of the Worldwide LHC Computing Grid for delivering so effectively the computing infrastructure essential to our analyses. Finally, we acknowledge the enduring support for the construction and operation of the LHC and the CMS detector provided by the following funding agencies: BMBWF and FWF (Austria); FNRS and FWO (Belgium); CNPq, CAPES, FAPERJ, FAPERGS, and FAPESP (Brazil); MES (Bulgaria); CERN; CAS, MoST, and NSFC (China); COLCIENCIAS (Colombia); MSES and CSF (Croatia); RPF (Cyprus); SENESCYT (Ecuador); MoER, ERC IUT, and ERDF (Estonia); Academy of Finland, MEC, and HIP (Finland); CEA and CNRS/IN2P3 (France); BMBF, DFG, and HGF (Germany); GSRT (Greece); NKFIA (Hungary); DAE and DST (India); IPM (Iran); SFI (Ireland); INFN (Italy); MSIP and NRF (Republic of Korea); MES (Latvia); LAS (Lithuania); MOE and UM (Malaysia); BUAP, CINVESTAV, CONACYT, LNS, SEP, and UASLP-FAI (Mexico); MOS (Montenegro); MBIE (New Zealand); PAEC (Pakistan); MSHE and NSC (Poland); FCT (Portugal); JINR (Dubna); MON, RosAtom, RAS, RFBR, and NRC KI (Russia); MESTD (Serbia); SEIDI, CPAN, PCTI, and FEDER (Spain); MOSTR (Sri Lanka); Swiss Funding Agencies (Switzerland); MST (Taipei); ThEPCenter, IPST, STAR, and NSTDA (Thailand); TUBITAK and TAEK (Turkey); NASU and SFFR (Ukraine); STFC (United Kingdom); DOE and NSF (USA).

\hyphenation{Rachada-pisek} Individuals have received support from the Marie-Curie programme and the European Research Council and Horizon 2020 Grant, contract No. 675440 (European Union); the Leventis Foundation; the A.P.\ Sloan Foundation; the Alexander von Humboldt Foundation; the Belgian Federal Science Policy Office; the Fonds pour la Formation \`a la Recherche dans l'Industrie et dans l'Agriculture (FRIA-Belgium); the Agentschap voor Innovatie door Wetenschap en Technologie (IWT-Belgium); the F.R.S.-FNRS and FWO (Belgium) under the ``Excellence of Science -- EOS" -- be.h project n.\ 30820817; the Ministry of Education, Youth and Sports (MEYS) of the Czech Republic; the Lend\"ulet (``Momentum") Programme and the J\'anos Bolyai Research Scholarship of the Hungarian Academy of Sciences, the New National Excellence Program \'UNKP, the NKFIA research grants 123842, 123959, 124845, 124850, and 125105 (Hungary); the Council of Science and Industrial Research, India; the HOMING PLUS programme of the Foundation for Polish Science, cofinanced from European Union, Regional Development Fund, the Mobility Plus programme of the Ministry of Science and Higher Education, the National Science Center (Poland), contracts Harmonia 2014/14/M/ST2/00428, Opus 2014/13/B/ST2/02543, 2014/15/B/ST2/03998, and 2015/19/B/ST2/02861, Sonata-bis 2012/07/E/ST2/01406; the National Priorities Research Program by Qatar National Research Fund; the Programa Estatal de Fomento de la Investigaci{\'o}n Cient{\'i}fica y T{\'e}cnica de Excelencia Mar\'{\i}a de Maeztu, grant MDM-2015-0509 and the Programa Severo Ochoa del Principado de Asturias; the Thalis and Aristeia programmes cofinanced by EU-ESF and the Greek NSRF; the Rachadapisek Sompot Fund for Postdoctoral Fellowship, Chulalongkorn University and the Chulalongkorn Academic into Its 2nd Century Project Advancement Project (Thailand); the Welch Foundation, contract C-1845; and the Weston Havens Foundation (USA). 
\end{acknowledgments}

\bibliography{auto_generated}

\cleardoublepage \appendix\section{The CMS Collaboration \label{app:collab}}\begin{sloppypar}\hyphenpenalty=5000\widowpenalty=500\clubpenalty=5000\vskip\cmsinstskip
\textbf{Yerevan Physics Institute, Yerevan, Armenia}\\*[0pt]
A.M.~Sirunyan, A.~Tumasyan
\vskip\cmsinstskip
\textbf{Institut f\"{u}r Hochenergiephysik, Wien, Austria}\\*[0pt]
W.~Adam, F.~Ambrogi, E.~Asilar, T.~Bergauer, J.~Brandstetter, M.~Dragicevic, J.~Er\"{o}, A.~Escalante~Del~Valle, M.~Flechl, R.~Fr\"{u}hwirth\cmsAuthorMark{1}, V.M.~Ghete, J.~Hrubec, M.~Jeitler\cmsAuthorMark{1}, N.~Krammer, I.~Kr\"{a}tschmer, D.~Liko, T.~Madlener, I.~Mikulec, N.~Rad, H.~Rohringer, J.~Schieck\cmsAuthorMark{1}, R.~Sch\"{o}fbeck, M.~Spanring, D.~Spitzbart, W.~Waltenberger, J.~Wittmann, C.-E.~Wulz\cmsAuthorMark{1}, M.~Zarucki
% ...
\vskip\cmsinstskip
% ...
1:  Also at Vienna University of Technology, Vienna, Austria\\
% ...
\end{sloppypar}
\end{document}